\documentclass[11pt]{article}
\usepackage{amsmath,amssymb}
\usepackage{framed}

\usepackage{graphicx}

\usepackage{amsmath,latexsym}

\begin{document}

\setcounter{page}{1}

\pagestyle{plain}

\begin{center}
\Large{\bf Gauss-Bonnet Inflation after Planck2018 }\\
\small \vspace{1cm}  { Narges
Rashidi}$^{}$\footnote{n.rashidi@umz.ac.ir}\quad and \quad {
Kourosh Nozari}$^{}$\footnote{knozari@umz.ac.ir} \\
\vspace{0.5cm} $^{}$Department of Theoretical Physics, Faculty of Basic
Sciences,
University of Mazandaran,\\
P. O. Box 47416-95447, Babolsar, IRAN\\
$^{}$ Research Institute for Astronomy and Astrophysics of Maragha (RIAAM),\\
P. O. Box 55134-441, Maragha, Iran\\
\end{center}

\begin{abstract}
We study the primordial perturbations and reheating process in the
models where the Gauss-Bonnet term is non-minimally coupled to the
canonical and non-canonical (DBI and tachyon) scalar fields. We
consider several potentials and Gauss-Bonnet coupling terms as
power-law, dilaton-like, $\cosh$-type, E-model and T-model. To
seek the observational viability of these models, we study the
scalar perturbations numerically and compare the results with the
Planck2018 TT, TE, EE+lowE+lensing+BK14+BAO joint data at $68\%$
CL and $95\%$ CL. We also study the tensor perturbations in
confrontation with the Planck2018 TT, TE,
EE+lowE+lensing+BK14+BAO+ LIGO$\&$Virgo2016 joint data at $68\%$
CL and $95\%$ CL. In this regard, we obtain some constraints on
the Gauss-Bonnet coupling parameter $\beta$. Another important
process in the early universe is the reheating phase after
inflation which is necessary to reheat the universe for subsequent
evolution. In this regard, we study the reheating process in these
models and find some expressions for the e-folds number and
temperature during that era. Considering that from Planck
TT,TE,EE+lowEB+lensing data and BICEP2/Keck Array 2014, based on
the $\Lambda$CDM$+r+\frac{dn_{s}}{d\ln k}$ model, we have
$n_{s}=0.9658\pm 0.0038$ and $r<0.072$, we obtain some constraints
on the e-folds number and temperature. From the values of the
e-folds number and the effective equation of state and also the
observationally viable value of the scalar spectral index, we
explore the capability of the models in explaining the reheating phase. 
\end{abstract}
\newpage

\section{Introduction}

One simplest way to solve some problems of the standard model of
cosmology is to consider a single canonical scalar field
(inflaton) with a flat potential leading to the slow-roll of the
inflaton. The slow-rolling of the inflaton causes enough
exponential expansion of the early universe. The primordial
perturbations in this single field model would have an adiabatic,
scale invariant and gaussian dominant
modes~\cite{Gut81,Lin82,Alb82,Lin90,Lid00a,Lid97,Rio02,Lyt09,Mal03}.
However, some cosmologists are interested in the extended
inflation models predicting the non-Gaussian distributed
perturbations\\~\cite{Mal03,Bar04,Che10,Fel11a,Fel11b,Noz13a,Noz15,NoJ11,NoJ18}.

Thinking over the very early time in the history of the universe,
approaching the Planck scale, and studying that epoch, it seems
necessary to incorporate some quantum corrections into the
Einstein gravity. The quantum theory of gravity at the low-energy
limit leads to the Einstein theory of gravity~\cite{Bur04}. There
is this belief that, as a promising candidate for the quantum
gravity, we can consider the string theory. To import the quantum
effects of gravity by using the higher-order curvature correction
to the gravitational action, the string theory suggests to
consider the Gauss-Bonnet (GB) term~\cite{Gro87}. This term is a
quadratic term defined by
\begin{equation}\label{eq1}
{\cal{L}}_{GB}=R_{abcd}\,R^{abcd}-4R_{ab}\,R^{ab}+R^{2}\,
\end{equation}
which is part of Lovelock's theorem~\cite{Lov71} and its role in
the dynamics of the early universe is
significant~\cite{Zwi85,Bou85}. By adding this term to the action
of the theory, which makes the action ghost-free, we don't face
the unitarity problem. However, it turns out that, when we deal
with the GB term in dimensions less than five, this term behaves
just like a topological term and therefore has no influence on the
background dynamics. To import the GB effect on the background
dynamics, one way is to consider the GB term in higher
dimensions~\cite{Bro07,Bam07,And07,Noz08,Noz09a,Noz09b,Noz09c}.
Another way, if we look for the GB effect in 4 dimensions, is to
couple it non-minimally to a scalar filed or adopt a function of
GB term in the 4-dimensional
action~\cite{Noj05a,Noj05b,Noj07,Guo09,Guo10,Koh14,Koh16,Wu17,El18,Od18a,Od18b,Noji19}.
Among the works done on this issue, we focus on the models in
which the GB term is non-minimally coupled to the scalar field in
the theory. Some authors have studied this type of GB inflation
models and found some interesting observational results. In this
regard, the Gauss-Bonnet inflation models with power-law, inverse
power-law and exponential potentials and GB coupling have been
studied~\cite{Guo09,Guo10,Koh14,Od18a,Jia13} and the results have
been compared with different data sets such as WMAP
5-year~\cite{Kom09}, Planck+WP~\cite{Pl13},
Planck+WP+highL+BICEP2\\~\cite{Ade14} and
BICEP2/Keck-Array~\cite{Ade16b} data.

Another interesting case in studying the inflation models is the
idea of ``cosmological attractor''. In this regard,
$\alpha$-attractor models are one class of the models
incorporating the idea of cosmological attractors which have
attracted a lot of
attention~\cite{Kai14,Fer13,Kal13c,Kal14,Cec14,Kal14b,Lin15,Jos15a,Jos15b,Kal16,Sha16,Odi16,Ras18a,Noz18}.
In the $\alpha$-attractor models, the E-model potential is given
by
$V\sim\Big[1-\exp\big(-\sqrt{\frac{2\kappa^{2}}{3\alpha}}\phi\big)\Big]^{2n}$.
It is interesting to consider the GB coupling term as E-model and
study the inflation and perturbations~\cite{Noz17,Yi18}. T-Model
potential in the $\alpha$-attractor models is given by
$V\sim\tanh^{2n}(\frac{\kappa\phi}{\sqrt{6\alpha}})$. It is also
possible to take the GB term in the inflation models as T-model,
which gives cosmologically viable results~\cite{Yi18}.

On the other hand, we should notice that the scalar field
responsible for the inflation can be a canonical as well as a
non-canonical scalar field like as
tachyon\\~\cite{Sen99,Sen02a,Sen02b} or Dirac-Born-Infeld
(DBI)~\cite{Sil04,Ali04}. Studying the inflation in tachyon and
DBI models gives interesting
results~\cite{Noz13a,Car06,Spa07,Des09,Cam09,Miz10,Noz13b,Ras18b,Noz19}.
In this regard, one can consider the coupling between these
non-canonical scalar fields and the GB term. Also, it is possible
in the GB inflation models to consider the nonminimal coupling or
nonminimal derivative coupling between the scalar field and
gravity (or any generalized inflationary
models)~\cite{Fel11,Noz16}.

Although a lot of works have been done on the GB inflation issue,
the observational viability of those models depends on the newest
data released at any time. Recently, the Planck 2018 collaboration
have released the new results~\cite{Akr18a,Akr18b}. From
Planck2018 TT,TE,EE+lowE+lensing data\footnote[1]{TT, TE and EE
refer to temperature auto-power spectrum, temperature-E-mode
cross-power spectrum and E-mode polarization auto-power spectrum,
respectively. Planck2018 TT,TE,EE+lowEB denotes the combination of
the likelihood at multipole $l\geq 30$ using TT, TE, and EE
spectra, the low-$l$ SimAll EE likelihood and the low-$l$
temperature Commander likelihood~\cite{Akr18a}. When Planck2018
B-mode information is included, the abbreviation is Planck
TT,TE,EE+lowEB. Also, BK14 refers to BICEP2/Keck Array 2014 data
and BAO denotes Baryon Acoustic Oscillations.}, based on the
$\Lambda$CDM$+r+\frac{dn_{s}}{d\ln k}$ model which supports
quasi-de Sitter expansion of the universe during inflation epoch,
we have constraints on the scalar spectral index and
tensor-to-scalar ratio as $n_{s}=0.9647\pm 0.0044$ and
$r<0.16$~\cite{Akr18a}. However, when we consider the joint data
of Planck 2018, BAO and BICEP2/Keck Array 2014, that means
Planck2018 TT,TE,EE+lowE+lensing+BAO +BK14 data
(hereafter, base data), we have $n_{s}=0.9658\pm 0.0038$
and $r<0.072$~\cite{Akr18a}. By these new constraints on the
perturbation's parameters, some inflation models might be ruled
out and the constraints on some parameters of the other inflation
models might be changed. Another information that Planck2018 team
gives us is on the tensor spectral index. Planck2018 gives the
constraint on the tensor spectral index as $-0.62<n_{T}<0.53$ with
$r<0.080$, obtained from Planck2018 TT, TE, EE +lowE+lensing+BK14+
BAO+LIGO$\&$Virgo2016 joint data (hereafter, base+GW
data)~\cite{Akr18a}. It seems that, analyzing and studying the
tensor part of the perturbations in the Gauss-Bonnet models, which
has been less studied before, give us some more information about
the inflation models.

Another important issue in studying the inflation models is the
reheating process after inflation. As long as the potential is
sufficiently flat, meaning that the slow-roll parameters
$\epsilon$ and $\eta$ are very small, the universe inflates
exponentially. When one of the slow-roll parameters meets unity,
the inflation ends and the inflaton field rolls down to the
minimum of the potential. By reaching the minimum of the
potential, inflaton starts oscillating about that minimum and
loosing the energy. In this regard, according to the physics of
particles creation and non-equilibrium phenomena, it decays into
the plasma of the relativistic particles and the universe becomes
radiation-dominated~\cite{Ab82,Do82,Al82}. There are other
interesting but complicated reheating scenarios including the
non-perturbative processes, proposed by some authors. Some
examples of the non-perturbative reheating scenarios are as the
parametric resonance decay~\cite{Ko94,Tr90,Ko97}, tachyonic
instability~\cite{Gr97,Sh06,Du06,Ab10,Fel01a,Fel01b} and the
instant preheating\\~\cite{Fel99}. To analyze the reheating
process, we focus on two important parameters $N_{rh}$ (e-folds
number) and $T_{rh}$ (temperature) in this phase. Studying these
parameters, gives us some more constraints on the model's
parameters space~\cite{Dai14,Un15,Co15,Cai15,Ue16}. The effective
equation of state parameter, $\omega_{eff}$, is another important
parameter in exploring the reheating phase. For a massive
inflaton, domination of the potential over the kinetic energy
leads to $\omega_{eff}=-1$ and domination of the kinetic energy
over the potential leads to $\omega_{eff}=1$. Given that at the
initial epoch of the reheating process, the massive inflaton
oscillates with frequency very larger than the expansion rate, the
averaged effective pressure at that epoch is zero. This means
that, at the initial epoch of the reheating phase, we can assume
$\omega_{eff}=0$ which corresponds to the equation of state
parameter of the dust matter. Also, at the end of the reheating
phase, we have $\omega_{eff}=\frac{1}{3}$. Therefore, exploring
the effective equation of state gives us some more information
about the reheating phase.

Based on these preliminaries, in this paper we focus on the GB
inflation models and re-consider them to seek for their
observational viability in confrontation with base and base+GW
data sets. In this regard, in section 2, we study the inflation
and perturbations in the general Gauss-Bonnet model. In section 3,
we consider the Gauss-Bonnet model with a canonical scalar field.
In this respect, by adopting power-law potential and two types of
GB coupling as inverse power-law and dilaton-like couplings, we
obtain the tensor-to-scalar ratio, scalar and tensor spectral
indices and investigate the observational viability of the model.
In section 4, we perform analysis on the the Gauss-Bonnet natural
inflation, in which the potential of the scalar field is
$\cos$-type and the GB coupling is inverse of $\cos$. The
Gauss-Bonnet $\alpha$-attractor is studied in section 5, with both
E-Model and T-Model potential and GB coupling. In section 6, we
analyze a Gauss-Bonnet inflation in which the inflaton is tachyon
field. By adopting power-law potential and both inverse power-law
and dilaton-like GB coupling, we check the observational viability
of this model. The DBI Gauss-Bonnet inflation, with power-law
potential, inverse power-law DBI field and both inverse power-law
and dilaton-like GB coupling, is explored in section 7. The
reheating process after inflation for the GB model with canonical
scalar field is investigated in section 8. In this regard, we
obtain some expressions for the e-folds number and temperature
during reheating process. By using the observational constraint on
the scalar spectral index, we find some constraints on $N_{rh}$
and $T_{rh}$. We also study the effective equation of state during
this process. In section 9, we investigate the reheating phase in
the GB model with tachyon field. The reheating phase in DBI GB
model is studied in section 10. In section 11, we present a
summary of the paper. We emphasize that, although the GB inflation
models have been studied in several papers, however, the tensor
perturbations in the GB models have been less studied. Also, the
reheating phase is an interesting issue in studying the inflation
models, which for most of models we study here, have never been
studied.

\section{The general Gauss-Bonnet Inflation}

In this section, we present the inflation and perturbations in a
cosmological model in which a Gauss-Bonnet term is non-minimally
coupled to the scalar field. In this setup, the action is given by
\begin{eqnarray}
\label{eq2} S=\int d^{4}x\sqrt{-g}\Bigg[\frac{1}{2\kappa^{2}}
R+P(X,\phi)-{\cal{G}}(\phi){\cal{L}_{GB}} \Bigg]\,,\hspace{0.6cm}
\end{eqnarray}
where $\phi$ is the scalar field, $R$ is the Ricci scalar,
${\cal{L}_{GB}}$ is the Gauss-Bonnet term with the coupling function
${\cal{G}}(\phi)$, and
$X=-\frac{1}{2}g^{\mu\nu}\partial_{\mu}\phi\,\partial_{\nu}\phi$.
Action (\ref{eq2}) in a spatially flat FRW metric, gives the
following background equations
\begin{eqnarray}
\label{eq3}
H^{2}=\frac{\kappa^{2}}{3{\cal{F}}}\Bigg[-P+2XP_{,X}+24H^{3}\dot{{\cal{G}}}\Bigg]\,,
\end{eqnarray}

\begin{eqnarray}
\label{eq4}
\Big(P_{\,X}+2XP_{,XX}\Big)\ddot{\phi}+\Big(3HP_{,X}+\dot{\phi}P_{,\phi
X}\Big)\dot{\phi}
-P_{,\phi}+24H^{4}{\cal{G}}'+24H^{2}\dot{H}{\cal{G}}'=0\,,
\end{eqnarray}
where the subscript ``$,$'' shows derivative with respect to the
corresponding parameter, a dot denotes a derivative with respect
to the cosmic time and a prime shows a derivative with respect to
the scalar field.

The slow-roll parameters $\epsilon$ and $\eta$, defined as
\begin{eqnarray}
\label{eq5} \epsilon=-\frac{\dot{H}}{H^{2}}\,,\quad
\eta=-\frac{1}{H}\frac{\ddot{H}}{\dot{H}}\,,
\end{eqnarray}
under the conditions $\epsilon\ll 1$ and $|\eta|\ll 1$ show the
inflation phase. In this extended setup, with GB correction, the
slow-roll limits are as $\ddot{\phi}\ll |3H\dot{\phi}|$,
$\dot{\phi}^{2}\ll V(\phi)$, $H\dot{{\cal{G}}}\ll \kappa^{-2}$ and
$P_{,X}X\ll\kappa^{-2}H^{2}$ (see~\cite{Guo10,Noz16,Car15}).

The e-folds number, which is defined as
\begin{eqnarray}
\label{eq6} N=\int_{t_{hc}}^{t_{f}}H\,dt\,,
\end{eqnarray}
in this model and within the slow-roll conditions is given by
\begin{eqnarray}
\label{eq7}
N=\int_{\phi_{hc}}^{\phi_{f}}\frac{3H^{2}P_{,X}}{P_{,\phi}-
24H^{4}{\cal{G}}'}\,d\phi\,.
\end{eqnarray}
In equations (\ref{eq6}) and (\ref{eq7}) the subscript $hc$ and $f$
refer to the horizon crossing of the physical scales and the end of
the inflation, respectively.

By using the following Arnowitt-Deser-Misner (ADM) perturbed line
element
\begin{equation}
\label{eq8} ds^{2}=
-(1+2{\cal{R}})dt^{2}+2a(t){\cal{D}}_{i}\,dt\,dx^{i}
+a^{2}(t)\left[(1-2{\Phi})\delta_{ij}+2{\Theta}_{ij}\right]dx^{i}dx^{j}\,,
\end{equation}
we present the cosmological linear perturbation in this setup. In
the above perturbed metric, ${\cal{D}}^{i}$ is defined as
${\cal{D}}^{i}=\delta^{ij}\partial_{j}{\cal{D}}+v^{i}$ and the
parameters ${\cal{R}}$ and ${\cal{D}}$ are 3-scalars. Also, $v^{i}$
is a vector that satisfies the condition
$v^{i}_{,i}=0$~\cite{Muk92,Bau09}. In this metric, we have denoted
the spatial symmetric and traceless shear 3-tensor by
${\Theta}_{ij}$ and the spatial curvature perturbation by $\Phi$. To
study the scalar perturbation at the linear level, we consider the
scalar part of the the perturbed metric within the uniform-field
gauge (where, $\delta\phi=0$), as
\begin{eqnarray}
\label{eq9}
ds^{2}=-(1+2{\cal{R}})dt^{2}+2a(t){\cal{D}}_{,i}\,dt\,dx^{i}
+a^{2}(t)(1-2{\Phi})\delta_{ij}dx^{i}dx^{j}\,.
\end{eqnarray}
By using this perturbed metric, the action (\ref{eq2}) is expanded
up to the second order in perturbations, leading to the following
quadratic action
\begin{equation}
\label{eq10} S_{2}=\int
dt\,d^{3}x\,a^{3}{\cal{W}}_{s}\left[\dot{\Phi}^{2}-\frac{c_{s}^{2}}{a^{2}}(\partial
{\Phi})^{2}\right],
\end{equation}
where
\begin{eqnarray}
\label{eq11}
{\cal{W}}_{s}=\Bigg(\frac{1}{\kappa^{2}}-8H\dot{{\cal{G}}}\Bigg)
\Bigg[\Bigg(\frac{1}{\kappa^{2}}-8H\dot{{\cal{G}}}\Bigg)\Bigg(3\big(XP_{,X}+2X^{2}P_{,XX}\big)
+144H^{3}\dot{{\cal{G}}}\Bigg) +9\bigg(\frac{2H}{\kappa^{2}}
-24H^{2}\dot{{\cal{G}}}\bigg)^{2}\Bigg] \nonumber\\
\times\Bigg[3\bigg(\frac{2H}{\kappa^{2}}
-24H^{2}\dot{{\cal{G}}}\bigg)^{2}\Bigg]^{-1}\,,\hspace{2cm}
\end{eqnarray}
and the square of the sound speed is given by
\begin{eqnarray}
\label{eq12}
c_{s}^{2}=3\Bigg[2\bigg(\frac{1}{\kappa^{2}}-8H\dot{{\cal{G}}}\bigg)^{2}
\bigg(\frac{2H}{\kappa^{2}}
-24H^{2}\dot{{\cal{G}}}\bigg)H-\bigg(\frac{2H}{\kappa^{2}}
-24H^{2}\dot{{\cal{G}}}\bigg)^{2}
\bigg(\frac{1}{\kappa^{2}}-8\ddot{{\cal{G}}}\bigg)+4
\bigg(\frac{1}{\kappa^{2}}-8H\dot{{\cal{G}}}\bigg)\nonumber\\
\bigg(\frac{d}{dt}
\Big(\frac{1}{\kappa^{2}}-8H\dot{{\cal{G}}}\Big)\bigg)
\bigg(\frac{2H}{\kappa^{2}}
-24H^{2}\dot{{\cal{G}}}\bigg)-2\bigg(\frac{1}{\kappa^{2}}-8H\dot{{\cal{G}}}\bigg)^{2}
\bigg(\frac{d}{dt}\Big(\frac{2H}{\kappa^{2}} -24H^{2}\dot{{\cal{G}}}
\Big)\bigg)\Bigg]\Bigg[\frac{1}{\kappa^{2}}-8H\dot{{\cal{G}}}\Bigg]\nonumber\\
\Bigg[\Bigg(\frac{1}{\kappa^{2}}-8H\dot{{\cal{G}}}\Bigg)\Bigg(3\big(XP_{,X}+2X^{2}P_{,XX}\big)
-\frac{9}{\kappa^{2}}H^{2}
+144H^{3}\dot{{\cal{G}}}\Bigg)+9\bigg(\frac{2H}{\kappa^{2}}
-24H^{2}\dot{{\cal{G}}}\bigg)^{2}\Bigg]\,.
\end{eqnarray}

The following two-point correlation function is used to survey the
power spectrum of the curvature perturbation
\begin{equation}
\label{eq13} \langle
0|{\Phi}(0,\textbf{k}_{1}){\Phi}(0,\textbf{k}_{2})|0\rangle
=(2\pi)^{3}\delta^{3}(\textbf{k}_{1}+\textbf{k}_{2})\frac{2\pi^{2}}{k^{3}}{\cal{A}}_{s}\,,
\end{equation}
with ${\cal{A}}_{s}$, the power spectrum, defined by
\begin{equation}
\label{eq14}
{\cal{A}}_{s}=\frac{H^{2}}{8\pi^{2}{\cal{W}}_{s}c_{s}^{3}}\,.
\end{equation}
The scalar spectral index is obtained by using the power spectrum as
\begin{equation}
\label{eq15} n_{s}-1=\frac{d \ln {\cal{A}}_{s}}{d \ln
k}\Bigg|_{c_{s}k=aH},
\end{equation}
which is calculated at the time where the physical scales exit of
the sound horizon. In this setup the scalar spectral index is
obtained as
\begin{equation}
\label{eq16}
n_{s}-1=-2\epsilon-\frac{\frac{d\Big(\epsilon-\frac{4H\dot{{\cal{G}}}}{\kappa^{-2}}\Big)}{dt}}{H\Big(\epsilon
-\frac{4H\dot{{\cal{G}}}}{\kappa^{-2}}\Big)}-\frac{1}{Hc_{s}}\frac{d\,c_{s}}{dt}\,.
\end{equation}

By focusing on the tensor part of the perturbed metric (\ref{eq8}),
we can explore the tensorial perturbations. In this regard, we write
the 3-tensor ${\Theta}_{ij}$ as
\begin{equation}
\label{eq17}
{\Theta}_{ij}={\Theta}_{+}\vartheta_{ij}^{+}+{\Theta}_{\times}\vartheta_{ij}^{\times}\,,
\end{equation}
where $\vartheta_{ij}^{(+,\times)}$ are two polarization tensors
that satisfy the reality and normalization
conditions~\cite{Fel11a,Fel11b}. Now, the quadratic (second order)
action for the tensor mode is the following expression
\begin{eqnarray}
\label{eq18} S_{T}=\int dt\, d^{3}x\, a^{3}
{\cal{W}}_{T}\left[\dot{\Theta}_{+}^{2}-\frac{c_{T}^{2}}{a^{2}}(\partial
{\Theta}_{+})^{2}+\dot{\Theta}_{\times}^{2}-\frac{c_{T}^{2}}{a^{2}}(\partial
{\Theta}_{\times})^{2}\right]\,.
\end{eqnarray}
In this second order action, the parameters ${\cal{W}}_{T}$ and
$c_{T}^{2}$ are given by
\begin{equation}
\label{eq19}
{\cal{W}}_{T}=\frac{1}{4\kappa^{2}}\left(1-8\kappa^{2}H\dot{{\cal{G}}}+\frac{\kappa^{2}X{\cal{N}}}{M^{2}}\right),
\end{equation}
\begin{equation}
\label{eq20}
c_{T}^{2}=1+8\kappa^{2}H\dot{{\cal{G}}}-\frac{2\kappa^{2}X{\cal{N}}}{M^{2}}\,.
\end{equation}
For the tensor mode, the amplitude of the perturbations is defined
as
\begin{equation}
\label{eq21}
{\cal{A}}_{T}=\frac{H^{2}}{2\pi^{2}{\cal{W}}_{T}c_{T}^{3}}\,,
\end{equation}
and the tensor spectral index in this setup is given by
\begin{equation}
\label{eq22} n_{T}=\frac{d \ln {\cal{A}}_{T}}{d \ln k}=2\epsilon\,.
\end{equation}
Another important perturbation parameter, the tensor-to-scalar
ratio, is defined as the ratio of the amplitudes of the tensor
mode versus the scalar mode:
\begin{equation}
\label{eq23}
r=\frac{{\cal{A}}_{T}}{{\cal{A}}_{s}}\simeq16c_{s}\epsilon=-8c_{s}\Big(n_{T}+8\kappa^{2}H\dot{{\cal{G}}}\Big).
\end{equation}

For more details about obtaining the equations presented in this
section, see~\cite{Fel11a,Fel11,Noz16}. By having the required
equations, in the next sections we explore the observational
viability of some Gauss-Bonnet models.

\section{Gauss-Bonnet Inflation in a Model with the Canonical Scalar Field}

In this section, we consider the case where
\begin{equation}
\label{eq24} P(X,\phi)=X-V\,.
\end{equation}
This choice of $P(X,\phi)$ corresponds to the simple
inflation model where the inflaton rolls slowly down a nearly flat
potential~\cite{Gut81,Lin82,Alb82}. By this adoption, we have an
inflation model in which a Gauss-Bonnet term is non-minimally
coupled to the canonical scalar filed. In this case, the scalar
spectral index takes the following form
\begin{equation}
\label{eq25} n_{s}=1-{\frac {16\chi\,{V}^{3}{\cal{G}}''+8\chi'
V^{3}{\cal{G}}'+3\chi\,V''\,V-6\chi
\,{V'}^{2}+3\chi'VV'}{{V}(8V^{2}\alpha'+3V')}}\,,
\end{equation}
where
\begin{equation}
\label{eq26} \chi={\frac
{V'}{\kappa^{2}\,V}}+\frac{8\kappa^{2}}{3}\,{\cal{G}}'\,V\,.
\end{equation}
The tensor spectral index is given by
\begin{equation}
\label{eq27}
n_{T}=-2\Bigg(\frac{1}{2\kappa^{2}}\frac{V'^{2}}{V^{2}}+\frac{4}{3}\kappa^{2}{\cal{G}}'\,V'\Bigg)\,.
\end{equation}
Also, we have the following expression for the tensor-to-scalar
ratio
\begin{equation}
\label{eq28}
r=-8\Bigg(-\frac{8}{3}{\cal{G}}'\,\chi\,V-\frac{\chi\,V'}{V}\Bigg)\,.
\end{equation}

Note that in obtaining equations (\ref{eq25})-(\ref{eq28}), we
have used the slow-roll conditions. Now, we have to choose some
explicit functions for the potential and GB coupling function.
After adopting the functions, we study the model numerically and
compare the results with base\footnote[2]{ Planck2018
TT,TE,EE+lowE+lensing+BAO +BK14} and base+GW\footnote[3]{
Planck2018 TT, TE, EE +lowE+lensing+BK14+ BAO+LIGO$\&$Virgo2016}
observational data sets.

\subsection{Power-Law potential and Inverse Power-Law GB Coupling}
The model with the monomial potential and inverse monomial GB
coupling function has been considered as the simplest primordial
inflation model. In the absence of the GB effect, the simple
inflation model with $\phi^{2}$ and $\phi^{4}$ potentials is not
consistent with the base data~\cite{Akr18a}. We wonder whether the
presence of GB effect makes the model observationally viable. In
this regard, we adopt following potential and GB coupling function
\begin{equation}
\label{eq29} V=V_{0}\,\phi^{n}\,\quad \& \quad
{\cal{G}}={\cal{G}}_{0}\,\phi^{-n}\,.
\end{equation}

By this choice, we find the following expressions for the
perturbation parameters
\begin{equation}
\label{eq30} n_{s}=1-{\frac { \left( n+2 \right)  \left( 2\beta-1
\right) n}{{\phi}^{2}}}\,,
\end{equation}

\begin{equation}
\label{eq31} n_{T}={\frac {{n}^{2} \left( \beta-1 \right)
}{{\phi}^{2}}}\,,
\end{equation}
and
\begin{equation}
\label{eq32} r=8\, {\frac {{n}^{2} \left( \beta-1 \right)
^{2}}{{\phi}^{2}}}\,,
\end{equation}
where
\begin{equation}
\label{eq33} \beta=\frac{8}{3}V_{0}\,{\cal{G}}_{0}\,,
\end{equation}
and we have set $\kappa^{2}\equiv1$. We can use equation
(\ref{eq7}) to obtain the value of the scalar field at the time of
the horizon crossing of the physical scales. Now, we perform a
numerical analysis on the model's parameter space. In this regard,
we explore $r-n_{s}$ and $r-n_{t}$ in confrontation with
Planck2018 different data sets. To study $r-n_{s}$ behavior, we
use the base data. Note that, from this data set we have
$n_{s}=0.9658\pm 0.0038$ and $r<0.072$, based on the
$\Lambda$CDM$+r+\frac{dn_{s}}{d\ln k}$ model. These constraints on
the perturbation parameters imply the constraints $52.13\leq
N\leq 65.29$ and $0.407 \leq {\cal{G}} \leq 0.528$ on the GB model
with $V=V_{0}\,\phi^{n}$ and ${\cal{G}}={\cal{G}}_{0}\,\phi^{-n}$
and for $n=2$. In the top panels of Figure 1, we see $r-n_{s}$
plane in the background of the base data. In plotting this figure
(and all subsequent figures of this type), we have used $n=2,4$,
$50\leq N\leq 70$ and also $0<\beta<1$. As figure shows, $r-n_{s}$
plane in the GB model with $n=2$, in some ranges of the parameters
space is consistent with the observational data. However, for
$n=4$, there is no consistency of $r-n_{s}$ plane with the base
data.

In the sense that in studying the tensor spectral index we focus
on the tensor part of the perturbations (the gravitational waves),
we use the base+GW data to explore $r-n_{T}$. The results are
shown in the bottom panels of Figure 1. In this case, the $r-n_{T}$
plane for both $n=2$ and $n=4$ is consistent with the base+GW
data. By numerical analysis of the model in this case, we have
obtained some constraints on the model's parameter space which are
presented in Table 1. Note that, in~\cite{Akr18a} it has been used
the $68\%$ CL on measured parameter ($n_{s}$) and $95\%$ CL for
top bound on other parameters ($r$ and $n_{T}$). In this
respect, and as regards we study these three parameters to obtain
the constraints, both confidence levels are interesting to
consider.

In summary, our numerical analysis shows that the Gauss-Bonnet
inflation with $V=V_{0}\,\phi^{n}$ and
${\cal{G}}={\cal{G}}_{0}\,\phi^{-n}$ for $n=2$ is consistent with
observational data if $\beta\sim {\cal{O}} (10^{-1})$. Also, this
model with $V=V_{0}\,\phi^{n}$ and
${\cal{G}}={\cal{G}}_{0}\,\phi^{-n}$ for $n=4$ is ruled out. Note
that, as it can be seen from Table 1, any small variation of the
parameter $\beta$ can cause the model not to be consistent with
the observational data. This means that, physically, not only the
presence of the Gauss-Bonnet term but also the intensity of the
coupling between the Gauss-bonnet term (as the geometry side of
the model) and the scalar field (as the energy-momentum side of
the model) is very important in the viability of the model.

\begin{figure}[]
	\begin{center}
		\includegraphics[scale=0.47]{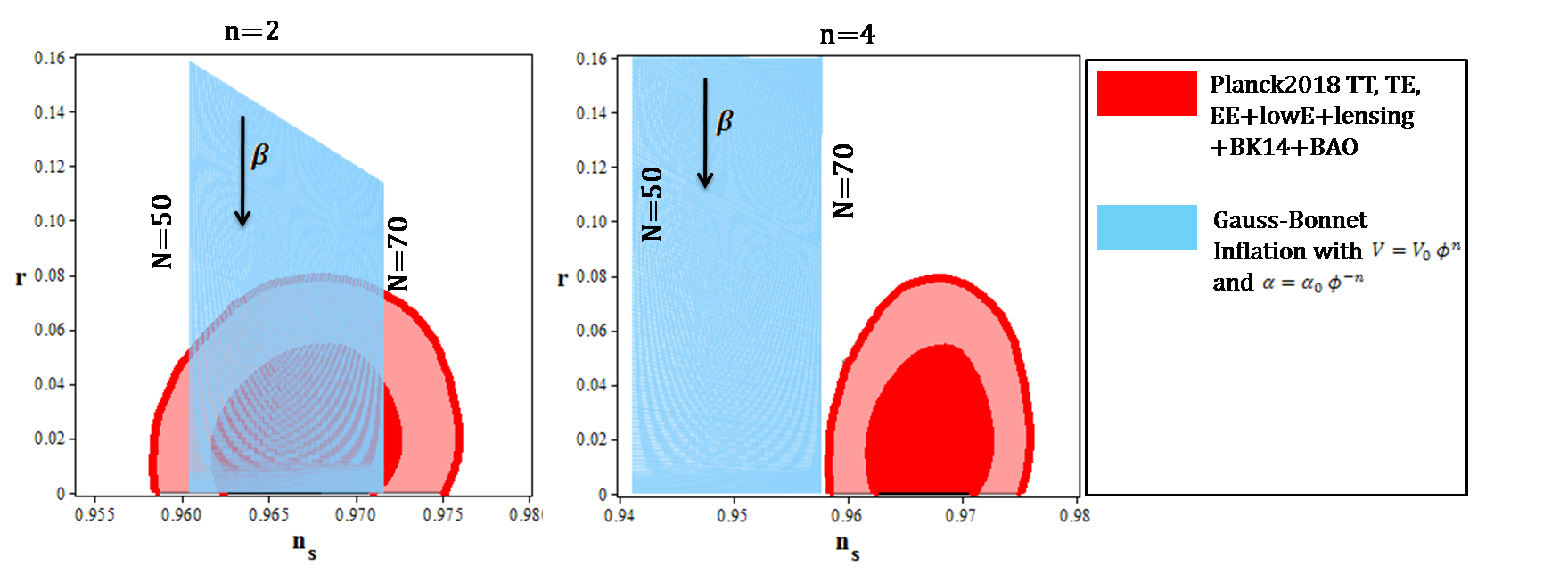}
		\includegraphics[scale=0.47]{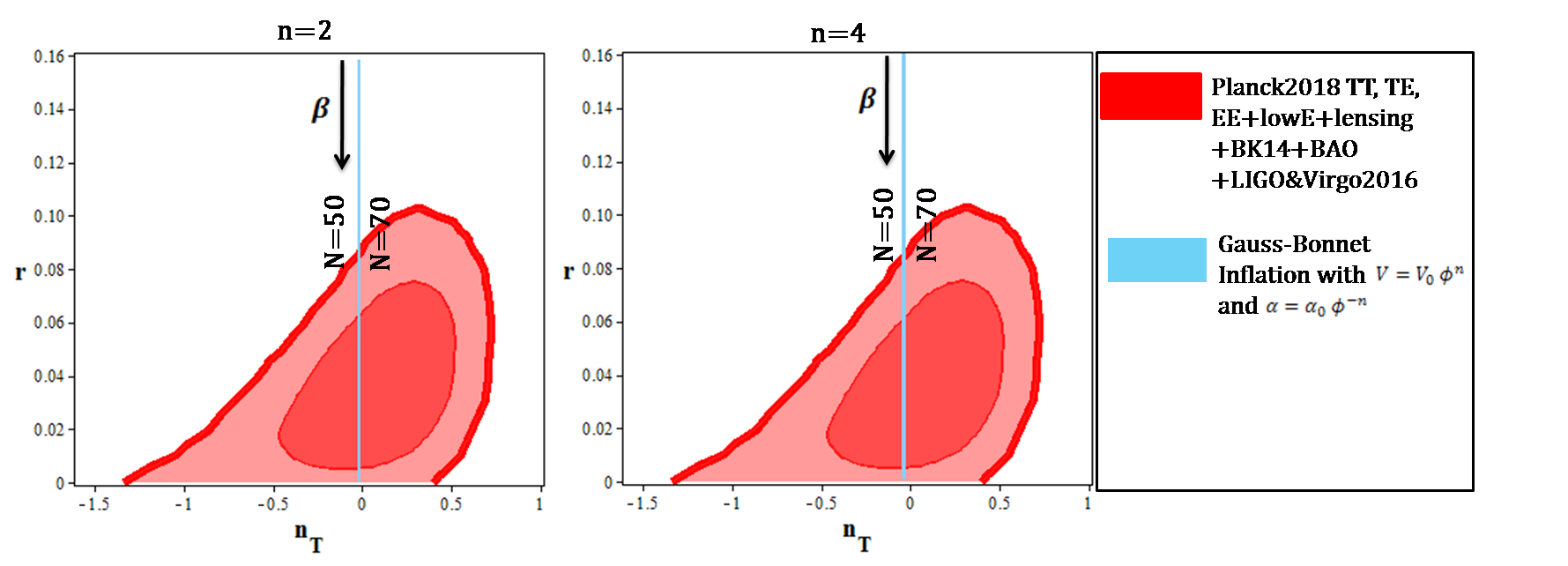}
	\end{center}
	\caption{\small Tensor-to-scalar ratio versus the
		scalar spectral index and tensor spectral index of the GB model
		with $V=V_{0}\,\phi^{n}$ and
		${\cal{G}}={\cal{G}}_{0}\,\phi^{-n}$.}
	\label{fig1}
\end{figure}

\begin{table*}
\tiny\caption{\small{\label{tab:1} The ranges of the parameter
$\beta$ in which tensor-to-scalar ratio, the scalar spectral index
and the tensor spectral index of the the GB model with
$V=V_{0}\,\phi^{n}$ and ${\cal{G}}={\cal{G}}_{0}\,\phi^{-n}$ are
consistent with different data sets.}}
\begin{center}
\begin{tabular}{ccccccc}
\\ \hline \hline \\ && Planck2018 TT,TE,EE+lowE & Planck2018 TT,TE,EE+lowE&Planck2018 TT,TE,EE+lowE&Planck2018 TT,TE,EE+lowE
\\
&$$& +lensing+BK14+BAO &
+lensing+BK14+BAO&lensing+BK14+BAO&lensing+BK14+BAO
\\
&$$&  & &+LIGO$\&$Virgo2016 &LIGO$\&$Virgo2016
\\
\hline \\&$N$& $68\%$ CL & $95\%$ CL &$68\%$ CL & $95\%$ CL
\\
\hline \\ &$50$&not consistent&$0.680\leq \beta< 1$&$0.610\leq
\beta \leq 0.970$&$0.470\leq \beta<
1$\\\
\\$n=2$&$60$&$0.601\leq \beta<1$&$0.410\leq \beta<
1$&$0.520\leq \beta \leq 0.964$&$0.340\leq \beta<
1$\\ \\
&$70$&$0.660\leq \beta <1$&$0.350\leq \beta <1$&$0.450\leq \beta
\leq 0.960$&$0.260\leq
\beta <1$\\
\hline  \\ &$50$&not consistent&not consistent&$0.810\leq \beta
\leq 0.985$&$0.730\leq
\beta <1$\\
\\$n=4$&$60$&not consistent&not consistent&$0.770\leq \beta \leq 0.982$&$0.680\leq
\beta <1$\\ \\
&$70$&not consistent&not consistent&$0.730\leq \beta \leq
0.980$&$0.630\leq
\beta <1$\\
\hline \hline
\end{tabular}
\end{center}
\end{table*}

\subsection{Power-Law Potential and Dilaton-Like GB Coupling}

Inspired from heterotic string theory, the GB term appears to be
coupled to the dynamical dilaton field with an exponential
coupling function. This issue has been studied in
Ref.~\cite{Bam07}. Therefore, in this subsection, we adopt
following potential and GB coupling function
\begin{equation}
\label{eq34} V=V_{0}\,\phi^{n}\,\quad \& \quad
{\cal{G}}={\cal{G}}_{0}\,e^{-\lambda\phi}\,.
\end{equation}
In this case, we have the perturbation parameters of the model as
\begin{equation}
\label{eq35}
n_{s}=1-\frac{-n(n+2)+\beta\lambda\,e^{-\lambda\phi}\,\phi^{n+1}(2\lambda\phi-n)}{\phi^{2}}\,,
\end{equation}

\begin{equation}
\label{eq36}
n_{T}=-\frac{n(n-\beta\lambda\,e^{-\lambda\phi}\,\phi^{n+1})}{\phi^{2}}\,,
\end{equation}
and
\begin{equation}
\label{eq37}
r=\frac{8(n-\beta\lambda\,e^{-\lambda\phi}\,\phi^{n+1})^{2}}{\phi^{2}}\,.
\end{equation}

Here also, we use equation (\ref{eq7}) to obtain the value of the
scalar field at the time of the horizon crossing of the physical
scales and then study $r-n_{s}$ and $r-n_{T}$ behaviors. top
panels of Figure 2 show $r-n_{s}$ plane in the background of the
base data for $N=60$. As figure shows, $r-n_{s}$ plane in the GB
model with both $n=2$ and $n=4$, in some ranges of the parameters
$\lambda$ and $\beta$ is consistent with the observational data.

In the bottom panels of Figure 2, we see $r-n_{T}$ plane in the
background of the base+GW data. Considering that the scalar
spectral index, tensor spectral index and the tensor-to-scalar
ratio depend on both $\lambda$ and $\beta$, to obtain the
observational constraints, we fix $\lambda$ in some sample values
and obtain the observationally viable ranges of $\beta$. The
results are shown in Table 2. In fact, according to our numerical
analysis, the Gauss-Bonnet inflation with $V=V_{0}\,\phi^{n}$ and
${\cal{G}}={\cal{G}}_{0}\,e^{-\lambda\phi}$ for $n=2$ is
consistent with observational data if $\beta\gtrsim {\cal{O}}
(10^{-2})$. Also, this model for $n=4$ is consistent with
observational data if $\beta\sim {\cal{O}} (10^{-1})$.

\begin{figure}[]
	\begin{center}
		\includegraphics[scale=0.47]{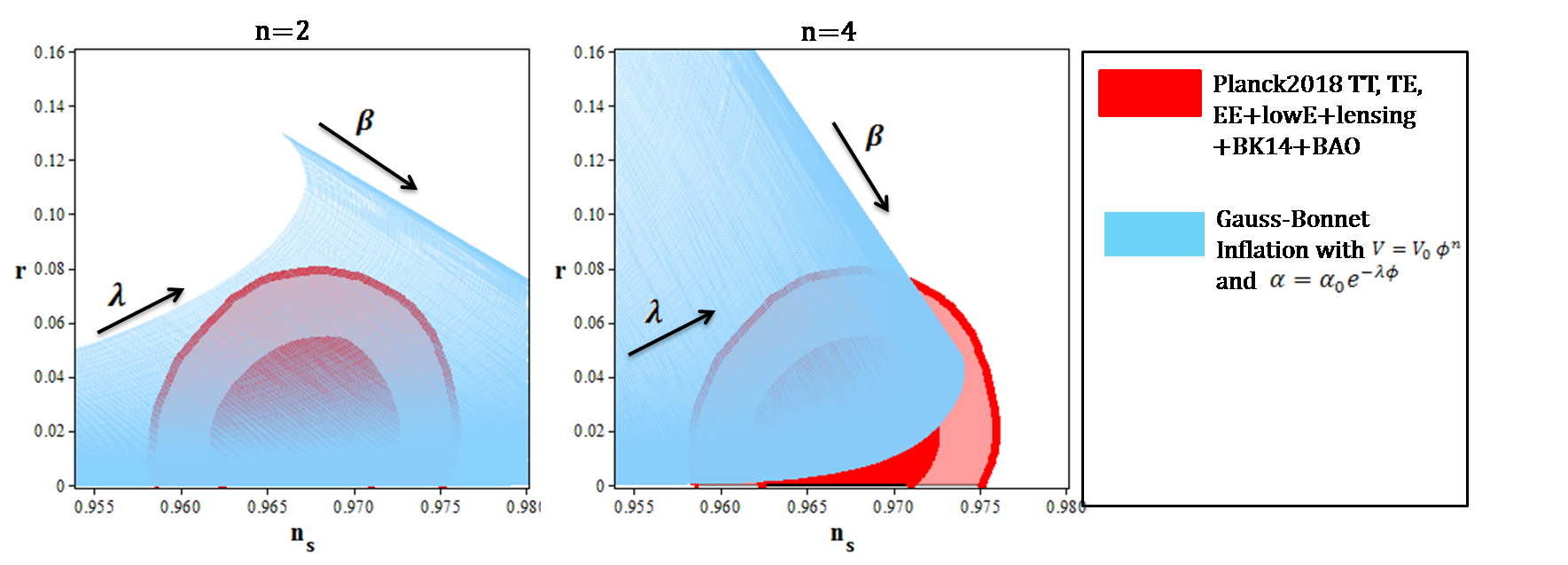}
		\includegraphics[scale=0.47]{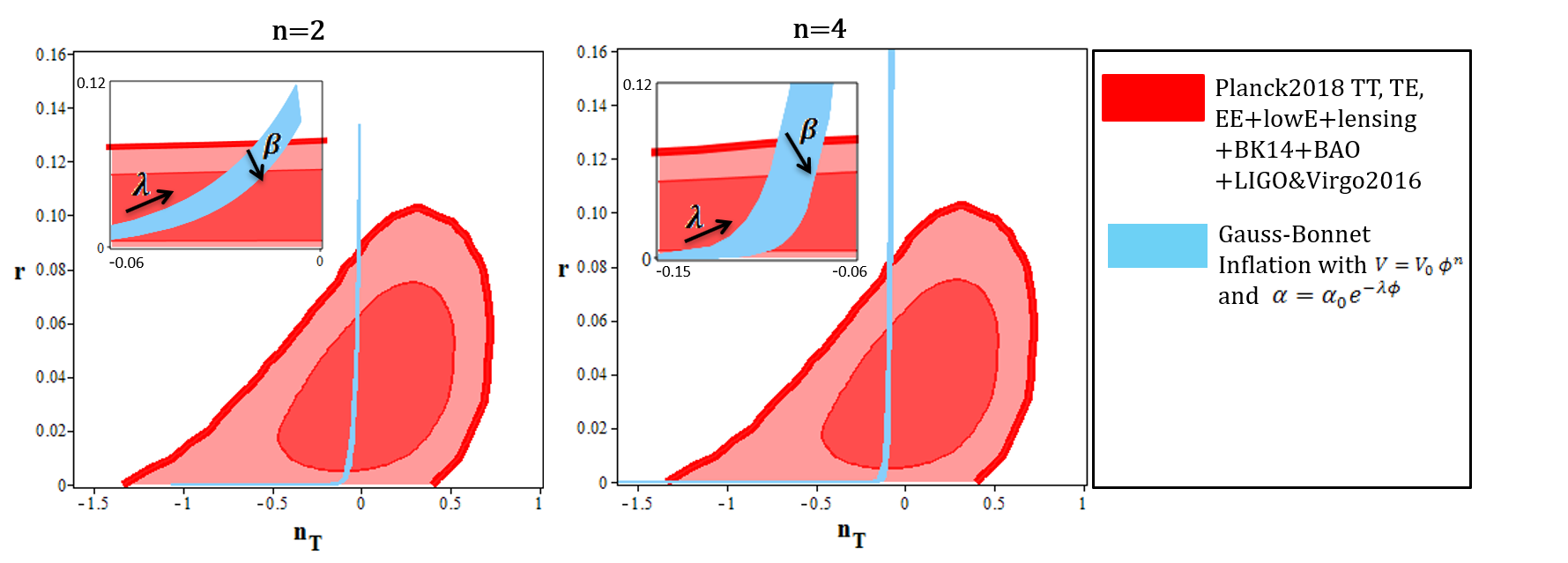}
	\end{center}
	\caption{\small Tensor-to-scalar ratio versus the
		scalar spectral index and tensor spectral index of the GB model
		with $V=V_{0}\,\phi^{n}$ and
		${\cal{G}}={\cal{G}}_{0}\,e^{-\lambda\phi}$.}
	\label{fig2}
\end{figure}

\begin{table*}
\tiny\caption{\small{\label{tab:2} The ranges of the parameter
$\beta$ in which the tensor-to-scalar ratio, the scalar spectral
index and the tensor spectral index of the the GB model with
$V=V_{0}\,\phi^{n}$ and
${\cal{G}}={\cal{G}}_{0}\,e^{-\lambda\phi}$ are consistent with
different data sets.}}
\begin{center}
\begin{tabular}{ccccccc}
\\ \hline \hline \\ && Planck2018 TT,TE,EE+lowE & Planck2018 TT,TE,EE+lowE&Planck2018 TT,TE,EE+lowE&Planck2018 TT,TE,EE+lowE
\\
&$$& +lensing+BK14+BAO &
+lensing+BK14+BAO&lensing+BK14+BAO&lensing+BK14+BAO
\\
&$$&  & &+LIGO$\&$Virgo2016 &LIGO$\&$Virgo2016
\\
\hline \\&$N$& $68\%$ CL & $95\%$ CL &$68\%$ CL & $95\%$ CL
\\
\hline \\ &$10$& $0.03\leq \beta\leq 0.05$ &$0.01\leq \beta\leq
0.072$&$0.01\leq \beta \leq 0.121$&$0.002\leq \beta<
1$\\
\\$n=2$&$10^{2}$&$0.062\leq \beta\leq 0.084$&$0.016\leq \beta\leq
0.095$&$0.031\leq \beta \leq 0.184$&$0.020\leq \beta<
1$\\ \\
&$10^{4}$&$0.041\leq \beta \leq 0.180 $&$0.031\leq \beta \leq
0.420$&$0.086\leq \beta \leq 0.269$&$0.260\leq
\beta <1$\\
\hline  \\ &$10$& $0.610\leq \beta \leq 0.683$ &$0.580\leq \beta
\leq 0.716$&$0.435\leq \beta \leq 0.812$&$0.016\leq
\beta \leq 0.871$\\
\\$n=4$&$10^2$&$0.642\leq
\beta \leq 0.711$&$0.621\leq \beta \leq 0.743$&$0.483\leq \beta
\leq 0.865$&$0.033\leq
\beta \leq 0.895$\\ \\
&$10^{4}$&$0.693\leq \beta \leq 0.789$&$0.670\leq \beta \leq
0.826$&$0.506\leq \beta \leq 0.884$&$0.081\leq
\beta \leq 0.910$\\
\hline \hline
\end{tabular}
\end{center}
\end{table*}

\section{GB Natural Inflation}

In the inflation models, to fit with the CMB anisotropy
measurements, there should be a large number of e-folds of the
scale factor. This means that the width of the potential in the
inflation models must be much larger than its height. In this
respect, the authors of Ref.~\cite{Ada91} have shown that, in
order to the potential be flat, the ratio between the height and
the fourth power of the width must satisfy the constraint $\Delta
V/ (\Delta\phi)^{4}\leq 10^{-6}$. In this constraint, $\Delta$
refers to the change in the corresponding parameters. In this
regard, in 1990, Freese, Frieman, and Olinto have proposed the
natural inflation model~\cite{Fre90}. In their model, they have
considered an axion-like particle (a pseudo-Nambu-Goldstone boson)
as the field responsible for the primordial inflation. Invariance
of the potential under a transformation as
$\phi\rightarrow\phi+constant$ (a shift symmetry) ensures flatness
of the potential~\cite{Fre90,Fre04}. The symmetry is broken after
enough inflation and the inflation phase terminates. Based on
these preliminaries, we consider a GB model where the potential is
the natural potential type and the GB coupling is inverse of the
natural potential as
\begin{equation}
\label{eq38}
V=V_{0}\bigg[1+\cos\Big(\frac{\phi}{f}\Big)\bigg]\,\quad\&\quad{\cal{G}}={\cal{G}}_{0}\bigg[1+\cos\Big(\frac{\phi}{f}\Big)\bigg]^{-1}\,.
\end{equation}
Note that, the natural inflation in its simplest realization, has
the above form of the potential. By these functions, the
perturbation parameters take the following forms
\begin{equation}
\label{eq39} n_{s}=1-\frac{\left( \cos \left( {\frac {\phi}{f}}
\right) -3 \right)  \Big( 1- \beta \Big) }{{f}^{2} \left( 1+\cos
\left( {\frac {\phi}{f}} \right) \right)}\,,
\end{equation}

\begin{equation}
\label{eq40} n_{T}= - \frac{\Big( -1+\beta \Big)  \left( \cos \left(
{\frac {\phi}{f}} \right) -1 \right)}{ {f}^{2} \left( 1+\cos \left(
{\frac {\phi}{f}} \right)  \right)}\,,
\end{equation}

\begin{equation}
\label{eq41} r=\frac{8\, \Big( -1+\beta \Big) ^{2} \left( 1-\cos
\left( {\frac {\phi}{f}} \right) \right) }{{f}^{2} \left( 1+\cos
\left( {\frac {\phi}{f}} \right)  \right) }\,.
\end{equation}

By using equation (\ref{eq7}), to obtain the value of the scalar
field at horizon crossing, and equations
(\ref{eq39})-(\ref{eq41}), we can study the model numerically. To
this end, we adopt $N=60$ and explore $r-n_{s}$ and $r-n_{T}$
planes for various values of $f$ and $\beta$. The top panel of
Figure 3 shows the tensor-to-scalar ratio versus the scalar
spectral index in the background of the base data. As Figure 3
shows, the natural GB inflation in some ranges of the parameter
space is observationally viable.

Tensor-to-scalar ratio versus the tensor spectral index in the
background of the base+GW data is shown in the bottom panel of
Figure 3. By performing a numerical analysis, we have found some
constraints on the model's parameters which have been summarized
in Table 3. Note that, the natural inflation with $N=50$ is not
consistent with new observational data. However, when we consider
the GB effect, this model for $\beta\geq 0.65$ is consistent with
base and base+GW data sets.

\begin{figure}[]
	\begin{center}
		\includegraphics[scale=0.55]{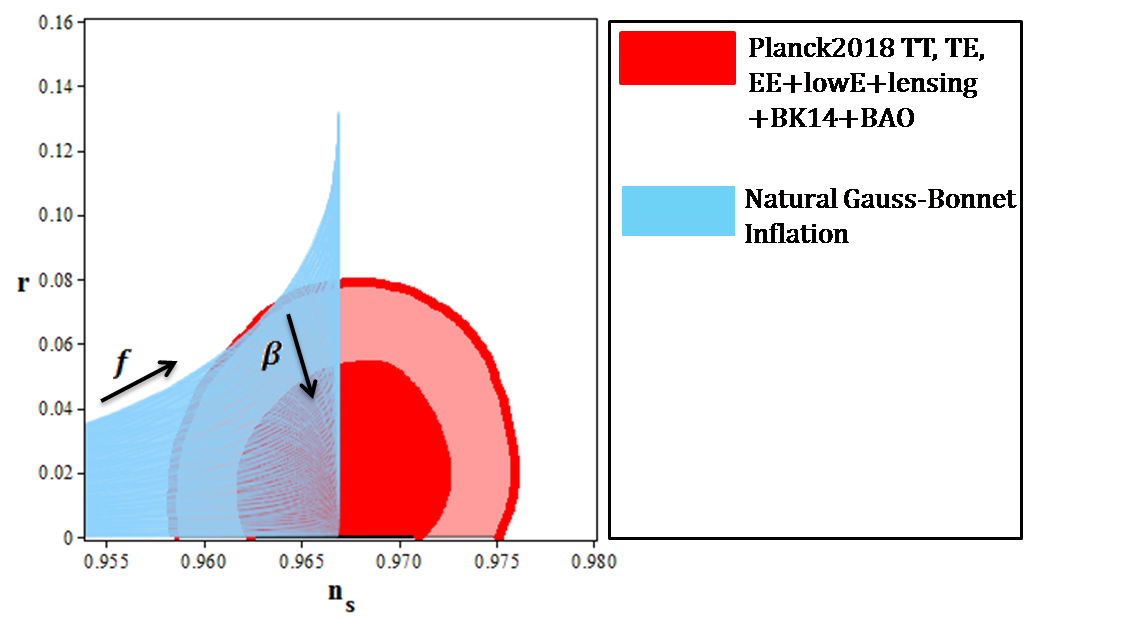}
		\includegraphics[scale=0.55]{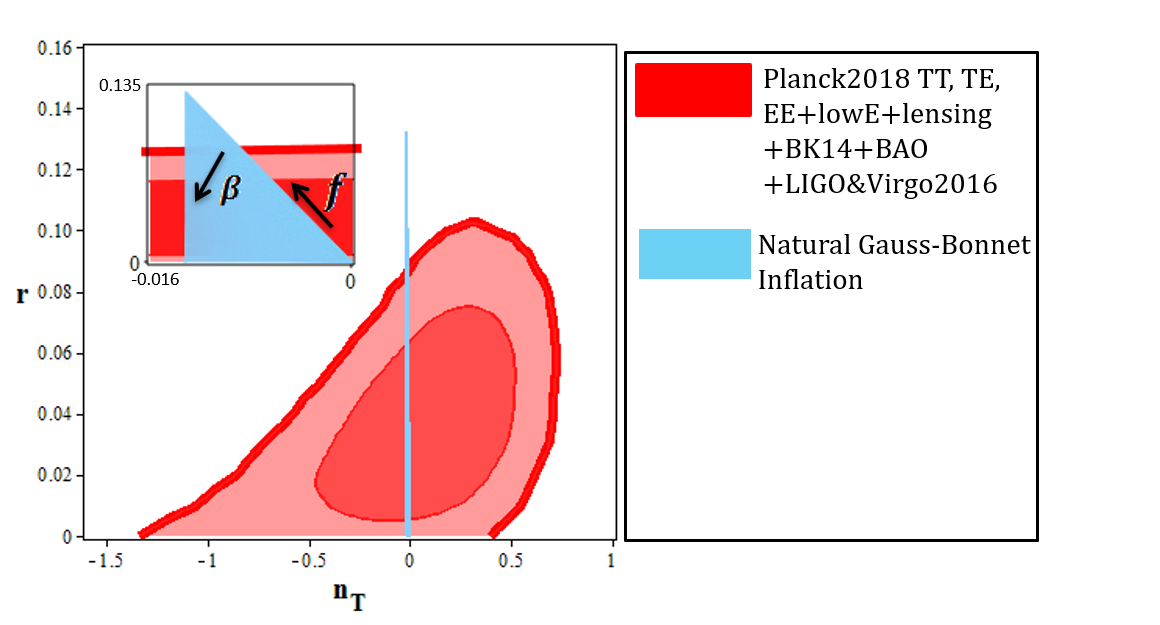}
	\end{center}
	\caption{\small Tensor-to-scalar ratio versus the
		scalar spectral index and tensor spectral index of the GB natural
		inflation.}
	\label{fig3}
\end{figure}

\begin{table*}
\tiny \caption{\small{\label{tab:3} The ranges of the parameter
$\beta$ in which the tensor-to-scalar ratio, the scalar spectral
index and the tensor spectral index of the GB natural inflation
are consistent with different data sets.}}
\begin{center}
\begin{tabular}{cccccc}
\\ \hline \hline \\ & Planck2018 TT,TE,EE+lowE & Planck2018 TT,TE,EE+lowE&Planck2018 TT,TE,EE+lowE&Planck2018 TT,TE,EE+lowE
\\
& +lensing+BK14+BAO &
+lensing+BK14+BAO&lensing+BK14+BAO&lensing+BK14+BAO
\\
&  & &+LIGO$\&$Virgo2016 &LIGO$\&$Virgo2016
\\
\hline \\$N$& $68\%$ CL & $95\%$ CL &$68\%$ CL & $95\%$ CL
\\
\hline \\  $4$& $0.635\leq \beta<1$ &$0.525\leq \beta<1$&$0\leq \beta\leq 0.962$ & all values of $\beta$\\
\\$15$&$0.578\leq \beta<1$&$0.345\leq \beta <1$ &$0.520\leq \beta \leq 0.962 $&$0.341\leq
\beta <1$
\\ \\
$35$&$0.596\leq \beta <1 $&$0.393\leq \beta  <1$&$0.520\leq \beta
\leq 0.962 $&$0.341\leq
\beta <1$\\ \\
$60$&$0.597\leq \beta <1 $&$0.400 \leq \beta <1 $&$0.524\leq \beta
\leq 0.962 $&$0.346 \leq
\beta  <1$\\
\hline \hline
\end{tabular}
\end{center}
\end{table*}

\section{Gauss-Bonnet $\alpha$-attractor}

In recent years, the idea of the ``cosmological attractor'' has
attracted \textbf{the attention of some cosmologists}. Among the
models that incorporate the idea of cosmological attractors, we
mention the conformal attractor~\cite{Kal13a,Kal13b} and
$\alpha$-attractor models~\cite{Fer13,Kal13c,Kal14}. The important
characteristic property in the conformal attractor model is that,
in the case of large e-folds number, it has the universal
prediction for the primordial curvature perturbations and the
tensor-to-scalar ratio as $n_{s}=1-\frac{2}{N}$ and
$r=\frac{12}{N^{2}}$. Considering the single field
$\alpha$-attractor model, the universal predictions for the
mentioned parameters are $n_{s}=1-\frac{2}{N}$ and
$r=\frac{12\alpha}{N^{2}}$. In this section, we consider the
Gauss-Bonnet effect on the $\alpha$-attractor model. We consider
Both potentials leading to the $\alpha$-attractor: E-model and
T-model. We also adopt E-model and T-model GB coupling function
and study the inflation in this setup.

\subsection{E-Model}

In Ref.~\cite{Kal13b}, the authors have considered a model with
two real scalar fields, $\varphi$ and $\psi$, which are
non-minimally coupled to the gravity. They have also considered a
potential term as
$V(\varphi,\psi)=\frac{\lambda}{4}\varphi^{2}(\varphi-\psi)^{2}$
by which the $SO(1,1)$ symmetry has been broken. By using the
conformal gauge as $\psi^{2}-\varphi^{2}=6$, and introducing a
canonically normalized field as
$\varphi=\sqrt{6}\cosh\frac{\phi}{\sqrt{6}}$ and
$\psi=\sqrt{6}\sinh\frac{\phi}{\sqrt{6}}$, they have obtained an
exponential-type potential named E-model which we use in this
subsection. In this case, we consider the following potential and
GB coupling
\begin{equation}\label{eq42}
V=V_{0}\Bigg[1-\exp\bigg(-\sqrt{\frac{2\kappa^{2}}{3\alpha}}\phi\bigg)\Bigg]^{2n}\,\quad\&\quad
{\cal{G}}={\cal{G}}_{0}\Bigg[1-\exp\bigg(-\sqrt{\frac{2\kappa^{2}}{3\alpha}}\phi\bigg)\Bigg]^{2n}\,,
\end{equation}
where the potential $V$ is an E-model potential. With these
functions, we obtain the perturbation parameters from equations
(\ref{eq25})-(\ref{eq28}). In this case, the scalar spectral index
takes the following form
\begin{eqnarray}\label{eq43}
n_{s}=1-\frac {8}{9}\,\frac {nZ \Big(
256\,{\beta}^{2}\,{Y}^{8\,n}Zn{
\kappa}^{4}-96\,{\beta}^{2}\,{Y}^{8\,n}\,{\kappa}^{4}+24\,{\kappa}^{4}\beta\,
{Y}^{4\,n}\,nZ-24\,{\kappa}^{4}\beta\,{Y}^{4\,n} \Big)
}{{Y}^{2}\alpha\, \left( 8\,\beta\,{Y}^{4\,n}+3
\right) }\nonumber\\
-{\frac {8}{9}}\,{\frac {nZ \Big(
48\,\beta\,{Y}^{4\,n}\,Zn-36\,\beta\,{Y} ^{4\,n}-9\,nZ-9 \Big)
}{{Y}^{2}\alpha\, \left( 8\,\beta\,{Y}^{4\,n}+3 \right) }}\,.
\end{eqnarray}
The tensor spectral index in the E-Model GB inflation is given by
\begin{eqnarray}\label{eq44}
n_{T}=-\frac {8}{9}\,\frac {{n}^{2}{Z}^{2} \left(
8\,{\kappa}^{4}\beta\,{Y}^{4 \,n}+3 \right) }{\alpha\,{Y}^{2}}\,.
\end{eqnarray}
Also, we obtain the following expression for the tensor-to-scalar
ratio in this case
\begin{eqnarray}\label{eq45}
r=\frac {64}{27}\,\frac {{n}^{2}{Z}^{2} \left(
64\,{\kappa}^{4}{\beta}^{2
}\,{Y}^{8\,n}+24\,{\kappa}^{4}\beta\,{Y}^{4\,n}+24\,\beta\,{Y}^{4\,n}+9
\right) }{\alpha\,{Y}^{2}}\,.
\end{eqnarray}
In these equations, we have defined the following parameters
\begin{eqnarray}\label{eq46}
Y=1-Z\,\quad and \quad Z={{\rm e}^{-\frac{\sqrt {6}}{3}\,\sqrt
{{\frac {{\kappa}^{2}}{\alpha}}}\phi}}\,.
\end{eqnarray}

\subsection{T-Model}
Another interesting case in the $\alpha$-attractor model is
T-model potential. Authors of Ref.\\~\cite{Kal13a} have studied a
model with two non-minimally coupled scalar fields and the
potential term as
$V(\varphi,\psi)=\frac{1}{36}F(\varphi/\psi)(\varphi^{2}-\psi^{2})^{2}$,
which breaks the $SO(1,1)$ symmetry. Note that $F(\varphi/\psi)$
is an arbitrary function. By using the gauge and canonically
normalized field used in obtaining the E-model potential, they
have found the potential as $V(\phi)=F(\tanh(\phi/\sqrt{6}))$. In
the case of the simplest set of functions as
$F(\varphi/\psi)=\lambda(\varphi/\psi)^{2n}$, the T-model
potential has been obtained. Now, in this subsection, we consider
the case where the potential and GB coupling function are T-model
type, as
\begin{equation}\label{eq47}
V=V_{0}\tanh^{2n}\bigg(\sqrt{\frac{\kappa^{2}}{6\alpha}}\phi\bigg)\,\,,\quad
{\cal{G}}={\cal{G}}_{0}\tanh^{2n}\bigg(\sqrt{\frac{\kappa^{2}}{6\alpha}}\phi\bigg)\,.
\end{equation}
By substituting these potential and GB coupling function in
equations (\ref{eq25})-(\ref{eq28}), we obtain the scalar spectral
index as
\begin{eqnarray}\label{eq48}
n_{s}=1+\frac{2}{9}\,\frac {n \Big(
192\,{\beta}^{2}\,{\cal{U}}^{8n}{\kappa}^{4}{\cal{{\cal{V}}}}-256
\,{\beta}^{2}\,{{\cal{U}}}^{8n}\,
n{\kappa}^{4}-96\,\beta\,{{\cal{U}}}^{8n}{
\kappa}^{4}+48\,{\kappa}^{4}\beta\,{{\cal{U}}}^{4n}{\cal{V}}+9\,n-9
\Big) }{{\cal{V}} \left( {\cal{V}}-1 \right)  \left( 8\,
\beta\,{{\cal{U}}}^{4n}+3 \right) \alpha}\nonumber\\
+\frac{2}{9}\,\frac {n \Big(
-24{\kappa}^{4}\beta\,{{\cal{U}}}^{4n}\,n-24{\kappa}^{4}\beta\,{{\cal{U}}}^{4n}+
72\beta\,{{\cal{U}}}^{4n}\,{\cal{V}}-48\beta\,{{\cal{U}}}^{4n}\,n-36
\beta\,{{\cal{U}}}^{4n}+18{\cal{V}} \Big) }{{\cal{V}} \left(
{\cal{V}}-1 \right) \left( 8\, \beta\,{{\cal{U}}}^{4n}+3 \right)
\alpha}\,,
\end{eqnarray}
the tensor spectral index as
\begin{eqnarray}\label{eq49}
n_{T}=-\frac{2}{9}\,{\frac {{n}^{2} \left(
8\,{\kappa}^{4}\beta\,{{\cal{U}}}^{4n}+3 \right) }{
\alpha\,{\cal{V}} \left( {\cal{V}}-1 \right) }}\,,
\end{eqnarray}
and finally the tensor-to-scalar ratio as
\begin{eqnarray}\label{eq50}
r=\frac {16}{27}\,\frac {{n}^{2} \left(
64{\kappa}^{4}{\beta}^{2}\,{{\cal{U}}}^{8
n}+24\,{\kappa}^{4}\beta\,{{\cal{U}}}^{4n}+24\beta\,{{\cal{U}}}^{4n}+9
\right) }{ {\cal{V}} \left( {\cal{V}}-1 \right) \alpha}\,,
\end{eqnarray}
where the parameters ${\cal{U}}$ and ${\cal{V}}$ are given by
\begin{eqnarray}\label{eq51}
{\cal{U}}=\tanh \left(\sqrt {{\frac {{\kappa}^{2}}{6\alpha}}}\phi
\right)\,,
\end{eqnarray}
and
\begin{eqnarray}\label{eq52}
{\cal{V}}= \cosh^{2} \left( \sqrt {{\frac {{\kappa}^{2}}{6\alpha
}}}\phi \right)\,.
\end{eqnarray}

By substituting the value of the scalar field at horizon crossing
in equations (\ref{eq43})-(\ref{eq45}) and equations
(\ref{eq48})-(\ref{eq50}), we can study the GB $\alpha$-attractor
model numerically. The results are shown in \textbf{Figures} 7 and
8. The $\alpha$-attractor models with E-model and T-model
potential meet the model with $\phi^{n}$ potential in
$\alpha\rightarrow \infty$ limit. On the other hand, these models
in $\alpha\rightarrow 0$ and large $N$ limits reach an attractor
point characterized by following scalar spectral index and
tensor-to-scalar ratio
\begin{eqnarray}\label{eq53}
n_{s}=1-\frac{2}{N}\,,\quad r=\frac{12\alpha}{N^{2}}\,.
\end{eqnarray}

The top panels of Figure 4 show the tensor-to-scalar ratio
versus the scalar spectral index of a GB model in the background
of the base data, for E-model potential and coupling function
(left panel) and T-model potential and coupling function (right
panel). The black lines have been plotted to show that the GB
$\alpha$-attractor model in the small $\alpha$ limit reaches an
attractor (green star) and in the large $\alpha$ limit meets the
GB model with $\phi^{2}$ potential and GB coupling (pink
stars). This figure has been plotted with $n=2$ and $N=60$. As we
see from figure, the GB model with $\phi^{2}$ potential and GB
coupling function is not consistent with the base data. However,
when we consider E-model or T-model potential and GB coupling
function, it is possible to find some ranges in the parameters
space leading to observationally viable values of $r$ and $n_{s}$.
Our analysis on $r-n_{s}$ viability shows that E-model GB
inflation is consistent with base data $68\%$ CL if $\alpha<78$
and at $95\%$ CL if $\alpha<4.1\times10^{2}$. Also, T-model GB
inflation has consistency with base data at $68\%$ CL if
$\alpha<43$ and at $95\%$ CL if $\alpha<91$.

The bottom panels of Figure 4 show the tensor-to-scalar ratio
versus the tensor spectral index in the background of base+GW
data. By performing a numerical analysis on $r-n_{T}$ viability,
we find that E-model GB inflation is consistent with base+GW data
at $68\%$ CL if $2.12<\alpha< 1.6\times 10^{2}$ and at $95\%$ CL
if $\alpha<7.4\times 10^{2}$. Also, in T-model GB inflation we
find the constraints $2.43<\alpha<73$ at $68\%$ CL and
$\alpha<2\times 10^{2}$ at $95\%$ CL. However, the values of
$\alpha$ imply some constraints on the GB coupling parameter which
have been summarized in Table 4, for some sample values of
$\alpha$.

By these considerations, we conclude that the Gauss-Bonnet
inflation with both E-model and T-model potentials and GB coupling
functions is consistent with the observational data if
$\beta\lesssim {\cal{O}} (10^{-3})$.

\begin{figure}[]
	\begin{center}
		\includegraphics[scale=0.47]{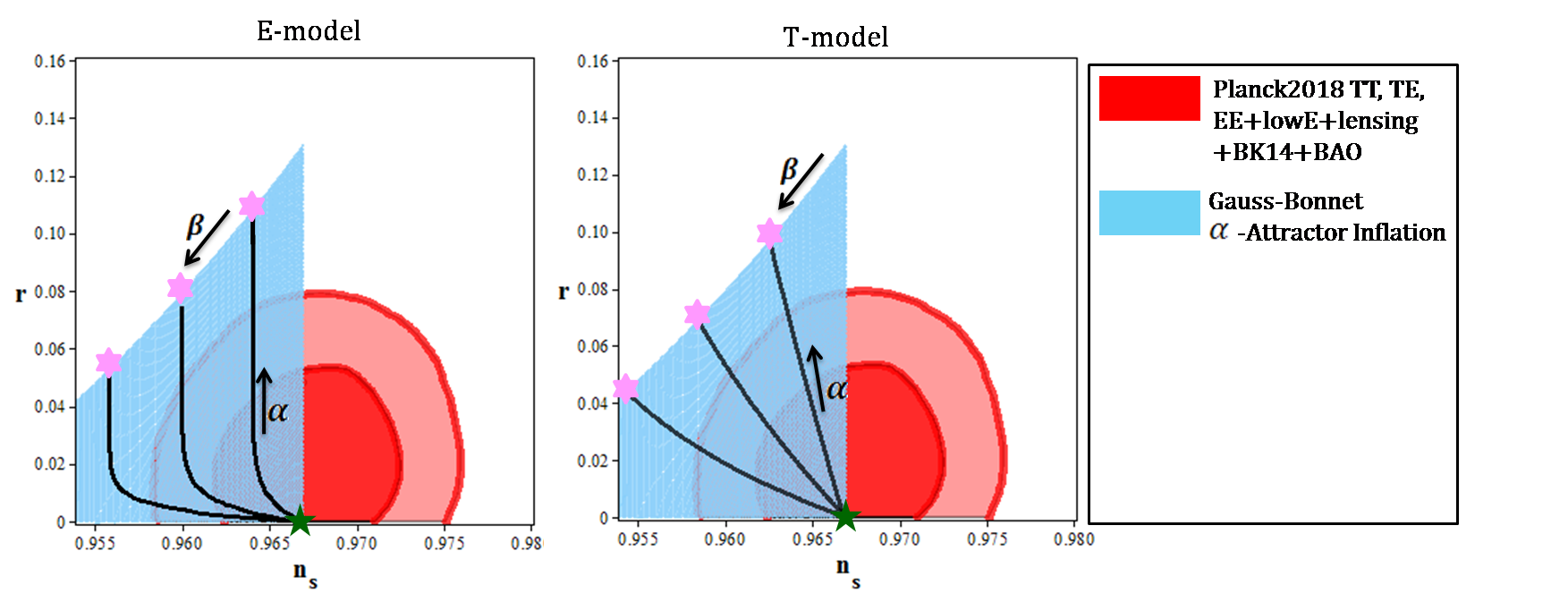}
		\includegraphics[scale=0.47]{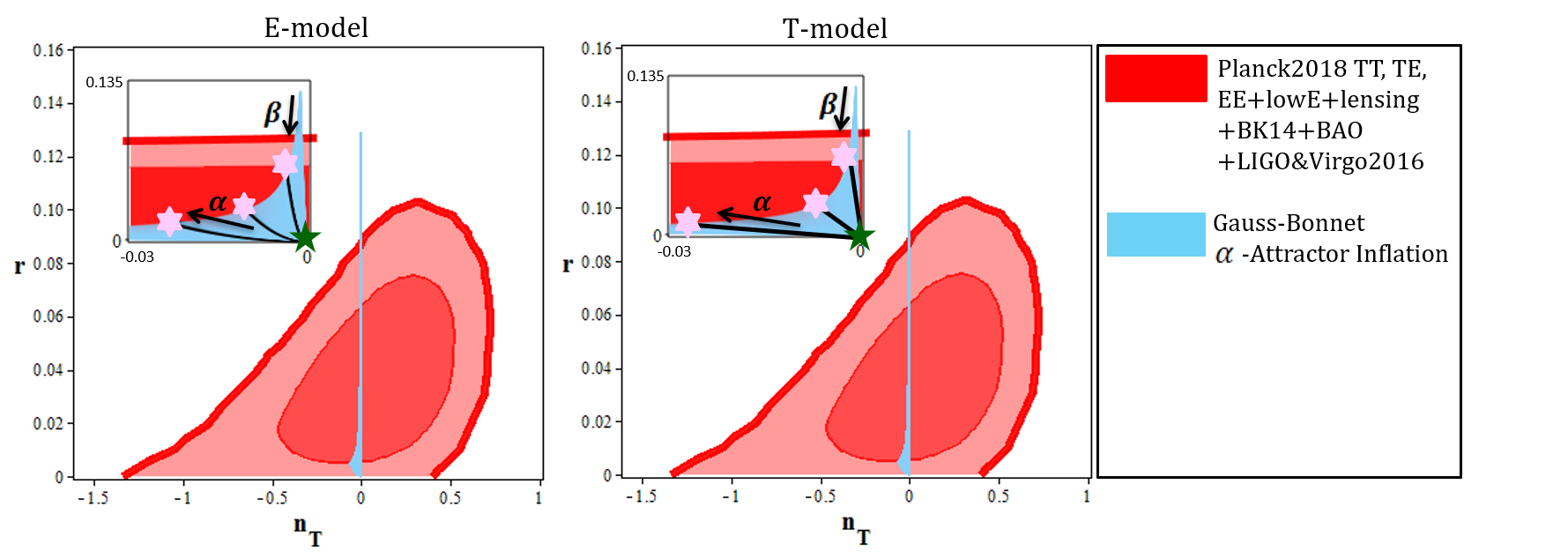}
	\end{center}
	\caption{\small Tensor-to-scalar ratio versus the
		scalar spectral index and tensor spectral index of the GB
		$\alpha$-attractor. The left panels are corresponding to the case
		where both potential and GB coupling are E-model. The right panels
		are corresponding to the case where both potential and GB coupling
		are T-model. The black lines have been plotted to show that the GB
		$\alpha$-attractor model in the small $\alpha$ limit reaches an
		attractor (green star) and in the large $\alpha$ limit meets the GB
		model with $\phi^{2}$ potential and GB coupling (pink
		stars).}
	\label{fig4}
\end{figure}

\begin{table*}
\tiny\caption{\small{\label{tab:4} The ranges of the parameter
$\beta$ in which the tensor-to-scalar ratio, the scalar spectral
index and the tensor spectral index of the GB $\alpha$-attractor
model with $n=2$ and $N=60$ are consistent with different data
sets.}}
\begin{center}
\begin{tabular}{ccccccc}
\\ \hline \hline \\ && Planck2018 TT,TE,EE+lowE & Planck2018 TT,TE,EE+lowE&Planck2018 TT,TE,EE+lowE&Planck2018 TT,TE,EE+lowE
\\&& +lensing+BK14+BAO &
+lensing+BK14+BAO&lensing+BK14+BAO&lensing+BK14+BAO
\\& & & &+LIGO$\&$Virgo2016 &LIGO$\&$Virgo2016
\\
\hline \\&$N$& $68\%$ CL & $95\%$ CL &$68\%$ CL & $95\%$ CL
\\
\hline \\ &$20$& $\beta\leq 5.17\times 10^{-2}$ & $\beta\leq 5.31\times 10^{-2}$&$ \beta< 3.10\times 10^{-3}$ & all values of $\beta$ \\
\\E-model&$50$& $ \beta\leq 4.92\times 10^{-2}$ &$\beta\leq 5.04\times 10^{-2}$&$ \beta< 4.52\times 10^{-3}$ &all values of $\beta$\\ \\
&$80$&$3.23\times 10^{-4} \leq\beta\leq 4.56\times 10^{-2} $&$\beta\leq 4.81\times 10^{-2}$&$\beta<8.22\times 10^{-3} $&all values of $\beta$\\
\hline  \\ &$20$& $\beta\leq 3.30\times 10^{-1}$ &$ \beta\leq 4.01\times 10^{-1}$&$\beta< 8.11\times 10^{-3}$ &all values of $\beta$\\
\\T-model&$50$&$ 3.03\times 10^{-4} \leq\beta\leq 1.42\times 10^{-1}$&$ \beta\leq 9.03\times 10^{-2}$&$ \beta< 9.82\times 10^{-3}$&all values
 of $\beta$\\ \\
&$80$&$4.28\times 10^{-4} \leq\beta\leq 1.14\times 10^{-1}$&$
\beta\leq 7.01\times 10^{-2}$&$2.37\times 10^{-4}<\beta<
1.43\times
10^{-2}$&all values of $\beta$\\
\hline \hline
\end{tabular}
\end{center}
\end{table*}

\section{Gauss-Bonnet Inflation in a Model with Tachyon Field}

Tachyon field is a scalar field associated with D-branes in string
theory~\cite{Sen99,Sen02a,Sen02b}. This field, which is described
by the Dirac-Born-Infeld action, has interesting cosmological
applications. The early time inflation in the history of the
Universe might be caused by a slow-rolling tachyon
field~\cite{Sam02,Fei02,Noz14}. Also, it is possible that the
current acceleration phase of the universe is due to the presence
of the tachyon field as the dark energy
component~\cite{Pad02,Gor04,Cop05}. These features make the the
tachyon field cosmologically interesting. In the case of tachyon
field, $P(X,\phi)$ is given by the following expression
\begin{equation}
\label{eq54} P(X,\phi)=-V\sqrt{1-2X}\,,
\end{equation}
Now, the scalar spectral index takes the following form
\begin{equation}
\label{eq55} n_{s}=1-{\frac {16\chi\,{V}^{3}{\cal{G}}''+8\chi'
V^{3}{\cal{G}}'+3\chi\,V''\,V-6\chi
\,{V'}^{2}+3\chi'VV'}{{V}(8V^{2}\alpha'+3V')}}\,.
\end{equation}
The tensor spectral index is given by
\begin{equation}
\label{eq56}
n_{T}=-2\Bigg(\frac{1}{2\kappa^{2}}\frac{V'^{2}}{V^{2}}+\frac{4}{3}\kappa^{2}{\cal{G}}'\,V'\Bigg)\,,
\end{equation}
and we have the following expression for the tensor-to-scalar ratio
\begin{equation}
\label{eq57}
r=-8\Bigg(-\frac{8}{3}{\cal{G}}'\,\chi\,V-\frac{\chi\,V'}{V}\Bigg)\,.
\end{equation}
In equations (\ref{eq55})-(\ref{eq57}), the parameter $\chi$ is
given by
\begin{equation}
\label{eq58} \chi={\frac
{V'}{\kappa^{2}\,V^{2}}}+\frac{8\kappa^{2}}{3}\,{\cal{G}}'\,.
\end{equation}

Now, by choosing some explicit functions for the potential and GB
coupling, we study this model numerically.

\subsection{Power-Law potential and Inverse Power-Law GB Coupling}
Our first choices for tachyon GB model are the following potential
and GB coupling function
\begin{equation}
\label{eq59} V=V_{0}\,\phi^{n}\,\quad \&\quad
{\cal{G}}={\cal{G}}_{0}\,\phi^{-n}\,.
\end{equation}
By these adoptions, we find
\begin{equation}
\label{eq60} n_{s}=1-2\,{\frac { \left(
2\,n\beta\,{\kappa}^{4}+2\,\beta\,{\kappa}^{4}-n-1 \right)
n}{{\phi}^{n+2}}} \,,
\end{equation}

\begin{equation}
\label{eq61} n_{T}=-{\phi}^{-2-n}{n}^{2} \left( 1-2\,\beta
\right)\,,
\end{equation}
and
\begin{equation}
\label{eq62} r=8\, \left[ {\phi}^{-2-n}{n}^{2} \left( 2\,\beta-1
\right) ^{2} \right]\,.
\end{equation}

\begin{figure}[]
	\begin{center}
		\includegraphics[scale=0.47]{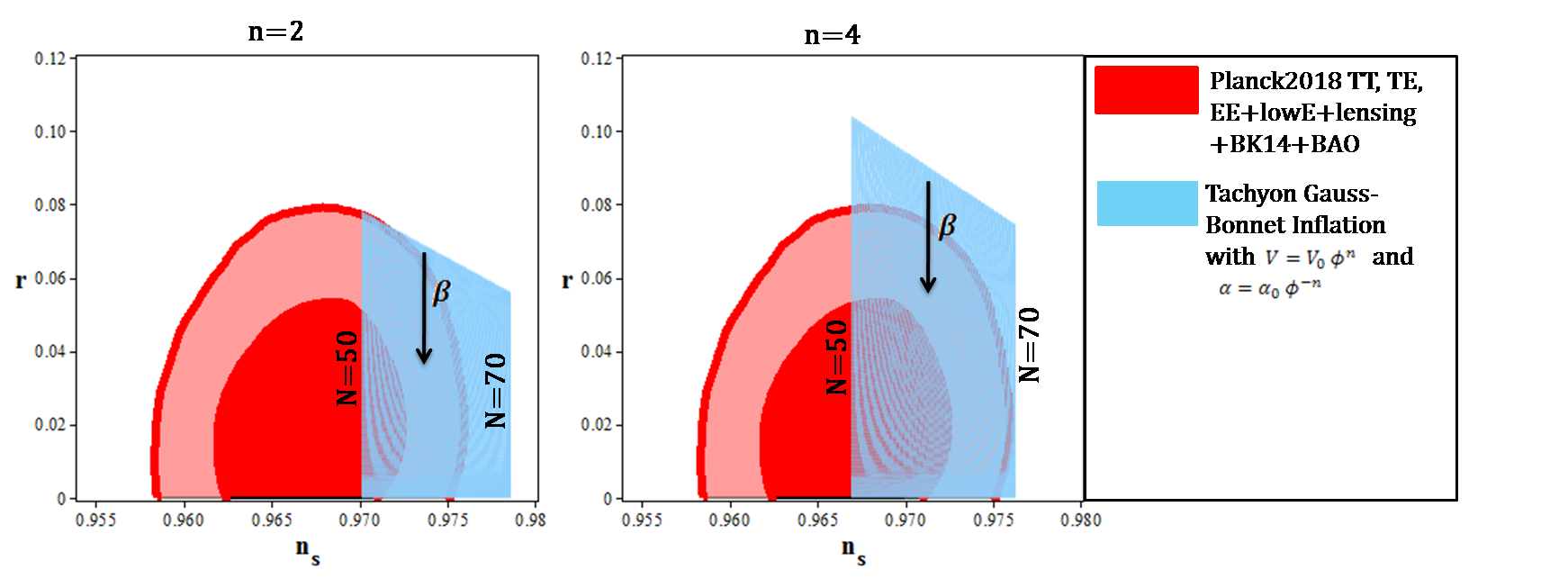}
		\includegraphics[scale=0.47]{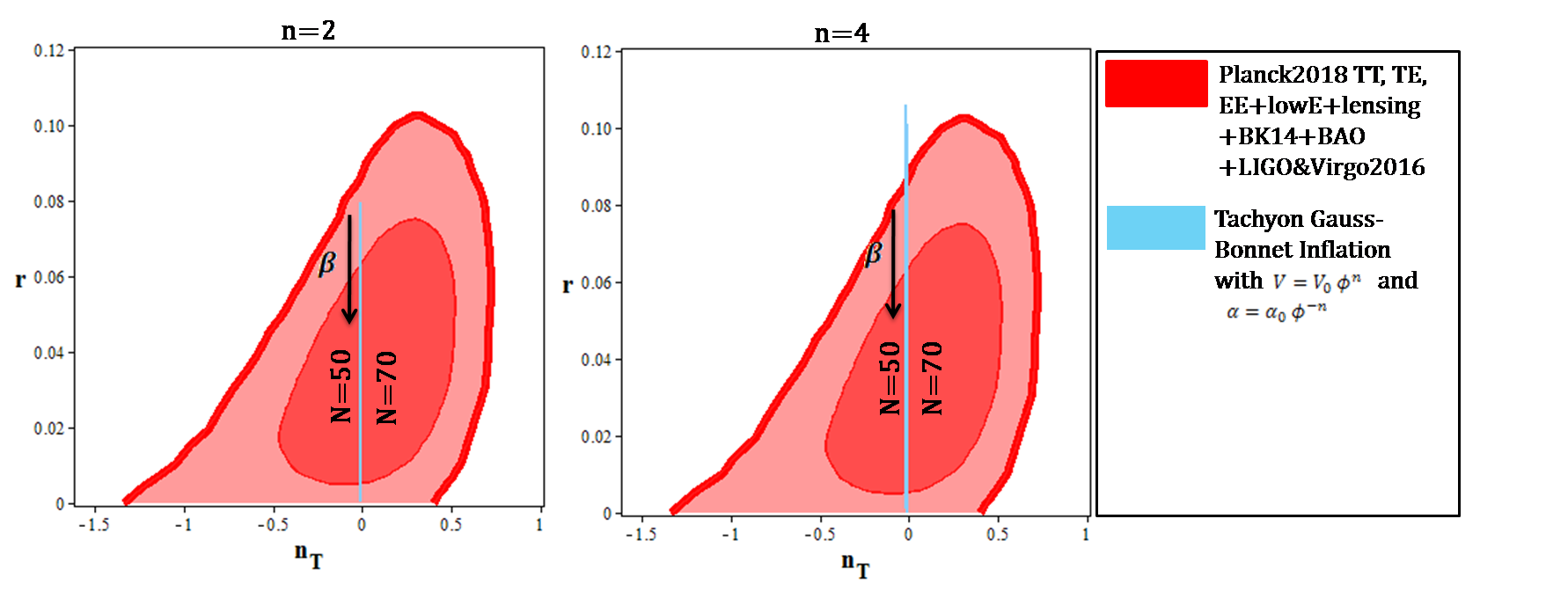}
	\end{center}
	\caption{\small Tensor-to-scalar ratio versus the
		scalar spectral index and tensor spectral index of the tachyon GB
		model with $V=V_{0}\,\phi^{n}$ and
		${\cal{G}}={\cal{G}}_{0}\,\phi^{-n}$.}
	\label{fig5}
\end{figure}

To compare the model with observational data, we substitute the
value of the scalar field at the time of the horizon crossing of
the physical scales in the above equations. Then, we perform the
numerical analysis on the parameters. In the top panels of Figure 5,
we see $r-n_{s}$ plane in the background of the base data for
$50\leq N\leq 70$. As figure shows, including the GB effect makes
the tachyon model more observationally viable. Our numerical
analysis shows that $r-n_{s}$ in the tachyon GB inflation with
$V=V_{0}\,\phi^{n}$ and ${\cal{G}}={\cal{G}}_{0}\,\phi^{-n}$, is
consistent with $68\%$ CL of the base data if $N<54.2$ for $n=2$
and $N<60.5$ for $n=4$. In this case, $r-n_{s}$ is consistent with
$95\%$ CL of base data if $N<62.4$ for $n=2$ and $N<70$ for $n=4$.

The bottom panels of Figure 5 show the tensor-to-scalar ratio
versus the tensor spectral index of the tachyon GB model with
$V=V_{0}\,\phi^{n}$ and ${\cal{G}}={\cal{G}}_{0}\,\phi^{-n}$. The
tachyon GB model is consistent with observational data in some
ranges of $\beta$. In Table 5, we present the constraints on
$\beta$ for $N=50,60,70$ and $n=2,4$ which makes the model
observationally viable. In this regard, we find that the tachyon
Gauss-Bonnet inflation with $V=V_{0}\,\phi^{n}$ and
${\cal{G}}={\cal{G}}_{0}\,\phi^{-n}$ for $n=2$ and $n=4$ is
consistent with observational data if $\beta\sim {\cal{O}}
(10^{-1})$.

\begin{table*}
\tiny\caption{\small{\label{tab:5} The ranges of the parameter
$\beta$ in which the tensor-to-scalar ratio, the scalar spectral
index and the tensor spectral index of the  tachyon GB model with
$V=V_{0}\,\phi^{n}$ and ${\cal{G}}={\cal{G}}_{0}\,\phi^{-n}$ are
consistent with different data sets.}}
\begin{center}
\begin{tabular}{ccccccc}
\\ \hline \hline \\ && Planck2018 TT,TE,EE+lowE & Planck2018 TT,TE,EE+lowE&Planck2018 TT,TE,EE+lowE&Planck2018 TT,TE,EE+lowE
\\&& +lensing+BK14+BAO &
+lensing+BK14+BAO&lensing+BK14+BAO&lensing+BK14+BAO
\\& & & &+LIGO$\&$Virgo2016 &LIGO$\&$Virgo2016
\\
\hline \\&$N$& $68\%$ CL & $95\%$ CL &$68\%$ CL & $95\%$ CL
\\
\hline \\ &$50$&$0.360\leq \beta<1$& all values&$0.201\leq \beta \leq 0.938$& all values\\
\\$n=2$&$60$&not consistent&$0.251\leq \beta<1$&$0.042\leq \beta \leq 0.924$& all values\\ \\
&$70$&not consistent&not consistent&$\beta \leq 0.911$& all values\\
\hline  \\ &$50$& $0.498\leq \beta< 1$&$0.254\leq \beta< 1$&$0.403\leq \beta \leq 0.953$&$0.188\leq\beta <1$\\
\\$n=4$&$60$&$0.685\leq \beta< 0.892$& $0.205\leq \beta< 1$&$0.283\leq \beta \leq 0.943$&$0.020\leq\beta <1$\\ \\
&$70$&not consistent&not consistent&$0.164\leq \beta \leq 0.937$& all values\\
\hline \hline
\end{tabular}
\end{center}
\end{table*}

\subsection{Power-Law Potential and Dilaton-Like GB Coupling}
In this subsection, we consider the following potential and GB
coupling
\begin{equation}
\label{eq63} V=V_{0}\,\phi^{n}\,\quad \&\quad
{\cal{G}}={\cal{G}}_{0}\,e^{-\lambda\phi}\,.
\end{equation}
Now, we have
\begin{equation}
\label{eq64} n_{s}=1-{\frac {-{{\rm
e}^{\lambda\,\phi}}\chi\,{n}^{2}+\beta\,{\lambda}^{2}
\chi\,{\phi}^{n+2}-\beta\,\lambda\,{\phi}^{n+2}\chi'+{{\rm e}^{
\lambda\,\phi}}n\chi'\,\phi-{{\rm e}^{\lambda\,\phi}}\chi\,n}{\phi\,
\left( {\phi}^{n+1}\beta\,\lambda-n{{\rm e}^{\lambda\,\phi}} \right)
}}\,,
\end{equation}

\begin{equation}
\label{eq65} n_{T}=-{\frac { \left(
-{\kappa}^{4}\beta\,\lambda\,{{\rm e}^{-\lambda\,\phi
}}+n{\phi}^{-n-1} \right) n}{\phi}} \,,
\end{equation}
and
\begin{equation}
\label{eq66} r=8\,{\frac {-{{\rm
e}^{-2\,\lambda\,\phi}}{\phi}^{n+1}{\beta}^{2}{
\lambda}^{2}+2\,{{\rm
e}^{-\lambda\,\phi}}n\beta\,\lambda-{\phi}^{-n-1 }{n}^{2}}{\phi}}
\,.
\end{equation}

By using these perturbation parameters we perform a numerical
analysis on the model. The top panels of Figure 6 show the
behavior of $r-n_{s}$ of the tachyon GB model with functions given
by equation (\ref{eq63}), in the background of the base data for
$N=60$. Our numerical analysis shows that, in this case, the
tachyon GB model with $0.15<\lambda$ (for $n=2$) and
$0.10<\lambda$ (for $n=4$) is consistent with the base data at
$95\%$ CL. Also, this model with $0.19<\lambda<0.37$ (for $n=2$)
and $0.18<\lambda<0.52$ (for $n=4$) is consistent with the base
data at $68\%$ CL.

The tensor spectral index versus the tensor-to-scalar ratio of the
tachyon GB model with $V=V_{0}\,\phi^{n}$ and
${\cal{G}}={\cal{G}}_{0}\,e^{-\lambda\phi}$, in the background of
the base+GW data, is shown in the bottom panels of Figure 6. As
figure shows, for all values of $\lambda$ and $\beta$, the
tensor-to-scalar ratio and the tensor spectral index of this model
are consistent with base+GW data at $95\%$ CL. However, there are
some constraints on $\lambda$ and $\beta$ at $68\%$ CL. At this
level, the constraints on $\lambda$ are $0.23<\lambda$ (for $n=2$)
and $0.19<\lambda$ (for $n=4$). For some sample values of
$\lambda$, the constraints on $\beta$ are summarized in Table 6.
According to this analysis, for the tachyon Gauss-Bonnet inflation
with $V=V_{0}\,\phi^{n}$ and
${\cal{G}}={\cal{G}}_{0}\,e^{-\lambda\phi}$ and for $n=2$, we
can't find any constraint on $\beta$. This is because, this model
for all values of $\beta$ is consistent with observational data at
$95\%$ CL. For $n=4$, if we assume large values of $\lambda$ as
$\lambda \sim {\cal{O}}(10^{4})$, we find that the constraint
$\beta\lesssim {\cal{O}}(10^{-2})$.

\begin{figure}[]
	\begin{center}
		\includegraphics[scale=0.47]{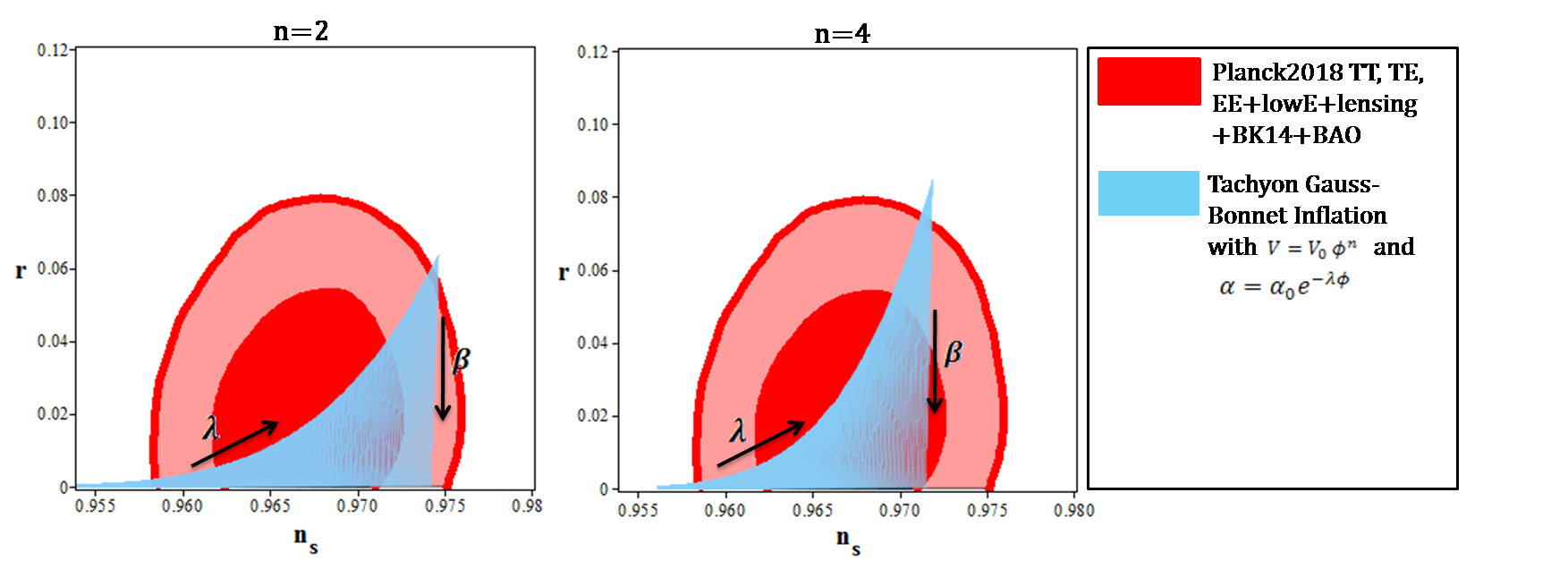}
		\includegraphics[scale=0.47]{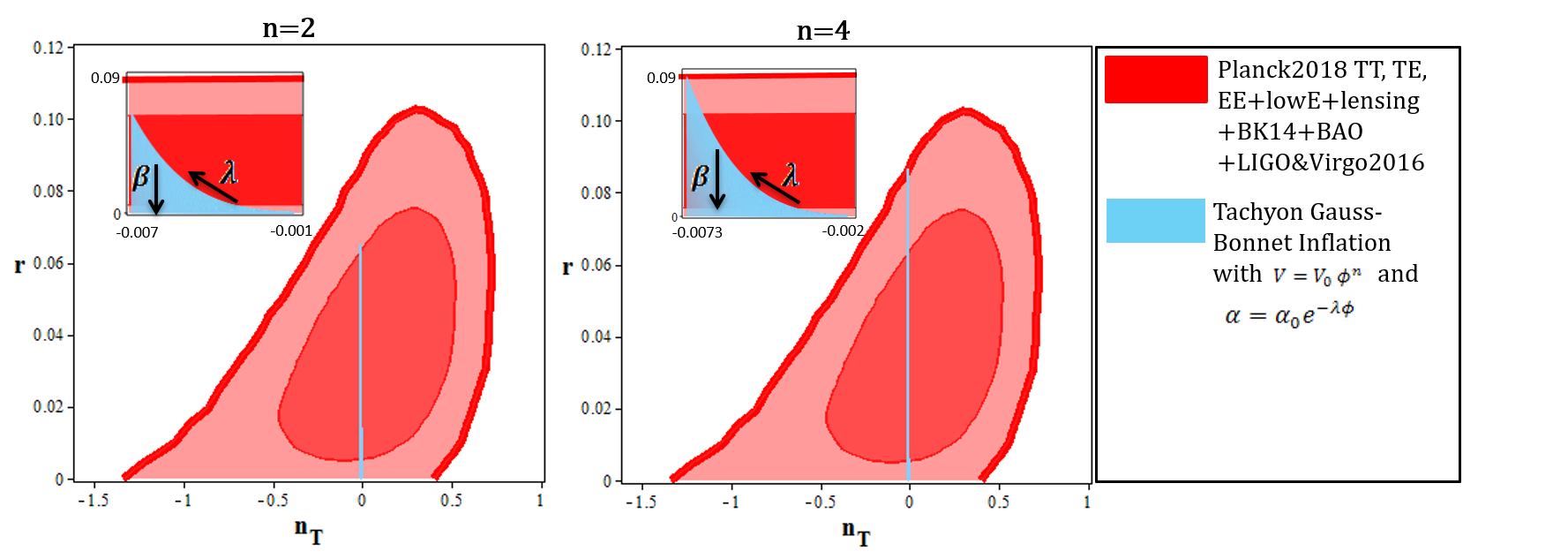}
	\end{center}
	\caption{\small Tensor-to-scalar ratio versus the
		scalar spectral index and tensor spectral index of the tachyon GB
		model with $V=V_{0}\,\phi^{n}$ and
		${\cal{G}}={\cal{G}}_{0}\,e^{-\lambda\phi}$.}
	\label{fig6}
\end{figure}

\begin{table*}
\tiny\caption{\small{\label{tab:6} The ranges of the parameter
$\beta$ in which the tensor-to-scalar ratio, the scalar spectral
index and the tensor spectral index of the tachyon GB model with
$V=V_{0}\,\phi^{n}$ and
${\cal{G}}={\cal{G}}_{0}\,e^{-\lambda\phi}$ are consistent with
different data sets.}}
\begin{center}
\begin{tabular}{ccccccc}
\\ \hline \hline \\ && Planck2018 TT,TE,EE+lowE & Planck2018 TT,TE,EE+lowE&Planck2018 TT,TE,EE+lowE&Planck2018 TT,TE,EE+lowE
\\&& +lensing+BK14+BAO &
+lensing+BK14+BAO&lensing+BK14+BAO&lensing+BK14+BAO
\\& & & &+LIGO$\&$Virgo2016 &LIGO$\&$Virgo2016
\\
\hline \\&$N$& $68\%$ CL & $95\%$ CL &$68\%$ CL & $95\%$ CL
\\
\hline \\  &$10$&  $\beta\leq 0.832 $ & all values of $\beta$& not consistent &all values of $\beta$\\
\\$n=2$&$10^{2}$& $0.421\leq \beta\leq 0.683$ & all values of $\beta$&$\beta \leq 0.742$&all values of $\beta$\\ \\
&$10^{4}$&not consistent& all values of $\beta$&$\beta \leq 0.839$&all values of $\beta$\\
\hline  \\ &$10$& all values of $\beta$ & all values of $\beta$&not consistent&all values of $\beta$\\
\\$n=4$&$10^{2}$& $0.093\leq\beta $ &all values of $\beta$&$\beta \leq 0.886$&all values of $\beta$\\ \\
&$10^{4}$& $ 0.411\leq\beta$& $0.046\leq\beta$&$0.235\leq \beta \leq 0.901$&all values of $\beta$\\
\hline \hline
\end{tabular}
\end{center}
\end{table*}

\section{DBI Gauss-Bonnet Inflation}

There is another proposal arisen from the string theory, which is
based on the Dirac-Born-Infeld action~\cite{Sil04,Ali04}. This
proposal suggests that the field responsible for inflation is
characterized by the radial coordinate of a D3 brane which moves
in a ``throat'' (often $AdS_{5}$ throat) region of a warped
compactified space. Both the speed of the brane and the warp
factor of the throat, set a speed limit upon the brane's motion.
In this model, besides the potential, there is a function of the
scalar field related to the local geometry of the compact manifold
traversed by the D3 brane. Also, the kinetic term of the field is
non-canonical~\cite{Sil04}.

In the DBI model, we have
\begin{equation}
\label{eq67}
P(X,\phi)=-{\cal{F}}^{-1}(\phi)\sqrt{1-2{\cal{F}}(\phi)\,X}-V(\phi)\,,
\end{equation}
By this definition, the scalar spectral index takes the following
forms
\begin{eqnarray}
\label{eq68} n_{s}=1-2\,M \chi- \Bigg[ 4\,M\chi^{2} \left(
{\cal{F}}^{-1}+V \right)  \left( {\frac { B}{M}}+{\frac
{\chi'}{\chi}} \right) -\frac{16}{3}\alpha'\,\chi^{2} \left(
{\cal{F}}^{-1}+V \right) ^{3} \nonumber\\ \bigg( {\frac
{\alpha''}{\alpha'}}+2\, \frac{ -{\frac
{{\cal{F}}'}{{\cal{F}}^{2}}}+V'  }{  {\cal{F}}^{-1}+V }+{\frac
{\chi'}{\chi}} \bigg)  \Bigg]  \Bigg[ 2\,M
\chi-\frac{8}{3}\,\alpha'\,\chi \left( {\cal{F}}^{-1}+V \right) ^{2}
\Bigg] ^{-1}\,.
\end{eqnarray}
The tensor spectral index in DBI GB model is given by
\begin{equation}
\label{eq69} n_{T}=-2M\chi\,,
\end{equation}
and the tensor-to-scalar ratio is obtained as follows
\begin{equation}
\label{eq70} r=-16\,M\,\chi+{\frac {64}{3}}\,\alpha'\,\chi \left(
V+{\cal{F}}^{-1} \right) ^{2}\,,
\end{equation}
where
\begin{equation}
\label{eq71} \chi=\frac{1}{2}\,\frac{  {\frac
{{\cal{F}}'}{{\cal{F}}^{2}}}-V'-24\,{H}^{4}\alpha' }{ {\kappa}^{2}
\left( {\cal{F}}^{-1}+V \right) ^{2}}\,,
\end{equation}
\begin{eqnarray}
\label{eq72} M={\frac {{\cal{F}}'}{{\cal{F}}^{2}}}-4\,{\frac
{{\cal{F}}'\, \left( {\cal{F}}^{-1}+V \right)
^{4}{\chi}^{2}}{{\cal{F}}}}-V'\,,
\end{eqnarray}
and
\begin{eqnarray}
\label{eq73} B={\frac {{\cal{F}}''}{{\cal{F}}^{2}}}-2\,{\frac
{{{\cal{F}}'}^{2}}{{\cal{F}}^{3}}}-4\,{ \frac
{\left({\cal{F}}''{\cal{F}}-{\cal{F}}'^{2}\right) \left(
{\cal{F}}^{-1}+V \right) ^{4}{\chi}^{2}}{{\cal{F}}^{2}}}
-4\,\frac{{\cal{F}}'\, \left( -4\,{\frac
{{\cal{F}}'}{{\cal{F}}^{2}}}+4\,V' \right) \left( {\cal{F}}^{-1}+V
\right) ^{3}{\chi}^{2}}{{\cal{F}}}\nonumber\\-8\,{\frac
{{\cal{F}}'\, \left( {\cal{F}}^{-1}+V \right)
^{4}\chi\chi'}{\cal{F}}}-V''\,.
\end{eqnarray}
As pervious sections, by choosing some explicit functions for the
potential and coupling function, we study this model numerically.

\subsection{Power-Law potential and Inverse Power-Law GB Coupling}
We start this subsection by adopting the following potential and
GB coupling function
\begin{equation}
\label{eq74} V=V_{0}\,\phi^{n}\,,\quad
\,{\cal{F}}={\cal{F}}_{0}\phi^{-n}\quad \&\quad
{\cal{G}}={\cal{G}}_{0}\,\phi^{-n}\,.
\end{equation}
In this case, we get
\begin{eqnarray}
\label{eq75} n_{s}=1-{\frac {4\,\mu\,\phi\,\beta\,{\varphi
}^{2}+16\,{\phi}^{n}n\beta \,\varphi -8\,{\phi}^{n}\varphi
+8\,{\phi}^{n+1}\mu-2\,{\phi}^{-n}{ \varphi
}^{5}+2\,{\phi}^{-n}n\beta\,{\varphi }^{5}-10\,{\phi}^{-n}n{ \varphi
}^{5}} {4\phi\, \left( 2\,\beta\,{\phi}^{n}+{\varphi
}^{2}-2\,{\phi}^{n} \right)
\left( {\varphi }^{2}-2\,{\phi}^{n} \right) }}\nonumber\\
-{\frac {-8\,{\phi}^{n+1}\mu\,\beta+{\phi}^{-2\,n}n{\varphi }^{7}
+6\,{\phi}^{-n+1}{\varphi }^{4}\mu+24\,{\varphi }^{3}n-12\,n\beta\,{
\varphi }^{3}-16\,{\phi}^{n}n\varphi +8\,{\phi}^{n}\beta\,\varphi
-16 \,{\varphi }^{2}\mu\,\phi} {4\phi\, \left(
2\,\beta\,{\phi}^{n}+{\varphi }^{2}-2\,{\phi}^{n} \right)
\left( {\varphi }^{2}-2\,{\phi}^{n} \right) }}\nonumber\\
-{\frac {-4\,\beta\,{\varphi }^{3}+8\,{\varphi }^{3}} {4\phi\,
\left( 2\,\beta\,{\phi}^{n}+{\varphi }^{2}-2\,{\phi}^{n} \right)
\left( {\varphi }^{2}-2\,{\phi}^{n} \right) }}\,,
\end{eqnarray}

\begin{equation}
\label{eq76} n_{T}=-{\frac {{n}^{2} \left(
16\,{\phi}^{n-2}{n}^{2}{\beta}^{3}-24\,{\phi}^
{n-2}{n}^{2}{\beta}^{2}+12\,{\phi}^{n-2}{n}^{2}\beta-2\,{\phi}^{n-2}{n
}^{2}-2\,\beta+1 \right) }{{\phi}^{2}}} \,,
\end{equation}
and
\begin{equation}
\label{eq77} r=8\,{\frac {{n}^{2} \left( 2\,{\kappa}^{4}\beta-1
\right)  \left( 8\,{
\phi}^{n-2}{n}^{2}{\kappa}^{8}{\beta}^{2}-8\,{\phi}^{n-2}{n}^{2}\beta
\,{\kappa}^{4}+{\kappa}^{4}\beta-{\kappa}^{4}+2\,{\phi}^{n-2}{n}^{2}
\right) }{{\kappa}^{6}{\phi}^{2}}} \,,
\end{equation}
with
\begin{eqnarray}
\label{eq78} \varphi=2\,{\phi}^{n-1}n \left( 2\,\beta-1 \right)\,,
\end{eqnarray}
and
\begin{eqnarray}
\label{eq79} \mu=2\,{\phi}^{n-2}n \left( 2\,\beta\,n-n-2\,\beta+1
\right)\,.
\end{eqnarray}

Performing an analysis on the above scalar spectral index and
tensor-to-scalar ratio gives Figure 7. Our numerical analysis
shows that although the GB effect makes the DBI model
observationally viable, but this viability happens with $N>52$ for
$n=2$ and $N>59$ for $n=4$. The bottom panels of Figure 7 show the
tensor spectral index versus the tensor-to-scalar ratio of the DBI
GB model with $V=V_{0}\,\phi^{n}$ and
${\cal{G}}={\cal{G}}_{0}\,\phi^{-n}$, in the background of the
base+GW data. The ranges of the parameter $\beta$ in which the
scalar spectral index, the tensor spectral index and tensor to
scalar ratio of this model are consistent with the base data at
$68\%$ CL and $95\%$ CL, are summarized in Table 7. These
considerations show that, the DBI Gauss-Bonnet inflation with
$V=V_{0}\,\phi^{n}$ and ${\cal{G}}={\cal{G}}_{0}\,\phi^{-n}$ for
both $n=2$ and $n=4$ cases is consistent with observational data
if $\beta\lesssim {\cal{O}}(10^{-1})$ and $N\gtrsim 60$.

\begin{figure}[]
	\begin{center}
		\includegraphics[scale=0.47]{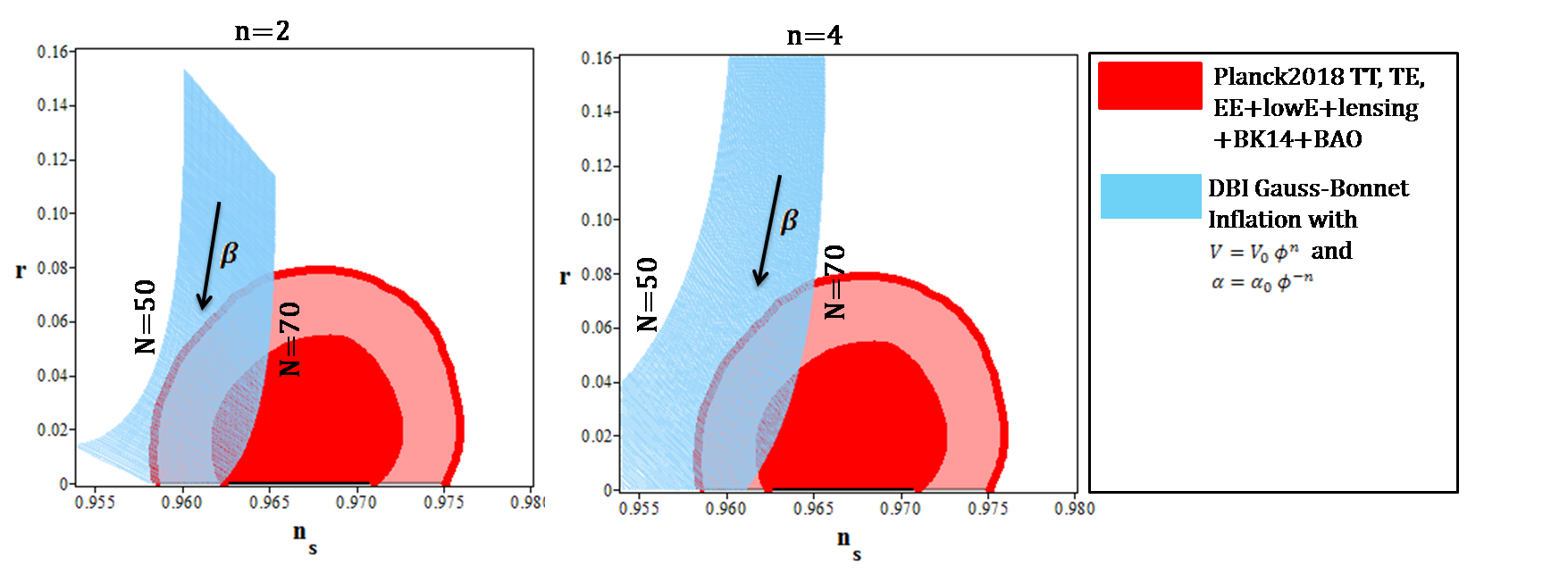}
		\includegraphics[scale=0.47]{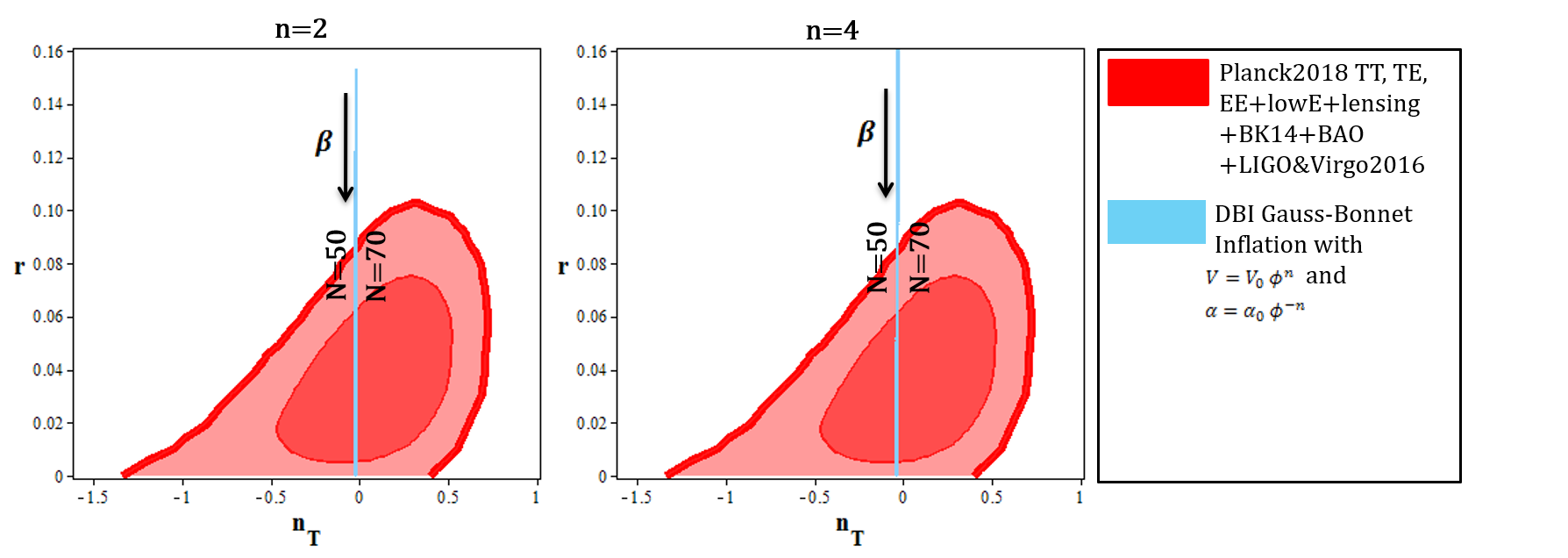}
	\end{center}
	\caption{\small Tensor-to-scalar ratio versus the
		scalar spectral index and tensor spectral index of the DBI GB
		model with $V=V_{0}\,\phi^{n}$ and
		${\cal{G}}={\cal{G}}_{0}\,\phi^{-n}$.}
	\label{fig7}
\end{figure}

\begin{table*}
\tiny\caption{\small{\label{tab:7} The ranges of the parameter
$\beta$ in which the tensor-to-scalar ratio, the scalar spectral
index and the tensor spectral index of the DBI GB model with
$V=V_{0}\,\phi^{n}$ and ${\cal{G}}={\cal{G}}_{0}\,\phi^{-n}$ are
consistent with different data sets.}}
\begin{center}
\begin{tabular}{ccccccc}
\\ \hline \hline \\ && Planck2018 TT,TE,EE+lowE & Planck2018 TT,TE,EE+lowE&Planck2018 TT,TE,EE+lowE&Planck2018 TT,TE,EE+lowE
\\&& +lensing+BK14+BAO &
+lensing+BK14+BAO&lensing+BK14+BAO&lensing+BK14+BAO
\\& & & &+LIGO$\&$Virgo2016 &LIGO$\&$Virgo2016
\\
\hline \\&$N$& $68\%$ CL & $95\%$ CL &$68\%$ CL & $95\%$ CL
\\
\hline \\ &$50$&not consistent&not consistent&$0.617\leq \beta <1$&$0.411\leq \beta<1$\\
\\$n=2$&$60$&not consistent&$0.613\leq \beta\leq0.948$&$0.607\leq \beta <1$&$0.352\leq \beta<1$\\ \\
&$70$&$0.705\leq \beta \leq0.981$&$0.326\leq \beta <1$&$0.489\leq \beta \leq 0.971$&$0.361\leq \beta <1$\\
\hline  \\ &$50$&not consistent&not consistent&$0.733\leq \beta <1$&$0.674\leq \beta <1$\\
\\$n=4$&$60$&not consistent&$0.683\leq \beta \leq0.915$&$0.712\leq \beta \leq 0.980$&$0.676\leq\beta <1$\\ \\
&$70$&$0.774\leq \beta \leq0.865$&$0.643\leq \beta <1$&$0.702\leq \beta \leq 0.984$&$0.680\leq\beta <1$\\
\hline \hline
\end{tabular}
\end{center}
\end{table*}

\begin{figure}[]
	\begin{center}
		\includegraphics[scale=0.47]{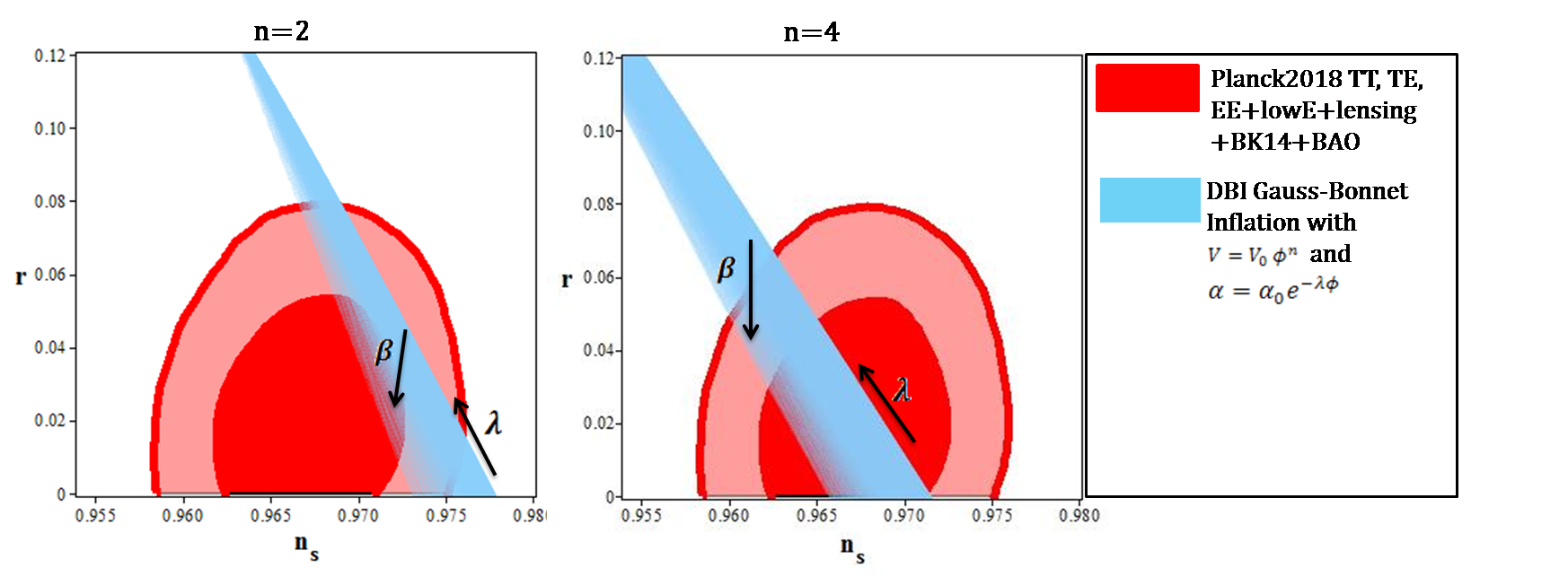}
		\includegraphics[scale=0.47]{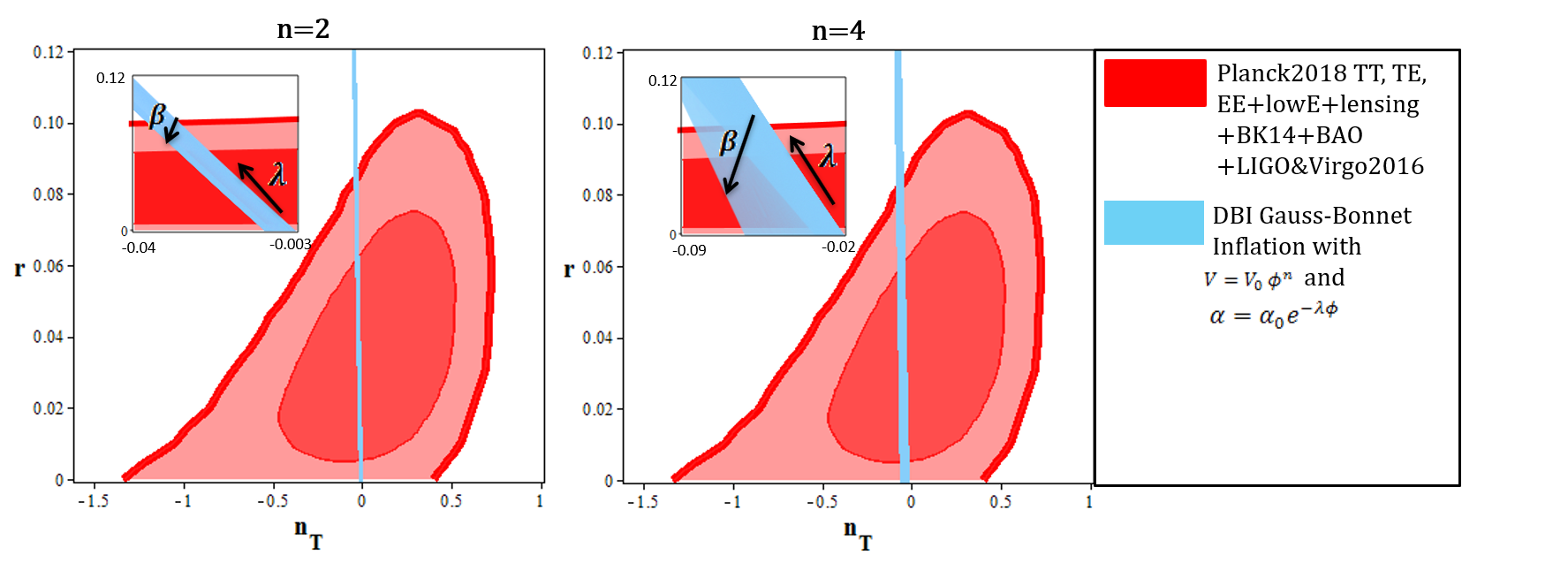}
	\end{center}
	\caption{\small Tensor-to-scalar ratio versus the
		scalar spectral index and tensor spectral index of the DBI GB
		model with $V=V_{0}\,\phi^{n}$ and
		${\cal{G}}={\cal{G}}_{0}\,e^{-\lambda\phi}$.}
	\label{fig8}
\end{figure}

\begin{table*}
\tiny\caption{\small{\label{tab:8} The ranges of the parameter
$\beta$ in which the tensor-to-scalar ratio, the scalar spectral
index and of the DBI GB model with $V=V_{0}\,\phi^{n}$ and
${\cal{G}}={\cal{G}}_{0}\,e^{-\lambda\phi}$ are consistent with
different data sets.}}
\begin{center}
\begin{tabular}{ccccccc}
\\ \hline \hline \\ && Planck2018 TT,TE,EE+lowE & Planck2018 TT,TE,EE+lowE&Planck2018 TT,TE,EE+lowE&Planck2018 TT,TE,EE+lowE
\\&& +lensing+BK14+BAO &
+lensing+BK14+BAO&lensing+BK14+BAO&lensing+BK14+BAO
\\& & & &+LIGO$\&$Virgo2016 &LIGO$\&$Virgo2016
\\
\hline \\&$N$& $68\%$ CL & $95\%$ CL &$68\%$ CL & $95\%$ CL
\\
\hline \\ &$10$&not consistent&$0.112\leq \beta \leq0.986$&$0.609\leq \beta \leq0.973$&$0.332\leq \beta<1$\\
\\$n=2$&$10^{2}$&$0.917\leq \beta<1$&all values&$0.607\leq \beta \leq0.970$&$0.329\leq \beta<1$\\ \\
&$10^{4}$&not consistent&$0.326\leq \beta <1$&$0.601\leq \beta \leq 0.966$&$0.321\leq \beta <1$\\
\hline  \\ &$10$&all values &all values&$0.521\leq \beta \leq0.971$&$0.388\leq \beta <1$\\
\\$n=4$&$10^{2}$&$0.182\leq \beta <1$&all values&$0.518\leq \beta \leq 0.969$&$0.381\leq\beta <1$\\ \\
&$10^{4}$&not consistent&$0.442\leq \beta <1$&$0.511\leq \beta \leq 0.965$&$0.370\leq\beta <1$\\
\hline \hline
\end{tabular}
\end{center}
\end{table*}

\subsection{Power-Law Potential and Dilaton-Like GB Coupling}
In this section, we consider the following potential and GB
coupling
\begin{equation}
\label{eq80} V=V_{0}\,\phi^{n}\,,\quad
\,{\cal{F}}={\cal{F}}_{0}\phi^{-n}\quad \&\quad
{\cal{G}}={\cal{G}}_{0}\,e^{-\lambda\phi}\,,
\end{equation}
which lead to
\begin{eqnarray}
\label{eq81} n_{s}=1- \frac{ -{\frac {nB}{\phi}}+{\frac
{nA}{\phi\,{\kappa}^{2}}} }{A^{-1}{\kappa}^{2}{B}^{2}}-2\,{\frac {n
\left( -B{\kappa}^{2}+A \right)  \left( -2\,{B}^{2}C{\phi
}^{n}{\kappa}^{2}+6\,A{B}^{2}C-A{\phi}^{n-1}n+A{\phi}^{n-1} \right)
}{ B \left( -2\,nB{\kappa}^{2}+2\,nA+\beta\,\lambda\,{{\rm
e}^{-\lambda\, \phi}}\phi\,{\kappa}^{2}{B}^{2} \right)  \left(
-{\phi}^{n}{\kappa}^{2 }+A \right) }}\nonumber\\
-{\frac {\beta\,\lambda\,{{\rm e}^{-\lambda\,\phi}} \left(
2\,C{\kappa }^{2}{B}^{2}\phi-\lambda\,\phi\,A+2\,nA \right)
B}{-2\,nB{\kappa}^{2}+ 2\,nA+\beta\,\lambda\,{{\rm
e}^{-\lambda\,\phi}}\phi\,{\kappa}^{2}{B}^ {2}}}\,,
\end{eqnarray}
with
\begin{eqnarray}
\label{eq82} B={\phi}^{n}+ \frac{1}{ {\phi}^{-n}}\,,
\end{eqnarray}
\begin{eqnarray}
\label{eq83} A=-{\frac
{nB}{\phi}}+{\kappa}^{4}{B}^{2}\beta\,\lambda\,{{\rm e}^{-
\lambda\,\phi}}\,,
\end{eqnarray}
and
\begin{eqnarray}
\label{eq84} C=\frac{1}{2} \left( -{\frac
{B{n}^{2}}{{\phi}^{2}}}+{\frac {nB}{{\phi}^{2}} }+2\,{\frac
{{\kappa}^{4}{B}^{2}\beta\,\lambda\,{{\rm e}^{-\lambda\,
\phi}}n}{\phi}}-{\kappa}^{4}{B}^{2}\beta\,{\lambda}^{2}{{\rm e}^{-
\lambda\,\phi}} \right) {\kappa}^{-2}{B}^{-2}-{\frac
{An}{{B}^{2}\phi \,{\kappa}^{2}}}\,.
\end{eqnarray}

The tensor spectral index is given by
\begin{equation}
\label{eq85} n_{T}=-{\frac {n \left(
n\phi\,{\kappa}^{2}-2\,{\phi}^{n+2}{\kappa}^{6}\beta
\,\lambda\,{{\rm
e}^{-\lambda\,\phi}}+{n}^{2}-4\,{\phi}^{n+1}n{\kappa}
^{4}\beta\,\lambda\,{{\rm e}^{-\lambda\,\phi}}+4\,{\phi}^{2\,n+2}{
\kappa}^{8}{\beta}^{2}{\lambda}^{2}{{\rm e}^{-2\,\lambda\,\phi}}
\right) }{{\kappa}^{4}{\phi}^{3}}} \,.
\end{equation}
The tensor-to-scala ratio takes the following form
\begin{eqnarray}
\label{eq86} r=8\,{\frac
{{n}^{2}\phi\,{\kappa}^{2}-2\,{\phi}^{n+2}n{\kappa}^{6}
\beta\,\lambda\,{{\rm
e}^{-\lambda\,\phi}}+{n}^{3}-4\,{\phi}^{n+1}{n}^
{2}{\kappa}^{4}\beta\,\lambda\,{{\rm
e}^{-\lambda\,\phi}}+4\,{\phi}^{2
\,n+2}n{\kappa}^{8}{\beta}^{2}{\lambda}^{2}{{\rm
e}^{-2\,\lambda\,\phi }}}{{\kappa}^{4}{\phi}^{3}}}\nonumber\\
+8\,{\frac {2\,{\beta}^{2}{\lambda}^{2}{{\rm
e}^{-2\,\lambda\,\phi}}{\kappa}^{6
}{\phi}^{3+2\,n}-\beta\,\lambda\,{{\rm
e}^{-\lambda\,\phi}}{\kappa}^{2
}{\phi}^{n+2}n}{{\kappa}^{4}{\phi}^{3}}}\,.
\end{eqnarray}

Now, we analyze these perturbation parameter numerically and the
results are shown in Figure 8. Our numerical analysis shows that
the GB effect makes the DBI model observationally viable. The
ranges of the parameter $\beta$ in which both the scalar spectral
index and tensor to scalar ratio of this model are consistent with
the base and base+GW data at $68\%$ CL and $95\%$ CL, are
summarized in Table 8. These considerations show that, the DBI
Gauss-Bonnet inflation with $V=V_{0}\,\phi^{n}$ and
${\cal{G}}={\cal{G}}_{0}\,e^{-\lambda\phi}$ for both $n=2$ and
$n=4$ cases is consistent with observational data if
$\beta\lesssim {\cal{O}}(10^{-1})$.

\section{Reheating Phase in a Gauss-Bonnet Model with Canonical Scalar Field}
The reheating process after inflation is necessary to reheat the
universe for subsequent evolution. Actually, this process can
explain the cosmic origin of the matter component of the
universe~\cite{Ko94,Kof96}. The production of cosmic relics, such
as photons and neutrinos, can be explained by considering the
process of reheating in the universe~\cite{Gi99,Wa18}. Also, the
reheating phase accounts for the observed matter-antimatter
asymmetry in the universe~\cite{Di04,Lo14}. By studying the
reheating process in the Gauss-Bonnet models, we can find more
constraints on the model's parameter space. To study this process,
we focus on two important parameters $N_{rh}$ and $T_{rh}$ (where
subscript \emph{rh} stands for reheating). We obtain some
expressions for these parameters in terms of the scalar spectral
index, which let us to compare the model with observational data
(see Refs.~\cite{Dai14,Un15,Co15,Cai15,Ue16}). By using the
following expression
\begin{equation}
\label{eq87} N_{hc}=\ln \left(\frac{a_{e}}{a_{hc}}\right)\,,
\end{equation}
we define the e-folds number between the time when the physical
scales cross the horizon and the time when the inflation ends. The
subscripts $hc$ and $e$ show the value of the parameter at the
horizon crossing and end of inflation, respectively. For the energy
density during the reheating epoch, we have the relation $\rho\sim
a^{-3(1+\omega_{eff})}$, with $\omega_{eff}$ being the effective
equation of state corresponding to the dominant energy density in
the universe. Therefore, the e-folds number is written in terms of
$\rho$ and $\omega_{eff}$ as follows
\begin{eqnarray}\label{eq88}
N_{rh}=\ln\left(\frac{a_{rh}}{a_{e}}\right)=-\frac{1}{3(1+\omega_{eff})}\ln\left(\frac{\rho_{rh}}{\rho_{e}}\right)\,,
\end{eqnarray}
At the horizon crossing ($k=aH$) we have
\begin{eqnarray}\label{eq89}
0=\ln\left(\frac{k_{hc}}{a_{hc}\,H_{hc}}\right)=
\ln\left(\frac{a_{e}}{a_{hc}}\frac{a_{rh}}{a_{e}}\frac{a_{0}}{a_{rh}}\frac{k_{hc}}{a_{0}H_{hc}}\right)\,,
\end{eqnarray}
where the subscript $0$ shows the value of the scale factor at the
current time. From equations \eqref{eq42}, \eqref{eq43} and
\eqref{eq44} we obtain
\begin{eqnarray}\label{eq90}
N_{hc}+N_{rh}+\ln\left(\frac{k_{hc}}{a_{0}H_{hc}}\right)+\ln\left(\frac{a_{0}}{a_{rh}}\right)=0\,.
\end{eqnarray}
To rewrite $\frac{a_{0}}{a_{rh}}$ in terms of temperature and
density, we use the following expression \cite{Co15,Ue16}
\begin{equation}\label{eq91}
\rho_{rh}=\frac{\pi^{2}g_{rh}}{30}T_{rh}^{4}\,,
\end{equation}
where the parameter $g_{rh}$ is the effective number of the
relativistic species at the reheating era. Also, the conservation of
the entropy gives \cite{Co15,Ue16}
\begin{equation}\label{eq92}
\frac{a_{0}}{a_{rh}}=\left(\frac{43}{11g_{rh}}\right)^{-\frac{1}{3}}\frac{T_{rh}}{T_{0}}\,,
\end{equation}
where the subscript $0$ denotes the current value of the
temperature. Now, from equations \eqref{eq91} and \eqref{eq92}, we
find the following expression for the scale factor
\begin{eqnarray}\label{eq93}
\frac{a_{0}}{a_{rh}}=\left(\frac{43}{11g_{rh}}\right)^{-\frac{1}{3}}T_{0}^{-1}\left(\frac{\pi^{2}g_{rh}}{30\rho_{rh}}\right)^{-\frac{1}{4}}\,.
\end{eqnarray}
In the GB model with a canonical scalar field, we can write the
energy density as follows
\begin{equation}\label{eq94}
\rho=\left(1+\frac{\epsilon}{3}\right)V-\frac{160}{27}\kappa^{6}\alpha'^{2}\,V^{3}-\frac{20}{9}\kappa^{2}\alpha'
V' V\,.
\end{equation}
To obtain the energy density at the end of inflation era, we set
$\epsilon=1$. Then, we find
\begin{equation}\label{eq95}
\rho_{e}=\frac{4}{3}V_{e}-\frac{160}{27}\kappa^{6}\alpha_{e}'^{2}\,V_{e}^{3}-\frac{20}{9}\kappa^{2}\alpha_{e}'
V_{e}'\, V_{e}\,.
\end{equation}
Now, by using equations \eqref{eq88} and \eqref{eq95} we obtain
\begin{eqnarray}\label{eq96}
\rho_{rh}=\left[\frac{4}{3}V_{e}-\frac{160}{27}\kappa^{6}\alpha_{e}'^{2}\,V_{e}^{3}-\frac{20}{9}\kappa^{2}\alpha_{e}'
V_{e}'\, V_{e}\right]\times \exp\Big[-3N_{rh}(1+\omega_{eff})\Big].
\end{eqnarray}
By using equations \eqref{eq93} and \eqref{eq96}, we find the
following expression for the scale factor
\begin{eqnarray}\label{eq97}
\ln\left(\frac{a_{0}}{a_{rh}}\right)=-\frac{1}{3}\ln\left(\frac{43}{11g_{rh}}\right)
-\frac{1}{4}\ln\left(\frac{\pi^{2}g_{rh}}{30\rho_{rh}}\right)-\ln T_{0}-\frac{3}{4}N_{rh}(1+\omega_{eff})\nonumber\\
+\frac{1}{4}\ln\left(\frac{4}{3}V_{e}-\frac{160}{27}\kappa^{6}\alpha_{e}'^{2}\,V_{e}^{3}-\frac{20}{9}\kappa^{2}\alpha_{e}'
V_{e}'\,V_{e}\right)\,.
\end{eqnarray}
To obtain $N_{rh}$, we find $H_{hc}$ from equation \eqref{eq14}.
After that, by using equations \eqref{eq90} and \eqref{eq97}, we
obtain the e-folds number during reheating as follows
\begin{eqnarray}\label{eq98}
N_{rh}=\frac{4}{1-3\omega_{eff}}\Bigg[-N_{hc}-\ln\Big(\frac{k_{hc}}{a_{0}T_{0}}\Big)-\frac{1}{4}\ln\Big(\frac{40}{\pi^{2}g_{rh}}\Big)
+\frac{1}{2}\ln\Big(8\pi^{2}{\cal{A}}_{s}{\cal{W}}_{s}
c_{s}^{3}\Big)-\frac{1}{3}\ln\Big(\frac{11g_{rh}}{43}\Big)\nonumber\\
-\frac{1}{4}\ln\bigg(\frac{4}{3}V_{e}-\frac{160}{27}\kappa^{6}\alpha_{e}'^{2}\,V_{e}^{3}-\frac{20}{9}\kappa^{2}\alpha_{e}'
V_{e}'\,V_{e}\bigg)\Bigg]\,.
\end{eqnarray}
The temperature during reheating is obtained from equations
\eqref{eq88}, \eqref{eq92} and \eqref{eq95} as follows
\begin{equation}\label{eq99}
T_{rh}=\bigg(\frac{30}{\pi^{2}g_{rh}}\bigg)^{\frac{1}{4}}\,
\bigg[\frac{4}{3}V_{e}-\frac{160}{27}\kappa^{6}\alpha_{e}'^{2}\,V_{e}^{3}-\frac{20}{9}\kappa^{2}\alpha_{e}'
V_{e}'\,V_{e}\bigg]^{\frac{1}{4}}\,\times\exp\bigg[-\frac{3}{4}N_{rh}(1+\omega_{eff})\bigg]\,.
\end{equation}
To perform a numerical analysis, it is useful to write equations
\eqref{eq98} and \eqref{eq99} in terms of the scalar spectral index.
To this end, we should specify the potential and GB coupling. In the
following, we adopt the potential and GB coupling used in the
previous sections and explore each case separately.

\subsection{Power-Law potential and Inverse Power-Law GB Coupling}
In this case, we use the potential and GB coupling defined in
equation \eqref{eq29}. As we have seen before, in this case the GB
model only with $n=2$ is consistent with the base data. Therefore,
we use equation \eqref{eq29} with $n=2$, find the final values of
the potential and GB coupling in terms of the scalar field at the
horizon crossing and substitute them in equations \eqref{eq98} and
\eqref{eq99}. After that, we obtain the scalar field at horizon
crossing in terms of the scalar spectral index. By considering the
value of the scalar spectral index, obtained from the base data,
we find some constraints on the e-folds number and temperature
during the reheating phase. In studying the $r-n_{s}$ behavior
with equation \eqref{eq19}, the tightest constraint on $\beta$ has
been obtained as $0.680\leq \beta <1$. To numerically analyze of the
reheating phase, we use this constraint and four values of
$\omega$ as $\omega=-1,-\frac{1}{3},0$ and 1. The results are
presented in Table 9. The behavior of $N_{rh}$ and $T_{rh}$ versus
$n_{s}$, for $\beta=0.7$, has been shown in Figure 9. Note that,
as the top left panel of Figure 9 shows, all curves converge to
$N_{rh}=0$ (corresponding to instantaneous reheating process) and
$n_{s}=0.965$ which is observationally viable from base data.

The bottom panel of Figure 9 shows the range of $N_{rh}$ and
$\omega_{eff}$, in the case considered in this subsection, leading
to the observationally viable values of the scalar spectral index.
As figure shows, when $\omega_{eff}$ changes from $-1$ (field's
potential domination) to $\frac{1}{3}$ (radiation domination) the
values of $N_{rh}$ increase. This means that, the reheating phase
of the universe is not instantaneous and lasts some e-folds.

The parameter $N_{rh}$, which describes the duration of the
reheating phase, is related to the scalar spectral index of the
perturbations. To have viable reheating phase, $N_{rh}$ should be
consistent with observational value of $n_{s}$. Any small
variation of $N_{rh}$ can lead to the values of $n_{s}$ which are
not observationally viable. Therefore, the reheating phase should
last until some specific values and not more than it. In this
regard, we have tried to obtain some precise values of the
parameters leading to the viable GB models.

\begin{table*}
\begin{small}
\caption{\small{\label{tab:9} Constraints on the e-folds number
and temperature during the reheating phase in the GB model with
canonical scalar field and with $V=V_{0}\,\phi^{n}$ and
${\cal{G}}={\cal{G}}_{0}\,\phi^{-n}$, obtained from Planck2018 TT,
TE, EE+lowE+lensing+BK14+BAO joint data.}}
\begin{center}
\begin{tabular}{cccccc}
\\ \hline \hline \\ && $\omega=-1$& $\omega=-\frac{1}{3}$
&$\omega=0$&$\omega=1$
\\
\hline \\ $n=2$&$0.680\leq\beta < 1$&$N_{rh}<7.4$&$N_{rh}<12.37$&$N_{rh}<23.35$& all values of $N_{rh}$\\ \\
\hline  \hline \\
$n=2$&$0.680\leq\beta <
1$&$\log_{10}\left(\frac{T_{rh}}{GeV}\right)>14.16$&$\log_{10}\left(\frac{T_{rh}}{GeV}\right)>7.96$&
$\log_{10}\left(\frac{T_{rh}}{GeV}\right)>-0.73$&all values of $T_{rh}$\\ \\
\hline \hline
\end{tabular}
\end{center}
\end{small}
\end{table*}

\begin{figure}[]
	\begin{center}
		\includegraphics[scale=0.5]{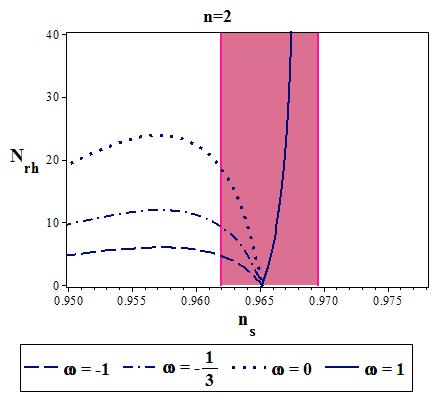}
		\includegraphics[scale=0.47]{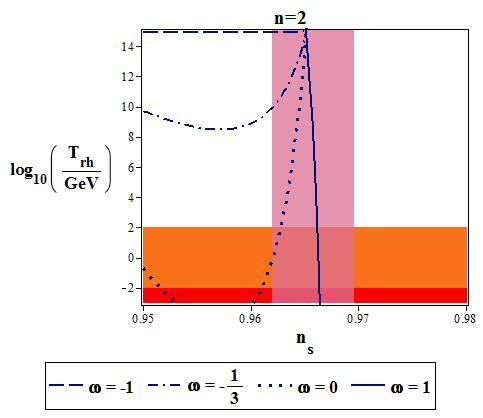}
		\includegraphics[scale=0.55]{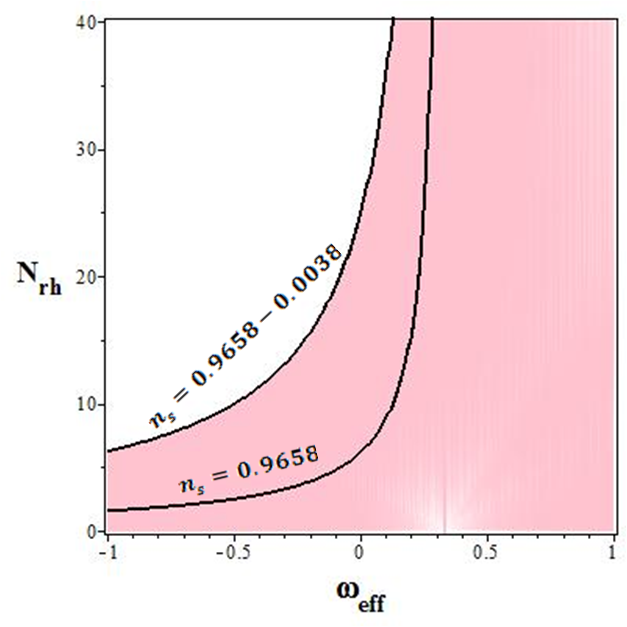}
	\end{center}
	\caption{\small Behavior of the e-folds number
		(top left panel) and temperature (top right panel) during the
		reheating phase versus , and the range of
		$N_{rh}$ and $\omega_{eff}$ leading to the observationally viable
		values of the scalar spectral index (bottom panel), in the GB model
		with canonical scalar field and with $V=V_{0}\,\phi^{2}$ and
		${\cal{G}}={\cal{G}}_{0}\,\phi^{-2}$. The magenta region in the top
		panels shows the values of the scalar spectral index released by
		Planck2018 TT, TE, EE+lowE+lensing+BK14+BAO joint data. In the
	top right panel, the orange region corresponds to the
		temperatures below the electroweak scale, $T<100$ GeV and the red
		region corresponds to the temperatures below the big bang
		nucleosynthesis scale, $T<10$ MeV.}
	\label{fig9}
\end{figure}

\begin{table*}
\begin{small}
\tiny\caption{\small{\label{tab:10} Constraints on the e-folds
number and temperature during the reheating phase in the GB model
with canonical scalar field and with $V=V_{0}\,\phi^{n}$ and
${\cal{G}}={\cal{G}}_{0}\,e^{-\lambda\phi}$, obtained from
Planck2018 TT, TE, EE+lowE+lensing+BK14+BAO joint data.}}
\begin{center}
\begin{tabular}{cccccc}
\\ \hline \hline \\ && $\omega=-1$& $\omega=-\frac{1}{3}$
&$\omega=0$&$\omega=1$
\\
\hline
\\$n=2$&$0.031\leq\beta<0.072$&$N_{rh}<6.85$&$N_{rh}<11.47$&$N_{rh}<22.66$&$N_{rh}<16.72$\\
\\
\hline
\\$n=4$&$0.670\leq\beta\leq
0.716$&$N_{rh}<0.56$&$N_{rh}<1.42$&$N_{rh}<2.37$&$N_{rh}<5.14$\\
\\
\hline  \hline
\\$n=2$&$0.031\leq\beta<0.072$&$\log_{10}\left(\frac{T_{rh}}{GeV}\right)>14.09$&$\log_{10}\left(\frac{T_{rh}}{GeV}\right)>8.65$&
$\log_{10}\left(\frac{T_{rh}}{GeV}\right)>-2.21$&all values of
$T_{rh}$\\
\hline \\
$n=4$&$0.670\leq\beta\leq
0.716$&$\log_{10}\left(\frac{T_{rh}}{GeV}\right)>14.18$&$\log_{10}\left(\frac{T_{rh}}{GeV}\right)>14.53$&
$\log_{10}\left(\frac{T_{rh}}{GeV}\right)>13.22$&$\log_{10}\left(\frac{T_{rh}}{GeV}\right)>6.27$\\ \\
\hline \hline
\end{tabular}
\end{center}
\end{small}
\end{table*}

\begin{figure}[]
	\begin{center}
		\includegraphics[scale=0.34]{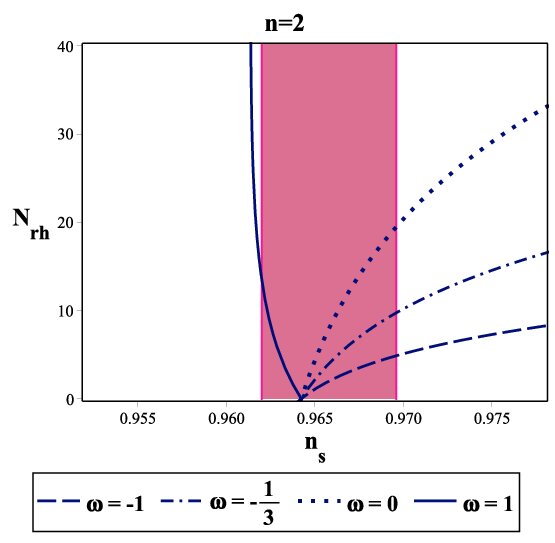}
		\includegraphics[scale=0.33]{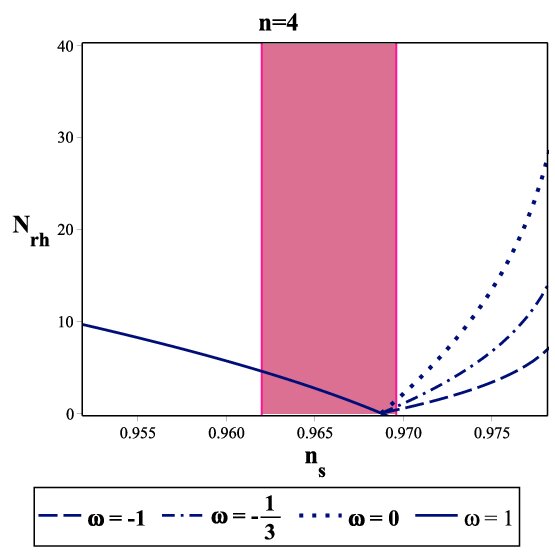}
		\includegraphics[scale=0.44]{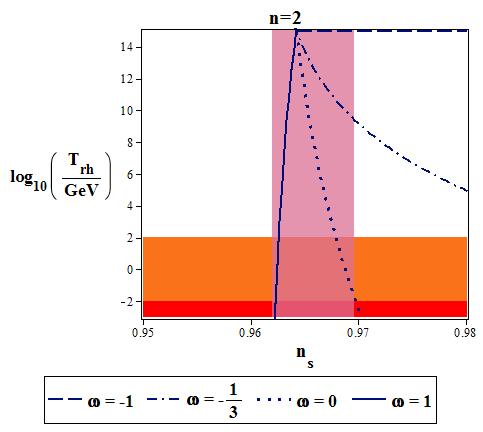}
		\includegraphics[scale=0.44]{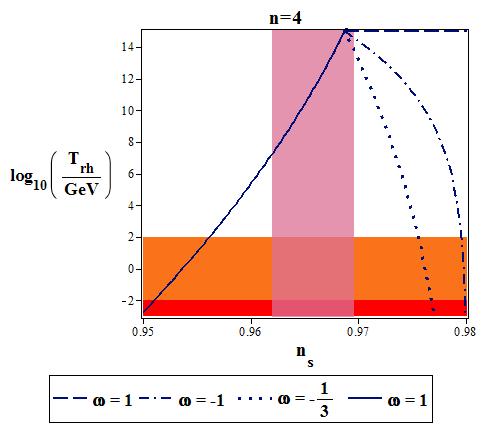}
		\includegraphics[scale=0.44]{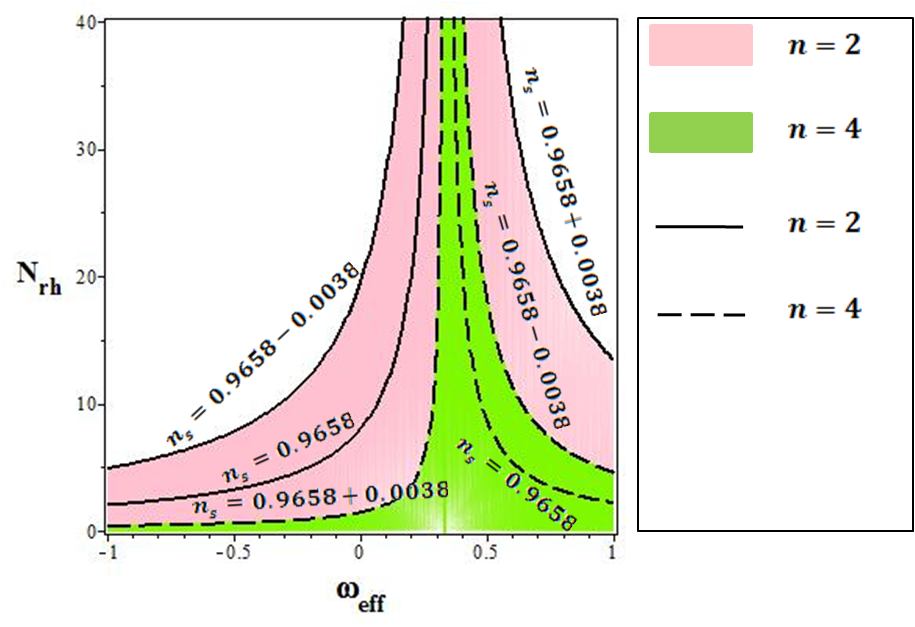}
	\end{center}
	\caption{\small Behavior of the e-folds number
		(top left panel) and temperature (top right panel) during the
		reheating phase versus the scalar spectral index, and the range of
		$N_{rh}$ and $\omega_{eff}$ leading to the observationally viable
		values of the scalar spectral index (bottom panel), in the GB model
		with canonical scalar field and with $V=V_{0}\,\phi^{2}$ and
		${\cal{G}}={\cal{G}}_{0}\,\phi^{-2}$. The magenta region in the top
		panels shows the values of the scalar spectral index released by
		Planck2018 TT, TE, EE+lowE+lensing+BK14+BAO joint data. In the
		top right panel, the orange region corresponds to the
		temperatures below the electroweak scale, $T<100$ GeV and the red
		region corresponds to the temperatures below the big bang
		nucleosynthesis scale, $T<10$ MeV.}
	\label{fig10}
\end{figure}

\subsection{Power-Law Potential and Dilaton-Like GB Coupling}
In this subsection, we use the potential and GB coupling defined
in equation \eqref{eq34}. We have shown that, in this case the GB
model with both $n=2$ and $n=4$, in some ranges of the parameters
is consistent with the observational data. By using the adopted
potential and GB coupling, we study the e-folds number and
temperature in the reheating phase numerically and find some
constraints on these parameters in confrontation with the base
data. To obtain the constraints, we use the ranges of $\beta$ from
Table 2. In $n=2$ case, the tightest constraint on $\beta$ is
$0.031\leq \beta \leq0.072$ and in $n=4$ case we have $0.670\leq
\beta \leq0.716$. With these ranges of $\beta$, we obtain the
constraints shown in Table 10. The behavior of $N_{rh}$ and
$T_{rh}$ versus $n_{s}$, for $\beta=0.7$, has been shown in the
top and middle panels of Figure 10.

The bottom panel of Figure 10 shows the range of $N_{rh}$ and
$\omega_{eff}$, in the case considered in this subsection, leading
to the observationally viable values of the scalar spectral index.
In this case also, when $\omega_{eff}$ changes from $-1$ to
$\frac{1}{3}$, the values of $N_{rh}$ increase. This means that,
the reheating phase of the universe is not instantaneous and lasts
some e-folds. However, this figure shows another point too. For
$\omega_{eff}>\frac{1}{3}$, by increasing the values of the
effective equation of state parameter the values of $N_{rh}$
decrease. Therefore, in this case, the value of $\omega_{eff}$
does not become larger than $\frac{1}{3}$, because the e-folds
number just increases and can not decrease.

\subsection{Gauss-Bonnet Natural Inflation}
Now, we study the reheating phase in the GB natural inflation
model. By using \eqref{eq38}, we study the e-folds number and
temperature in the reheating phase numerically and find some
constraints on these parameters. In this case, the tightest
constraint obtained from the base data at $95\%$ CL is
$0.525\leq\beta<1$ (see Table 3). The numerical results
corresponding to this constraint, are shown in Table 11. The
behavior of $N_{rh}$ and $T_{rh}$ versus $n_{s}$, for $\beta=0.7$,
has been shown in the top panels of Figure 11. As figure shows,
the instantaneous reheating in this case is corresponding to
$n_{s}=0.965$ which is observationally viable.

The bottom panel of Figure 11 shows the range of $N_{rh}$ and
$\omega_{eff}$, in the GB natural inflation model, leading to the
observationally viable values of the scalar spectral index. As
figure shows, the reheating phase of the universe described by GB
natural inflation is not instantaneous. In this case also, the
value of $\omega_{eff}$ does not become larger than $\frac{1}{3}$.

\begin{figure}[]
	\begin{center}
		\includegraphics[scale=0.36]{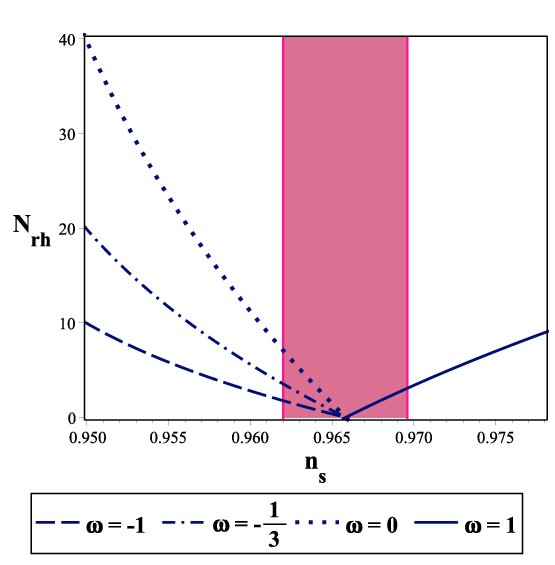}
		\includegraphics[scale=0.47]{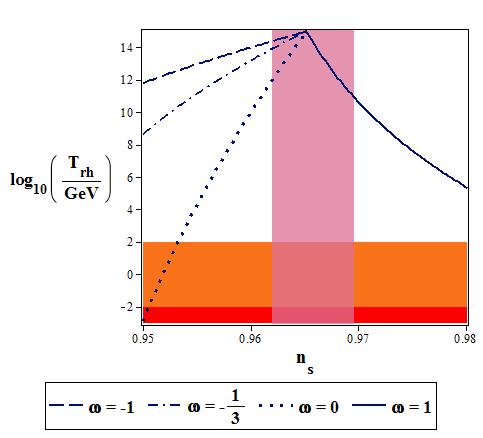}
		\includegraphics[scale=0.58]{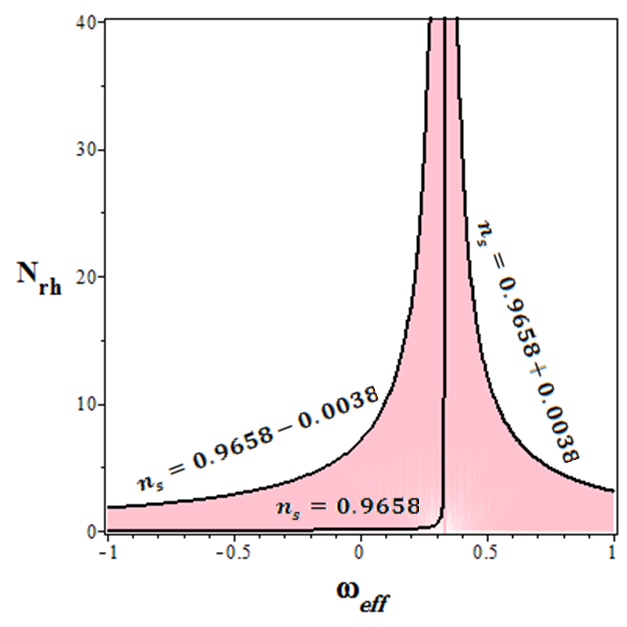}
	\end{center}
	\caption{\small Behavior of the e-folds number
		(top left panel) and temperature (top right panel) during the
		reheating phase versus the scalar spectral index, and the range of
		$N_{rh}$ and $\omega_{eff}$ leading to the observationally viable
		values of the scalar spectral index (bottom panel), in the GB
		natural inflation.}
	\label{fig11}
\end{figure}

\begin{table*}
\begin{small}
\caption{\small{\label{tab:11} Constraints on the e-folds number
and temperature during the reheating phase in the GB natural
inflation, obtained from Planck2018 TT, TE,
EE+lowE+lensing+BK14+BAO joint data.}}
\begin{center}
\begin{tabular}{cccccc}
\\ \hline \hline \\ & $\omega=-1$& $\omega=-\frac{1}{3}$
&$\omega=0$&$\omega=1$
\\
\hline \\ $0.525\leq\beta<
1$&$N_{rh}<2.85$&$N_{rh}<4.84$&$N_{rh}<10.42$&$N_{rh}<5.11$\\ \\
\hline  \hline \\
$0.525\leq\beta<
1$&$\log_{10}\left(\frac{T_{rh}}{GeV}\right)>14.04$&$\log_{10}\left(\frac{T_{rh}}{GeV}\right)>12.95$&
$\log_{10}\left(\frac{T_{rh}}{GeV}\right)>10.45$&$\log_{10}\left(\frac{T_{rh}}{GeV}\right)>9.71$\\
\\ \hline \hline
\end{tabular}
\end{center}
\end{small}
\end{table*}

\subsection{Gauss-Bonnet $\alpha$-attractor}

To study the reheating phase in the GB $\alpha$-attractor model,
we consider both E-model and T-Model types of potential and GB
coupling. In the following we present the results for each case.

\subsubsection{E-Model}
In this case we adopt the E-Model potential and GB coupling
defined in equation \eqref{eq42}. As before, by using these
functions, we study the e-folds number and temperature in
reheating phase numerically. With E-Model potential and GB
coupling, the tightest constraint obtained from the base data at
$95\%$ CL is $\beta\leq4.81\times 10^{-2}$. This range of $\beta$
leads to the constraints presented in Table 12. To obtain these
constraints, we have set $\alpha=50$. The behavior of $N_{rh}$ and
$T_{rh}$ versus $n_{s}$, for $\beta=0.03$, has been shown in the
top and middle panels of Figure 12. In this case, the
instantaneous reheating, corresponding to $n_{s}=0.965$, is
observationally viable. The bottom panel of Figure 12 shows the
range of $N_{rh}$ and $\omega_{eff}$ leading to the
observationally viable values of the scalar spectral index.

\begin{table*}
\begin{small}
\tiny\caption{\small{\label{tab:12} Constraints on the e-folds
number and temperature during the reheating phase in the GB
$\alpha$-attractor model, obtained from Planck2018 TT, TE,
EE+lowE+lensing+BK14+BAO joint data.}}
\begin{center}
\begin{tabular}{ccccccc}
\\ \hline \hline \\ && $\omega=-1$& $\omega=-\frac{1}{3}$
&$\omega=0$&$\omega=1$
\\
\hline
\\E-Model&$\beta\leq 4.81\times 10^{-2}$&$N_{rh}\leq16.825$&$N_{rh}\leq37.521$&$N_{rh}\leq52.647$&$N_{rh}\leq26.367$\\ \\
\hline
\\T-Model&$\beta\leq 7.01\times
10^{-2}$&$N_{rh}\leq4.110$&$N_{rh}\leq7.334$&$N_{rh}\leq12.45$&$N_{rh}\leq10.14$\\
\\
\hline  \hline \\
\\E-Model&$\beta\leq 4.81\times 10^{-2}$&$\log_{10}\left(\frac{T_{rh}}{GeV}\right)\geq12.49$&$\log_{10}\left(\frac{T_{rh}}{GeV}\right)\geq8.73$&
$\log_{10}\left(\frac{T_{rh}}{GeV}\right)\geq4.73$&$\log_{10}\left(\frac{T_{rh}}{GeV}\right)\geq0.266$\\ \\
\hline\\
\\T-Model&$\beta\leq 7.01\times 10^{-2}$&$\log_{10}\left(\frac{T_{rh}}{GeV}\right)\geq12.66$&$\log_{10}\left(\frac{T_{rh}}{GeV}\right)\geq10.21$&
$\log_{10}\left(\frac{T_{rh}}{GeV}\right)\geq7.12$&$\log_{10}\left(\frac{T_{rh}}{GeV}\right)\geq8.37$\\ \\
\hline \hline
\end{tabular}
\end{center}
\end{small}
\end{table*}

\begin{figure}[]
	\begin{center}
		\includegraphics[scale=0.32]{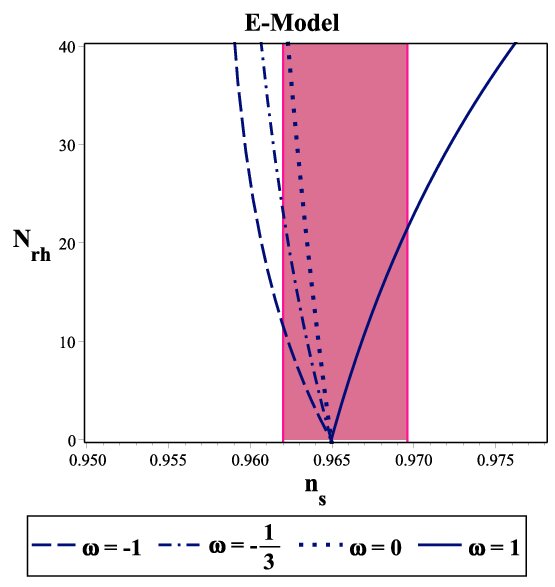}
		\includegraphics[scale=0.32]{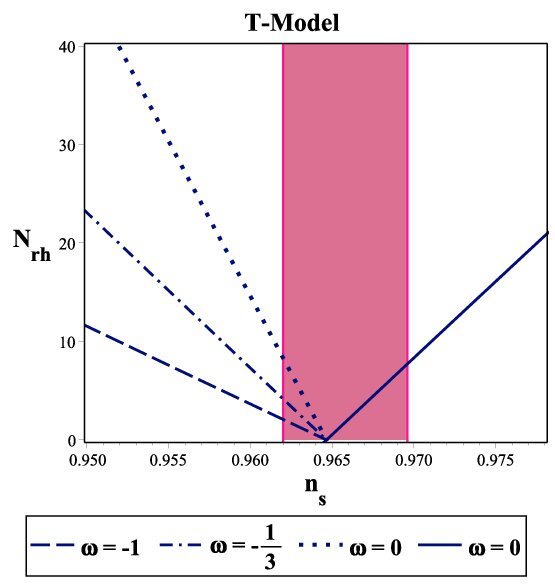}
		\includegraphics[scale=0.43]{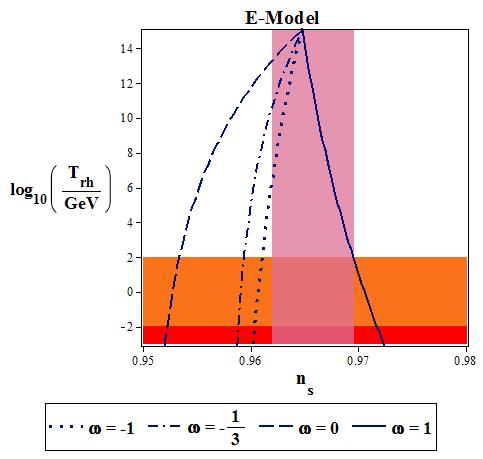}
		\includegraphics[scale=0.43]{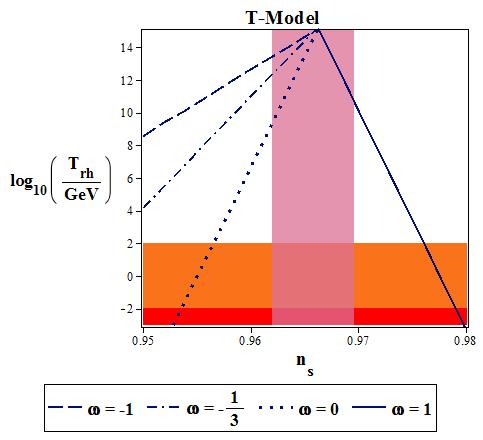}
		\includegraphics[scale=0.55]{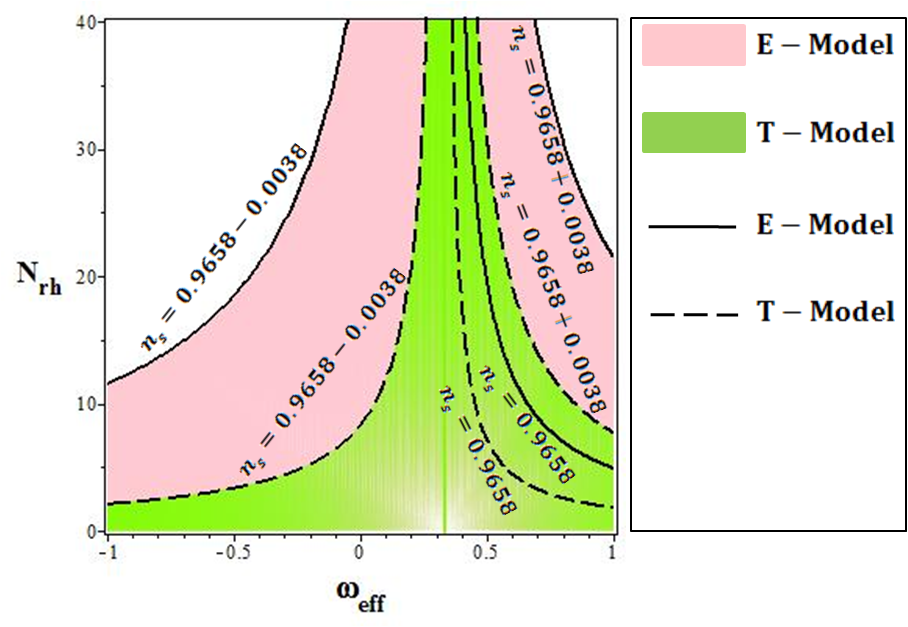}
	\end{center}
	\caption{\small Behavior of the e-folds number
		(top panels) and temperature (middle panels) the
		reheating phase versus the scalar spectral index, and the range of
		$N_{rh}$ and $\omega_{eff}$ leading to the observationally viable
		values of the scalar spectral index (bottom panel), in the GB
		$\alpha$-attractor model.}
	\label{fig12}
\end{figure}

\subsubsection{T-Model}
Now, we consider the T-Model potential and GB coupling defined in
equation \eqref{eq47}. In this case also, we analyze the e-folds
number and temperature in the reheating phase in confrontation
with observational data. With T-Model potential and GB coupling,
the tightest constraint on $\beta$, obtained from the base data at
$95\%$ CL, is $\beta\leq7.01\times 10^{-2}$ (see Table 4). The
numerical results of the T-model case with $\beta\leq7.01\times
10^{-2}$ and $\alpha=50$ are shown in Table 12. The behavior of
$N_{rh}$ and $T_{rh}$ versus $n_{s}$, for $\beta=0.03$, has been
shown in Figure 12. Here also, the instantaneous reheating is
favored by observational data. The bottom panel of Figure 12 shows
the range of $N_{rh}$ and $\omega_{eff}$, in the case considered
in this subsection, leading to the observationally viable values
of the scalar spectral index. According to our numerical analysis,
in the GB $\alpha$-attractor model with both E-model and T-model
functions, the reheating phase of the universe is not
instantaneous and the value of $\omega_{eff}$ does not become
larger than $\frac{1}{3}$.

\section{Reheating in a Gauss-Bonnet Model with Tachyon Field}

In this section, we study the reheating process in a Gauss-Bonnet
model with tachyon field. Here, we can use the equations
\eqref{eq88}-\eqref{eq93} of section 8. However, the energy density
and the equation of motion in the tachyon GB model are different
from the GB model with canonical scalar field. In this regard, we
have
\begin{equation}\label{eq100}
\rho=V\Bigg(1-\frac{2}{3}\epsilon-\frac{64}{27}\kappa^{6}V^{2}\alpha'^{2}-\frac{8}{3}\kappa^{2}\alpha'
V'\Bigg)^{-\frac{1}{2}}-\frac{64}{9}\kappa^{6}V^{2}\alpha'^{2}-\frac{8}{3}\kappa^{2}\alpha'
V'\,.
\end{equation}
At the end of inflation ($\epsilon=1$) we obtain
\begin{equation}\label{eq101}
\rho_{e}=V_{e}\Bigg(\frac{1}{3}-\frac{64}{27}\kappa^{6}V_{e}^{2}\alpha_{e}'^{2}-\frac{8}{3}\kappa^{2}\alpha_{e}'
V_{e}'\Bigg)^{-\frac{1}{2}}-\frac{64}{9}\kappa^{6}V_{e}^{2}\alpha_{e}'^{2}-\frac{8}{3}\kappa^{2}\alpha_{e}'
V_{e}'\,.
\end{equation}
Now, by using equations \eqref{eq88} and \eqref{eq101} we reach
\begin{eqnarray}\label{eq102}
\rho_{rh}=\left[V_{e}\Bigg(\frac{1}{3}-\frac{64}{27}\kappa^{6}V_{e}^{2}\alpha_{e}'^{2}-\frac{8}{3}\kappa^{2}\alpha_{e}'
V_{e}'\Bigg)^{-\frac{1}{2}}-\frac{64}{9}\kappa^{6}V_{e}^{2}\alpha_{e}'^{2}-\frac{8}{3}\kappa^{2}\alpha_{e}'
V_{e}'\right]\nonumber\\ \times
\exp\Big[-3N_{rh}(1+\omega_{eff})\Big].
\end{eqnarray}
Equations \eqref{eq93} and \eqref{eq102} give the following
expression for the scale factor
\begin{eqnarray}\label{eq103}
\ln\left(\frac{a_{0}}{a_{rh}}\right)=-\frac{1}{3}\ln\left(\frac{43}{11g_{rh}}\right)
-\frac{1}{4}\ln\left(\frac{\pi^{2}g_{rh}}{30\rho_{rh}}\right)-\ln T_{0}-\frac{3}{4}N_{rh}(1+\omega_{eff})\nonumber\\
+\frac{1}{4}\ln\left[V_{e}\Bigg(\frac{1}{3}-\frac{64}{27}\kappa^{6}V_{e}^{2}\alpha_{e}'^{2}-\frac{8}{3}\kappa^{2}\alpha_{e}'
V_{e}'\Bigg)^{-\frac{1}{2}}-\frac{64}{9}\kappa^{6}V_{e}^{2}\alpha_{e}'^{2}-\frac{8}{3}\kappa^{2}\alpha_{e}'
V_{e}'\right]\,.
\end{eqnarray}
To find $N_{rh}$, as before, we use $H_{hc}$ obtained from
equation \eqref{eq14}. Then, using \eqref{eq90} and \eqref{eq103}
gives the e-folds number during the reheating process as follows
\begin{eqnarray}\label{eq104}
N_{rh}=\frac{4}{1-3\omega_{eff}}\Bigg\{-N_{hc}-\ln\Big(\frac{k_{hc}}{a_{0}T_{0}}\Big)-\frac{1}{4}\ln\Big(\frac{40}{\pi^{2}g_{rh}}\Big)
+\frac{1}{2}\ln\Big(8\pi^{2}{\cal{A}}_{s}{\cal{W}}_{s}
c_{s}^{3}\Big)-\frac{1}{3}\ln\Big(\frac{11g_{rh}}{43}\Big)\nonumber\\
-\frac{1}{4}\ln\bigg[V_{e}\bigg(\frac{1}{3}-\frac{64}{27}\kappa^{6}V_{e}^{2}\alpha_{e}'^{2}-\frac{8}{3}\kappa^{2}\alpha_{e}'
V_{e}'\bigg)^{-\frac{1}{2}}-\frac{64}{9}\kappa^{6}V_{e}^{2}\alpha_{e}'^{2}-\frac{8}{3}\kappa^{2}\alpha_{e}'
V_{e}'\bigg]\Bigg\}\,.
\end{eqnarray}
From equations \eqref{eq88}, \eqref{eq92} and \eqref{eq101}, we
get the following expression for the temperature during the
reheating process
\begin{eqnarray}\label{eq105}
T_{rh}=\Bigg(\frac{30}{\pi
g_{rh}}\Bigg)^{\frac{1}{4}}\left[V_{e}\Bigg(\frac{1}{3}-\frac{64}{27}\kappa^{6}V_{e}^{2}\alpha_{e}'^{2}-\frac{8}{3}\kappa^{2}\alpha_{e}'
V_{e}'\Bigg)^{-\frac{1}{2}}-\frac{64}{9}\kappa^{6}V_{e}^{2}\alpha_{e}'^{2}-\frac{8}{3}\kappa^{2}\alpha_{e}'
V_{e}'\right]\,\nonumber\\ \times\exp\bigg[-\frac{3}{4}N_{rh}(1+\omega_{eff})\bigg]\,.
\end{eqnarray}
In the following, and to perform a numerical analysis, we adopt the
potential and GB function used in section 6 and explore each case
separately.

\begin{table*}
\begin{small}
\caption{\small{\label{tab:13} Constraints on the e-folds number
and temperature during the reheating phase in the tachyon GB model
with $V=V_{0}\,\phi^{n}$ and ${\cal{G}}={\cal{G}}_{0}\,\phi^{-n}$,
obtained from Planck2018 TT, TE, EE+lowE+lensing+BK14+BAO joint
data.}}
\begin{center}
\begin{tabular}{cccccc}
\\ \hline \hline \\ && $\omega=-1$& $\omega=-\frac{1}{3}$
&$\omega=0$&$\omega=1$
\\
\hline
\\$n=2$&$0.251\leq\beta < 1$&$N_{rh}\leq5.65$&$N_{rh}\leq9.86$&$N_{rh}\leq19.62$&not consistent\\ \\
\hline
\\$n=4$&$0.254\leq\beta < 1$&$N_{rh}\leq4.73$&$N_{rh}\leq7.88$&$N_{rh}\leq15.87$&not consistent\\ \\
\hline  \hline \\
$n=2$&$0.251\leq\beta <
1$&$\log_{10}\left(\frac{T_{rh}}{GeV}\right)\geq12.59$&$\log_{10}\left(\frac{T_{rh}}{GeV}\right)\geq8.62$&
$\log_{10}\left(\frac{T_{rh}}{GeV}\right)\geq2.54$&not consistent\\ \\
\hline
\\$n=4$&$0.254\leq\beta < 1$&$\log_{10}\left(\frac{T_{rh}}{GeV}\right)\geq12.03$&$\log_{10}\left(\frac{T_{rh}}{GeV}\right)\geq9.42$&
$\log_{10}\left(\frac{T_{rh}}{GeV}\right)\geq15.87$&not consistent\\ \\
\hline \hline
\end{tabular}
\end{center}
\end{small}
\end{table*}

\begin{figure}[]
	\begin{center}
		\includegraphics[scale=0.32]{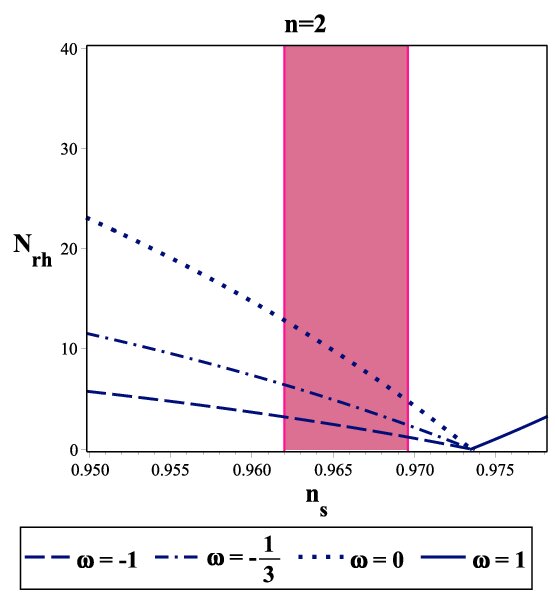}
		\includegraphics[scale=0.32]{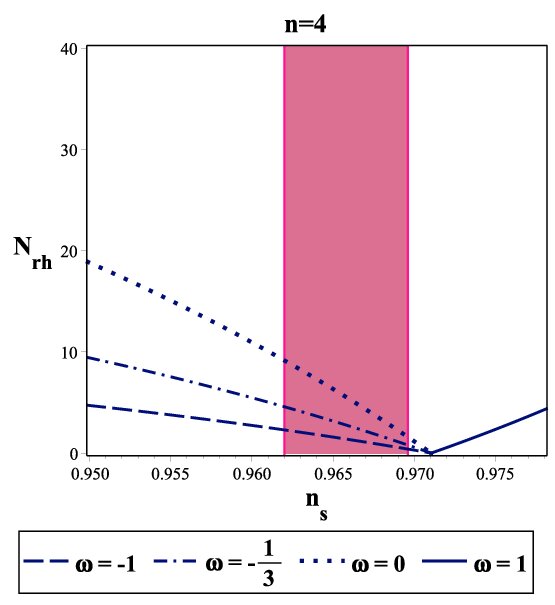}
		\includegraphics[scale=0.43]{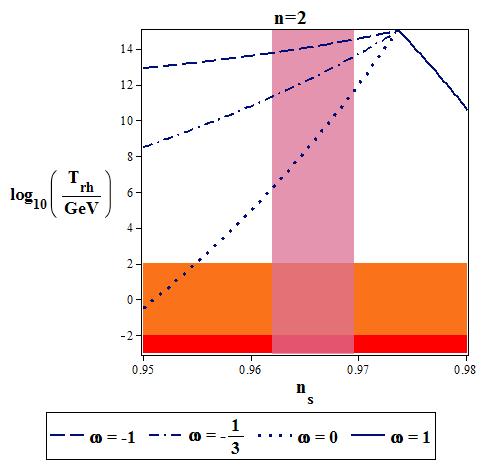}
		\includegraphics[scale=0.43]{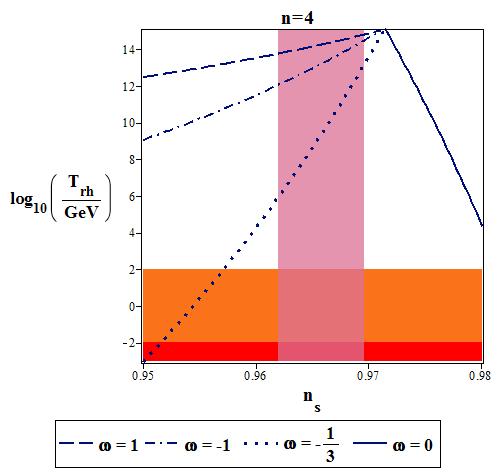}
		\includegraphics[scale=0.55]{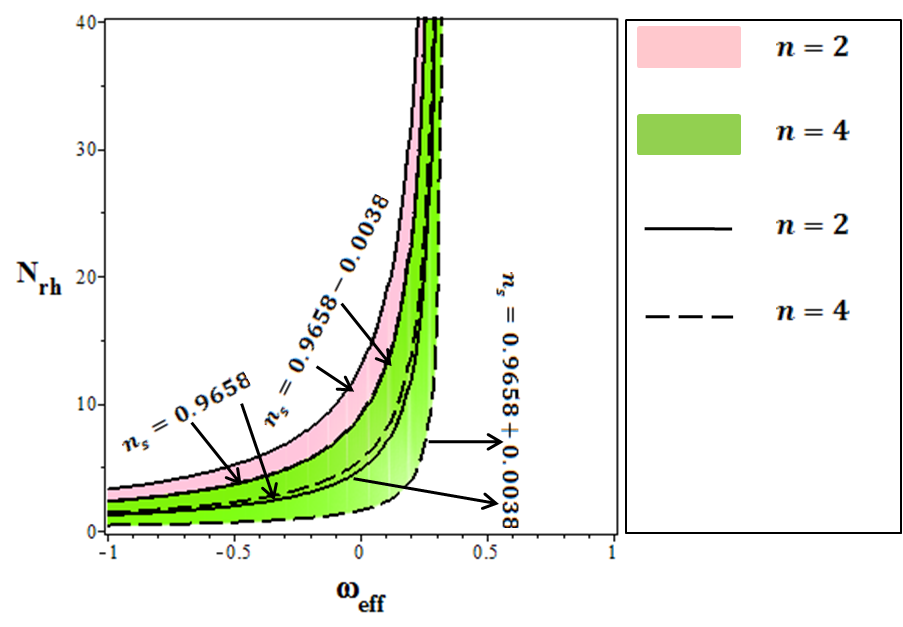}
	\end{center}
	\caption{\small Behavior of the e-folds number
		(top panels) and temperature (middle panels) during the
		reheating phase versus the scalar spectral index, and the range of
		$N_{rh}$ and $\omega_{eff}$ leading to the observationally viable
		values of the scalar spectral index (bottom panel), in the tachyon
		GB model with $V=V_{0}\,\phi^{n}$ and
		${\cal{G}}={\cal{G}}_{0}\,\phi^{-n}$.}
	\label{fig13}
\end{figure}

\subsection{Power-Law potential and Inverse Power-Law GB Coupling}
The first case we consider here is the tachyon GB model with
power-law potential and inverse power-law GB coupling (equation
\eqref{eq59}). We find the final values of these adopted functions
in terms of the scalar field at the horizon crossing and
substitute them in equations \eqref{eq104} and \eqref{eq105}.
Then, by obtaining the scalar field at horizon crossing in terms
of the scalar spectral index, we rewrite $N_{rh}$ and $T_{rh}$ in
terms of $n_{s}$ and perform a numerical analysis. As demonstrated
in Table 5, we have the tightest ranges on $\beta$ for $n=2$ as
$0.251\leq\beta<1$ and for $n=4$ as $0.254\leq\beta<1$. These
ranges of $\beta$ give the results summarized in Table 13.

The behavior of $N_{rh}$ and $T_{rh}$ versus $n_{s}$, for
$\beta=0.7$, has been shown in the top and middle panels of
Figure 13. As figure shows, the instantaneous reheating in this
case is corresponding to $n_{s}=0.971$ for $n=2$ and $n_{s}=0.974$
for $n=4$. Therefore, in this case the instantaneous reheating is
not observationally viable. The bottom panel of Figure 13 shows the
range of $N_{rh}$ and $\omega_{eff}$, in the case considered in
this subsection, leading to the observationally viable values of
the scalar spectral index.

\begin{figure}[]
	\begin{center}
		\includegraphics[scale=0.32]{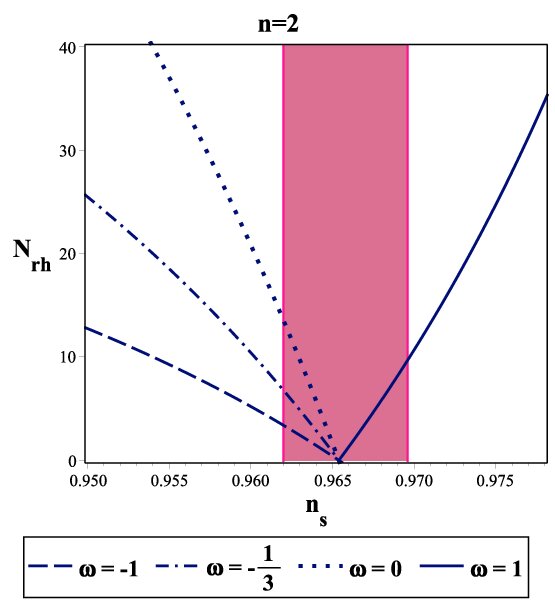}
		\includegraphics[scale=0.32]{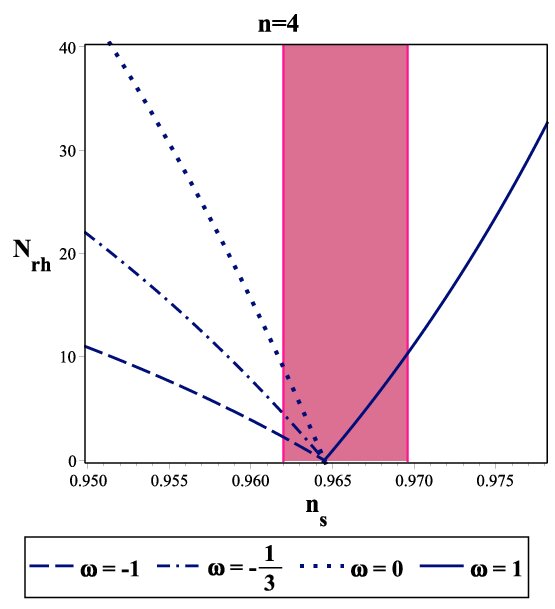}
		\includegraphics[scale=0.43]{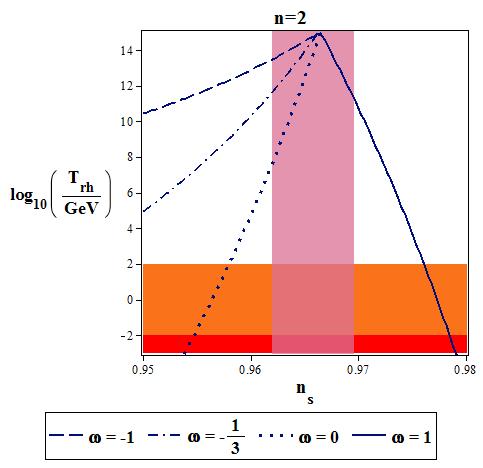}
		\includegraphics[scale=0.43]{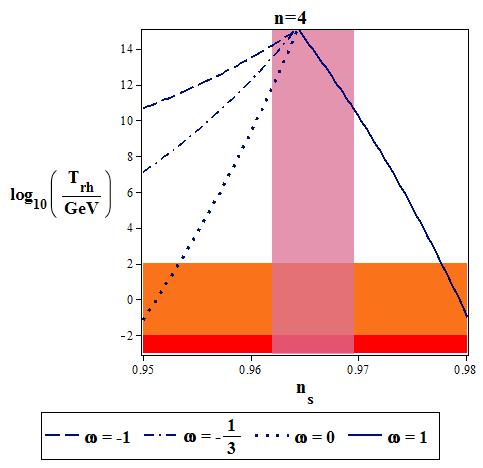}
		\includegraphics[scale=0.55]{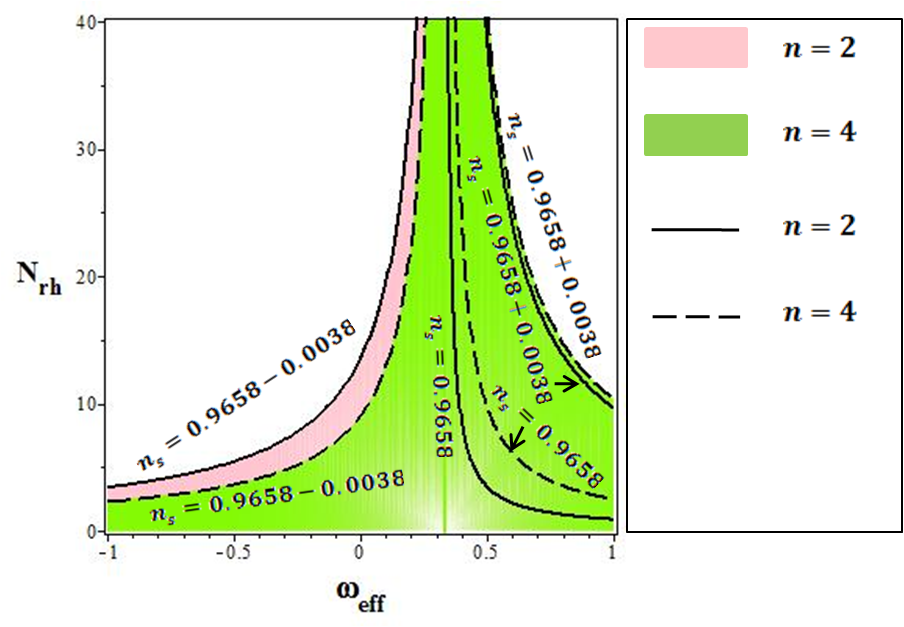}
	\end{center}
	\caption{\small Behavior of the e-folds number
		(top panels) and temperature (middle panels) during the
		reheating phase versus the scalar spectral index, and the range of
		$N_{rh}$ and $\omega_{eff}$ leading to the observationally viable
		values of the scalar spectral index (bottom panel), in the tachyon
		GB model with $V=V_{0}\,\phi^{n}$ and $V=V_{0}\,\phi^{n}$ and
		${\cal{G}}={\cal{G}}_{0}\,e^{-\lambda\phi}$.}
	\label{fig14}
\end{figure}

\begin{table*}
\begin{small}
\tiny\caption{\small{\label{tab:14} Constraints on the e-folds
number and temperature during the reheating phase in the tachyon
GB model with $V=V_{0}\,\phi^{n}$ and
${\cal{G}}={\cal{G}}_{0}\,e^{-\lambda\phi}$, obtained from
Planck2018 TT, TE, EE+lowE+lensing+BK14+BAO joint data.}}
\begin{center}
\begin{tabular}{cccccc}
\\ \hline \hline \\ &$\beta$& $\omega=-1$& $\omega=-\frac{1}{3}$
&$\omega=0$&$\omega=1$
\\
\hline \\ $n=2$&$0.023\leq\beta<
1$&$N_{rh}\leq5.73$&$N_{rh}\leq12.39$&$N_{rh}\leq20.04$&$N_{rh}\leq16.42$\\
\\ \hline \\
\\$n=4$&$0.046\leq\beta<1$&$N_{rh}\leq4.23$&$N_{rh}\leq8.11$&$N_{rh}\leq14.81$&$N_{rh}\leq17.76$\\ \\
\hline  \hline \\
$n=2$&$0.023\leq\beta<
1$&$\log_{10}\left(\frac{T_{rh}}{GeV}\right)\geq11.480$&$\log_{10}\left(\frac{T_{rh}}{GeV}\right)\geq7.41$&
$\log_{10}\left(\frac{T_{rh}}{GeV}\right)\geq3.11$&$\log_{10}\left(\frac{T_{rh}}{GeV}\right)\geq6.53$\\
\\ \hline \\
\\$n=4$&$0.046\leq\beta<1$&$\log_{10}\left(\frac{T_{rh}}{GeV}\right)\geq13.07$&$\log_{10}\left(\frac{T_{rh}}{GeV}\right)\geq11.90$&
$\log_{10}\left(\frac{T_{rh}}{GeV}\right)\geq8.91$&$\log_{10}\left(\frac{T_{rh}}{GeV}\right)\geq8.03$\\ \\
\hline \hline
\end{tabular}
\end{center}
\end{small}
\end{table*}

\subsection{Power-Law Potential and Dilaton-Like GB Coupling}
Now, by using equation \eqref{eq63}, we rewrite the e-folds number
and temperature during reheating in terms of the scalar spectral
index and perform a numerical analysis on the model. As we see
from Figure 6, this model in most ranges of the parameter's space
is consistent with observational data. In this case, for the
considered sample values of $\lambda$, all values of $\beta$ are
observationally viable. However if we adopt very large values of
$\lambda$, there would be some constraints on $\beta$. For
instance, we take $\lambda\sim 10^{5}$ and find $0.0023\leq
\beta<1$ for $n=2$ and $0.046\leq\beta<1$ for $n=4$. To analyze
the reheating phase numerically, we use this ranges of $\beta$
which lead to the constraints presented in Table 14.

The behavior of $N_{rh}$ and $T_{rh}$ versus $n_{s}$, for
$\beta=0.7$, has been shown in Figure 14. As figure shows, in this
case the instantaneous reheating is observationally viable. The
bottom panel of Figure 14 shows the range of $N_{rh}$ and
$\omega_{eff}$, in the case considered in this subsection, leading
to the observationally viable values of the scalar spectral index.

In summary, our study shows that in the tachyon GB model with
$V=V_{0}\,\phi^{n}$ and ${\cal{G}}={\cal{G}}_{0}\,\phi^{-n}$,
there is no chance to have $N_{rh}=0$ in a observationally viable
range. Therefore, considering that the reheating phase should
start with $N_{rh}=0$, this model is ruled out. Also, the tachyon
GB model with $V=V_{0}\,\phi^{n}$ and
${\cal{G}}={\cal{G}}_{0}\,e^{-\lambda\phi}$, predicts that by
increasing the value of $N_{rh}$, the value of $\omega_{eff}$ increases
until it reaches $\frac{1}{3}$. As before, the larger values of
$\omega_{eff}$ are not of interest.

\section{Reheating in the DBI-Gauss-Bonnet Model}

In this section, we study the reheating process in the DBI
Gauss-Bonnet model. Here also, we can use the equations
\eqref{eq88}-\eqref{eq93} of section 8, whereas, the energy density
and the equation of motion in the DBI GB model are different from
those in the GB model with canonical and also tachyon scalar fields.
In the DBI GB model we have
\begin{eqnarray}\label{eq106}
\rho=f^{-1}\Bigg(1+\frac{2}{3}\epsilon\big(1+fV\big)-\frac{64}{27}f\alpha'^{2}\kappa^{4}\Big(f^{-1}+V\Big)^{3}\Bigg)^{-\frac{1}{2}}
+V+\frac{8}{3}\alpha'\Big(f^{-1}+V\Big)\nonumber\\
\Bigg(f'f^{-2}-V'-\frac{8}{3}\kappa^{4}\alpha'\Big(f^{-1}+V\Big)^{2}\Bigg)\,.
\end{eqnarray}
We obtain the following expression at the end of inflation
($\epsilon=1$)
\begin{eqnarray}\label{eq107}
\rho_{e}=f_{e}^{-1}\Bigg(1+\frac{2}{3}\big(1+f_{e}V_{e}\big)
-\frac{64}{27}f_{e}\alpha_{e}'^{2}\kappa^{4}\Big(f_{e}^{-1}+V_{e}\Big)^{3}\Bigg)^{-\frac{1}{2}}
+V_{e}\nonumber\\+\frac{8}{3}\alpha_{e}'\Big(f_{e}^{-1}+V_{e}\Big)
\Bigg(f'_{e}f_{e}^{-2}-V_{e}'-\frac{8}{3}\kappa^{4}\alpha_{e}'\Big(f_{e}^{-1}+V_{e}\Big)^{2}\Bigg)\,.
\end{eqnarray}
Now, by using equations \eqref{eq88} and \eqref{eq107}, we find the
following expression for the energy density during the reheating
phase
\begin{eqnarray}\label{eq108}
\rho_{rh}=\Bigg[f_{e}^{-1}\Bigg(1+\frac{2}{3}\big(1+f_{e}V_{e}\big)
-\frac{64}{27}f_{e}\alpha_{e}'^{2}\kappa^{4}\Big(f_{e}^{-1}+V_{e}\Big)^{3}\Bigg)^{-\frac{1}{2}}
+V_{e}+\frac{8}{3}\alpha_{e}'\Big(f_{e}^{-1}+V_{e}\Big)\nonumber\\
\Bigg(f'_{e}f_{e}^{-2}-V_{e}'-\frac{8}{3}\kappa^{4}\alpha_{e}'\Big(f_{e}^{-1}+V_{e}\Big)^{2}\Bigg)\Bigg]
\times \exp\Big[-3N_{rh}(1+\omega_{eff})\Big].
\end{eqnarray}
From equations \eqref{eq93} and \eqref{eq108} we reach
\begin{eqnarray}\label{eq109}
\ln\left(\frac{a_{0}}{a_{rh}}\right)=-\frac{1}{3}\ln\left(\frac{43}{11g_{rh}}\right)
-\frac{1}{4}\ln\left(\frac{\pi^{2}g_{rh}}{30\rho_{rh}}\right)-\ln T_{0}-\frac{3}{4}N_{rh}(1+\omega_{eff})\nonumber\\
+\frac{1}{4}\ln\Bigg[f_{e}^{-1}\Bigg(1+\frac{2}{3}\big(1+f_{e}V_{e}\big)
-\frac{64}{27}f_{e}\alpha_{e}'^{2}\kappa^{4}\Big(f_{e}^{-1}+V_{e}\Big)^{3}\Bigg)^{-\frac{1}{2}}
+V_{e}+\frac{8}{3}\alpha_{e}'\Big(f_{e}^{-1}+V_{e}\Big)\nonumber\\
\Bigg(f'_{e}f_{e}^{-2}-V_{e}'-\frac{8}{3}\kappa^{4}\alpha_{e}'\Big(f_{e}^{-1}+V_{e}\Big)^{2}\Bigg)\Bigg]\,.
\end{eqnarray}
Then, by using \eqref{eq90} and \eqref{eq109}, we obtain the e-folds
number during reheating as follows
\begin{eqnarray}\label{eq110}
N_{rh}=\frac{4}{1-3\omega_{eff}}\Bigg\{-N_{hc}-\ln\Big(\frac{k_{hc}}{a_{0}T_{0}}\Big)-\frac{1}{4}\ln\Big(\frac{40}{\pi^{2}g_{rh}}\Big)
+\frac{1}{2}\ln\Big(8\pi^{2}{\cal{A}}_{s}{\cal{W}}_{s}
c_{s}^{3}\Big)-\frac{1}{3}\ln\Big(\frac{11g_{rh}}{43}\Big)\nonumber\\
-\frac{1}{4}\ln\bigg[f_{e}^{-1}\Bigg(1+\frac{2}{3}\big(1+f_{e}V_{e}\big)
-\frac{64}{27}f_{e}\alpha_{e}'^{2}\kappa^{4}\Big(f_{e}^{-1}+V_{e}\Big)^{3}\Bigg)^{-\frac{1}{2}}
+V_{e}+\frac{8}{3}\alpha_{e}'\Big(f_{e}^{-1}+V_{e}\Big)\nonumber\\
\Bigg(f'_{e}f_{e}^{-2}-V_{e}'-\frac{8}{3}\kappa^{4}\alpha_{e}'\Big(f_{e}^{-1}+V_{e}\Big)^{2}\Bigg)\bigg]\Bigg\}\,.
\end{eqnarray}
From equations \eqref{eq88}, \eqref{eq92} and \eqref{eq106}, we get
the temperature during reheating as follows
\begin{eqnarray}\label{eq111}
T_{rh}=\Bigg(\frac{30}{\pi
g_{rh}}\Bigg)^{\frac{1}{4}}\left[V_{e}\Bigg(\frac{1}{3}-\frac{64}{27}\kappa^{6}V_{e}^{2}\alpha_{e}'^{2}-\frac{8}{3}\kappa^{2}\alpha_{e}'
V_{e}'\Bigg)^{-\frac{1}{2}}-\frac{64}{9}\kappa^{6}V_{e}^{2}\alpha_{e}'^{2}-\frac{8}{3}\kappa^{2}\alpha_{e}'
V_{e}'\right]\,\nonumber\\
\times\exp\bigg[-\frac{3}{4}N_{rh}(1+\omega_{eff})\bigg]\,.
\end{eqnarray}
In the following, we study the reheating phase in the DBI GB model
numerically.

\subsection{Power-Law potential and Inverse Power-Law GB Coupling}
Here, we adopt power-law potential and inverse power-law GB
coupling (equation \eqref{eq74} and obtain the final values of
these adopted functions in terms of the scalar field at the
horizon crossing and substitute in equations \eqref{eq110} and
\eqref{eq111}. By obtaining the scalar field at the horizon
crossing in terms of the scalar spectral index, we rewrite
$N_{rh}$ and $T_{rh}$ in terms of $n_{s}$ and perform a numerical
analysis. As Table 7 shows, we have the tightest range on $\beta$
as $0.613\leq \beta\leq0.948$ for $n=2$ and
$0.683\leq\beta\leq0.915$  for $n=4$. In Table 15, we show the
observational constraints on $N_{rh}$ and $T_{rh}$, obtained from
the mentioned ranges of $\beta$.

In the top and middle panels of Figure 15 we see the behavior of
$N_{rh}$ and $T_{rh}$ versus $n_{s}$, for $\beta=0.7$. The
instantaneous reheating in this case, is not observationally
viable. The bottom panel of Figure 15 shows the range of $N_{rh}$
and $\omega_{eff}$, in the case considered in this subsection,
leading to the observationally viable values of the scalar
spectral index.

\begin{table*}
\tiny \caption{\small{\label{tab:15} Constraints on the e-folds
number and temperature during the reheating phase in the DBI GB
model with $V=V_{0}\,\phi^{n}$ and
${\cal{G}}={\cal{G}}_{0}\,\phi^{-n}$, obtained from Planck2018 TT,
TE, EE+lowE+lensing+BK14+BAO joint data.}}
\begin{center}
\begin{tabular}{ccccccc}
\\ \hline \hline \\ && $\omega=-1$& $\omega=-\frac{1}{3}$
&$\omega=0$&$\omega=1$
\\
\hline \\
\\$n=2$&$0.613\leq \beta\leq0.948$&$0.141<N_{rh}<9.08$&$0.411<N_{rh}<19.43$&$0.561<N_{rh}<30.04$&not consistent\\ \\
\hline \\
\\$n=4$&$0.683\leq\beta\leq0.915$&not consistent&not consistent&not consistent&$1.86<N<52.14$\\ \\
\hline  \hline \\
\\$n=2$&$0.613\leq \beta\leq0.948$&$11.86<\log_{10}\left(\frac{T_{rh}}{GeV}\right)<14.97$&$9.12<\log_{10}
\left(\frac{T_{rh}}{GeV}\right)<14.91$&
$6.11<\log_{10}\left(\frac{T_{rh}}{GeV}\right)<14.86$&not consistent\\ \\
\hline \\
\\$n=4$&$0.683\leq\beta\leq0.915$&not consistent&not consistent&
not consistent&$\log_{10}\left(\frac{T_{rh}}{GeV}\right)<7.42$\\ \\
\hline \hline
\end{tabular}
\end{center}
\end{table*}

\begin{figure}[]
	\begin{center}
		\includegraphics[scale=0.32]{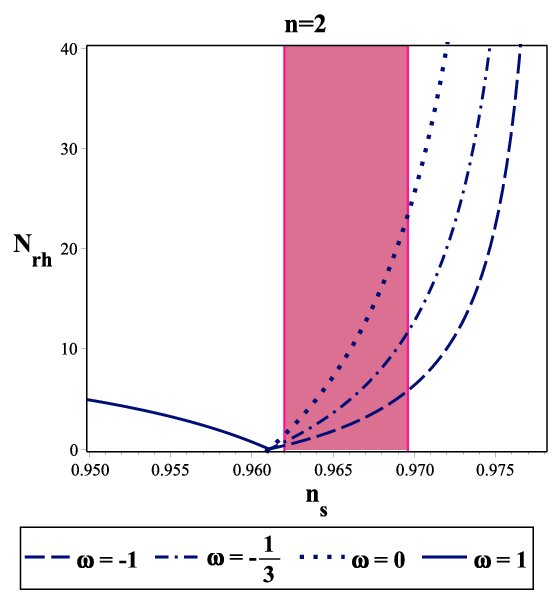}
		\includegraphics[scale=0.32]{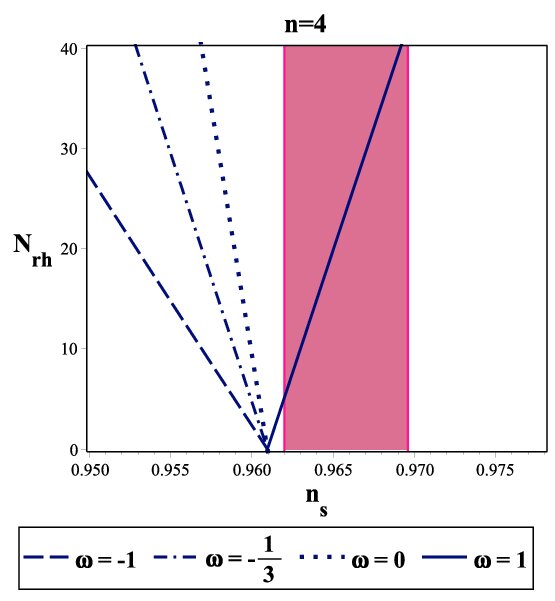}
		\includegraphics[scale=0.43]{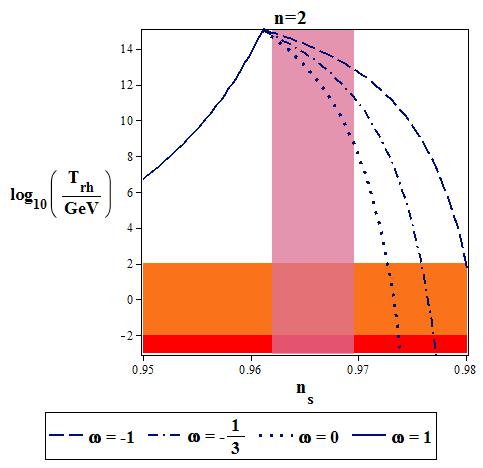}
		\includegraphics[scale=0.43]{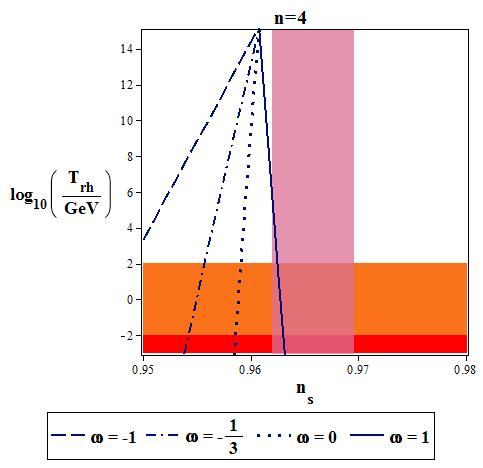}
		\includegraphics[scale=0.55]{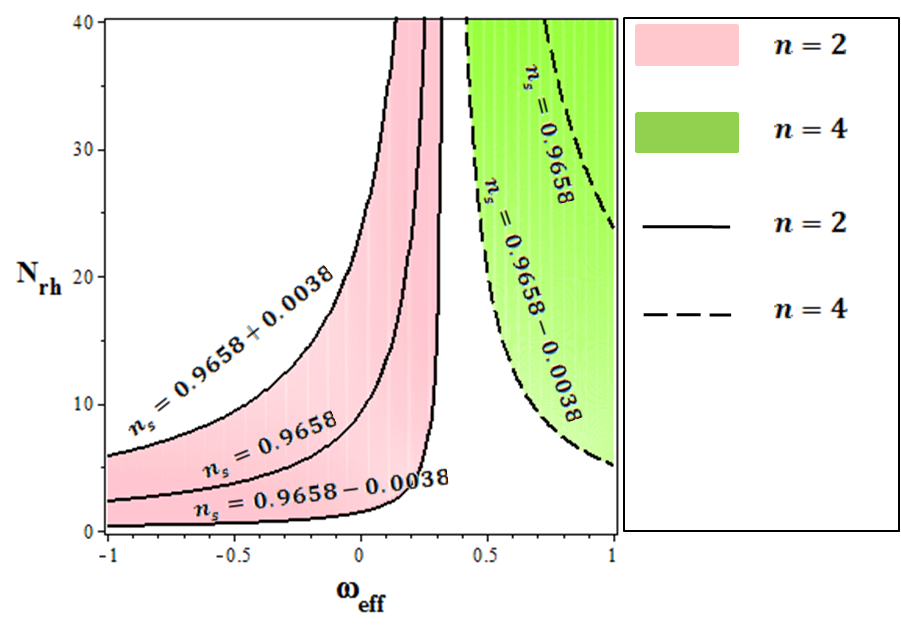}
	\end{center}
	\caption{\small Behavior of the e-folds number
		(top panels) and temperature (middle panels) during the
		reheating phase versus the scalar spectral index, and the range of
		$N_{rh}$ and $\omega_{eff}$ leading to the observationally viable
		values of the scalar spectral index (bottom panel), in the DBI GB
		model with $V=V_{0}\,\phi^{n}$ and
		${\cal{G}}={\cal{G}}_{0}\,\phi^{-n}$.}
	\label{fig15}
\end{figure}

\subsection{Power-Law Potential and Dilaton-Like GB Coupling}
Now, we consider the DBI Gauss-Bonnet model with power-law
potential and Dilaton-Like GB coupling (equation \eqref{eq80}).
This model in some ranges of the parameter's space is consistent
with observational data (see Table 8). In fact, ror $n=2$, we have
the tightest range on $\beta$ as $0.326\leq \beta\leq0.986$ and
for $n=4$ we have $0.404\leq\beta<1$. By these ranges of $\beta$
we perform numerical analysis which gives the constraints shown in
Table 16.

The behavior of $N_{rh}$ and $T_{rh}$ versus $n_{s}$, for
$\beta=0.6$, has been shown in Figure 16. As figure shows, with
$n=2$ the instantaneous reheating is no observationally viable and
with $n=4$ it is viable. The bottom panel of Figure 16 shows the
range of $N_{rh}$ and $\omega_{eff}$, in the case considered in
this subsection, leading to the observationally viable values of
the scalar spectral index.

According to these numerical considerations, in the DBI GB model
with $V=V_{0}\,\phi^{n}$ and ${\cal{G}}={\cal{G}}_{0}\,\phi^{-n}$,
for $n=2$, there is no chance to have observationally viable
$N_{rh}=0$. Therefore, this case is ruled out. For $n=4$, the
model doesn't predict $\omega_{eff}\leq \frac{1}{3}$ in the
observationally viable regions. So, this case is ruled out too. In
the DBI GB model with $V=V_{0}\,\phi^{n}$ and
${\cal{G}}={\cal{G}}_{0}\,e^{-\lambda\phi}$, the case with $n=2$
is not observationally viable. However, the case with $n=4$ is
consistent with observational data. In fact, this case also
predicts that by increasing the value of $N_{rh}$, the value of
$\omega_{eff}$ increases until it reaches $\frac{1}{3}$.

\begin{table}
\tiny \caption{\small{\label{tab:16} Constraints on the e-folds
number and temperature during the reheating phase in the DBI GB
model with $V=V_{0}\,\phi^{n}$ and
${\cal{G}}={\cal{G}}_{0}\,e^{-\lambda\phi}$, obtained from
Planck2018 TT, TE, EE+lowE+lensing+BK14+BAO joint data.}}
\begin{center}
\begin{tabular}{cccccc}
\\ \hline \hline \\ && $\omega=-1$& $\omega=-\frac{1}{3}$
&$\omega=0$&$\omega=1$
\\
\hline
\\$n=2$&$0.326\leq \beta\leq0.986$&$1.85\leq N_{rh}\leq11.93$&$4.21\leq N_{rh}\leq24.16$&$10.86\leq N_{rh}\leq41.22$&not
consistent\\ \\
\hline
\\$n=4$&$0.404\leq\beta<1$&$N\leq5.843$&$N\leq11.45$&$N\leq22.74$&$N\leq10.41$\\ \\
\hline  \hline
\\$n=2$&$0.326\leq \beta\leq0.986$&$9.03\leq\log_{10}\left(\frac{T_{rh}}{GeV}\right)\leq15.34$&$5.41\leq\log_{10}
\left(\frac{T_{rh}}{GeV}\right)\leq13.27$&
$0.121\leq\log_{10}\left(\frac{T_{rh}}{GeV}\right)\leq11.02$&not
consistent\\ \\
\hline
\\$n=4$&$0.404\leq\beta<1$&$\log_{10}\left(\frac{T_{rh}}{GeV}\right)\geq12.01$&$\log_{10}\left(\frac{T_{rh}}{GeV}\right)\geq7.868$&
$\log_{10}\left(\frac{T_{rh}}{GeV}\right)\geq0.442$&$\log_{10}\left(\frac{T_{rh}}{GeV}\right)\geq2.183$\\ \\
\hline \hline
\end{tabular}
\end{center}

\end{table}

\begin{figure}[]
	\begin{center}
		\includegraphics[scale=0.32]{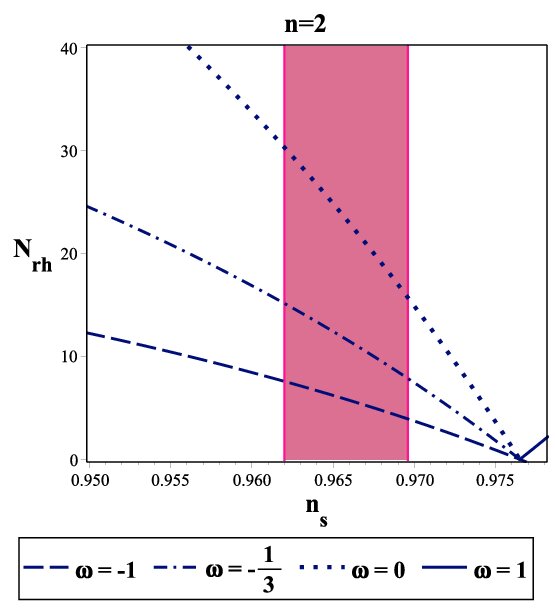}
		\includegraphics[scale=0.32]{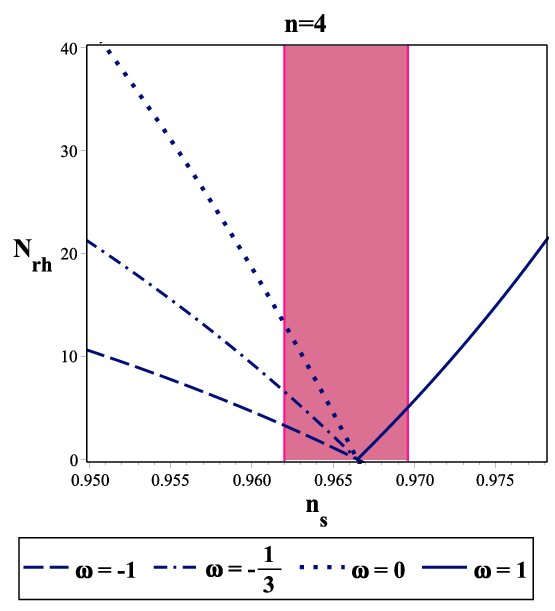}
		\includegraphics[scale=0.43]{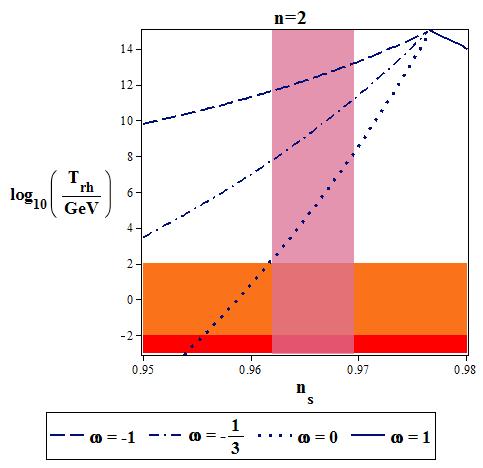}
		\includegraphics[scale=0.43]{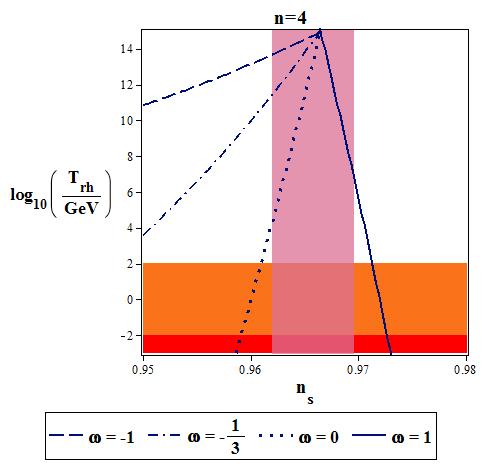}
		\includegraphics[scale=0.6]{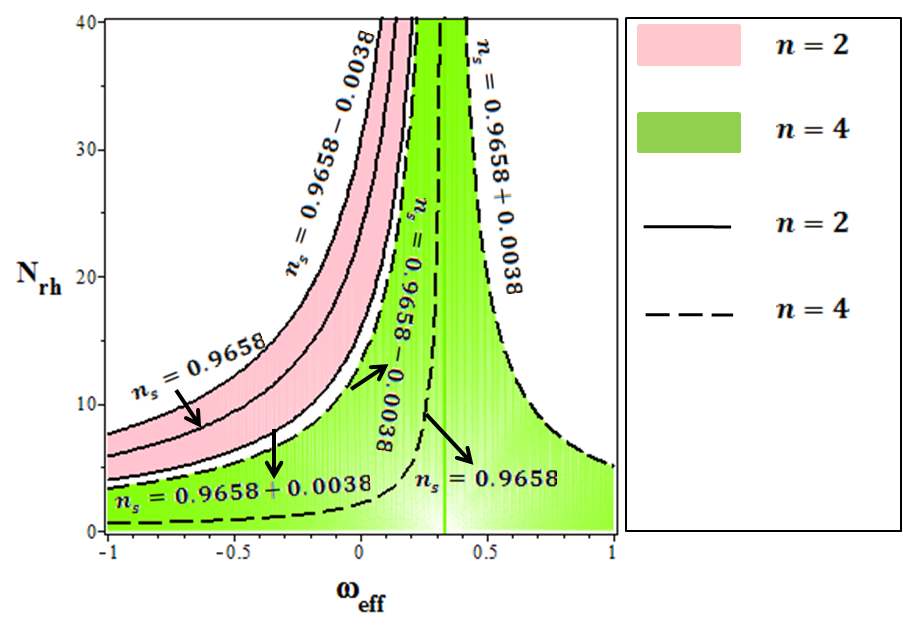}
	\end{center}
	\caption{\small Behavior of the e-folds number
		(top panels) and temperature (middle panels) during the
		reheating phase versus the scalar spectral index, and the range of
		$N_{rh}$ and $\omega_{eff}$ leading to the observationally viable
		values of the scalar spectral index (bottom panel), in the DBI GB
		model with $V=V_{0}\,\phi^{n}$ and
		${\cal{G}}={\cal{G}}_{0}\,e^{-\lambda\phi}$.}
	\label{fig16}
\end{figure}

\section{Summary}
In this paper, we have studied inflation and reheating in several
Gauss-Bonnet models. At first, we have considered a general GB
model and presented the main equations of the the model in the
inflation era. In this regard, we have obtained some important
perturbation parameters such as the scalar spectral index, tensor
spectral index and tensor-to-scalar ratio. Then, we have
considered several GB models and studied the perturbation
parameters numerically. By comparing the results with
observational data, we have obtained some constraints on the
model's parameter space. We have also analyzed the reheating epoch
in each model and explored the model's viability in this context
too. Our studies give the following results:

\begin{itemize}

\item Although the simple single filed inflation with $\phi^{2}$
potential is not consistent with base and base+GW data sets,
considering the GB effect with $\phi^{-2}$ coupling, makes the
model with $\phi^{2}$ potential observationally viable. In this
case, when the GB effect becomes larger, the tensor-to-scalar
ratio becomes smaller and lies in the base data region at $95\%$
CL. We have studied the cases with $N=50$, $N=60$ and $N=70$ for
both $r-n_{s}$ and $r-n_{T}$ trajectories. The constraint obtained
from these studied cases is as $0.680\leq \beta<1$. However, even
by including the GB effect, the model with $\phi^{4}$ potential is
not observationally viable. We have also analyzed the reheating
phase in the simple single filed inflation with $\phi^{2}$
potential. Our analysis shows that, in this model, it is possible
to have a viable reheating phase. In the simple single filed
inflation with $\phi^{2}$ potential, by increasing of $N_{rh}$,
the effective equation of state parameter changes from $-1$ and
reaches $\frac{1}{3}$ which is corresponding to a
radiation-dominated era.

\item Considering the GB effect with $e^{-\lambda\phi}$ coupling
causes the inflation models with both $\phi^{2}$ and $\phi^{4}$
potentials, in some ranges of the model's parameters, become
consistent with the base data at $95\%$ CL. In this case, have
studied the model with $N=50$, $\lambda=10$, $\lambda=10^{2}$ and
$\lambda=10^{4}$ for both $r-n_{s}$ and $r-n_{T}$ trajectories.
When we consider $n=2$ and $\lambda\lesssim 10^{2}$, we find the
constraint $0.020\leq \beta\leq 0.072$ which leads to
observational viability of both $r-n_{s}$ and $r-n_{T}$
trajectories. However, this constraint is not valid for $n=2$ and
$\lambda> 10^{2}$. By considering for $n=4$, we have $0.670\leq
\beta\leq 0.716$. Studying the reheating phase for both cases with
$n=2$ and $n=4$ shows that in these cases also, the effective
equation of state parameter changes from $-1$ and reaches
$\frac{1}{3}$, which is corresponding to a radiation-dominated
era.

\item Adding the GB effect to the natural inflation models, makes
it observationally viable. In this case also, the larger values of
$\beta$ lead to smaller values of the tensor-to-scalar ratio. By
exploring both $r-n_{s}$ and $r-n_{T}$ trajectories for $f=4$,
$f=15$, $f=35$ and $f=60$, we find $0.525\leq \beta <1$. For the
GB natural inflation, analyzing the reheating phase shows that
$\omega_{eff}$ in this epoch increases from $-1$ to $\frac{1}{3}$
which is corresponding to a radiation-dominated era. Therefore,
this model has capability to explain the reheating process after
inflation.

\item With both E-model and T-model potentials, the inflation
models are observationally viable, specially for pretty small
values of $\alpha$. When we consider the GB effect, for any values
of $\alpha$, the larger values of $\beta$ lead to the smaller
values of the tensor-to-scalar ratio. Therefore, the
$\alpha$-attractor GB inflation model is consistent with the base
and base+GW data sets at $95\%$ CL too. For both E-model and
T-model potentials, studying both $r-n_{s}$ and $r-n_{T}$
trajectories gives $\beta \leq 4.81 \times 10^{-2}$. In both
E-model and T-model cases, it is possible to explain the reheating
process. In both cases, the effective equation of state parameter
change from $-1$ to $\frac{1}{3}$.

\item Tachyon inflation with $\phi^{2}$ potential is consistent
with base data just for $N<52.7$. By considering the GB effect,
this model would be consistent with observational data for
$N<62.4$. Tachyon model with $\phi^{4}$ potential is not
consistent with the base and base+GW data sets at all. In this
case also, the GB effect causes the model becomes observationally
viable, for the considered range of $N$ as $50\leq N\leq70$. In
both cases, the viability arises because the GB effect makes the
tensor-to-scalar ratio of the model smaller. Exploring both
$r-n_{s}$ and $r-n_{T}$ trajectories gives $0.254 \leq \beta <1$
for $n=2$ and $0.188 \leq \beta <1$ for $n=4$. By studying the
reheating phase in the tachyon model with $\phi^{2}$ and
$\phi^{4}$ potentials, we have found that the reheating epoch cannot be explained in these models. This is because, in these models
the effective equation of state parameter doesn't reach
$\frac{1}{3}$ in an observationally viable range of the scalar
spectral index.

\item Considering the GB effect with $e^{-\lambda\phi}$ coupling
in the tachyon model leads to observationally viable tachyon model
with both $\phi^{2}$ and $\phi^{4}$ potentials. In this case, for
smaller values of $\lambda$ and larger values of $\beta$, we have
smaller values of tensor-to-scalar ratio which is consistent with
the base and base+GW data. This model with $n=2$, for $N=50$,
$\lambda=10$, $\lambda=10^{2}$ and $\lambda=10^{4}$ for all values
of $\beta$ is observationally viable. For $n=4$, and with $N=50$,
$\lambda=10$, $\lambda=10^{2}$ and $\lambda=10^{4}$, we have found
$0.046 \leq \beta $. We have studied the reheating phase in the
tachyon model with dilaton-like GB effect and found that in this
case the effective equation of state parameter reaches
$\frac{1}{3}$. This means that, this model has capability to
explain the reheating epoch.

\item DBI inflation with $\phi^{2}$ and $\phi^{4}$ potentials is
not consistent with the base and base+GW data sets. However,
considering the GB effect with $\phi^{-2}$ and $\phi^{-4}$
coupling functions makes the model observationally viable. Note
that, the DBI model with $\phi^{2}$ potential and $\phi^{-2}$ GB
coupling, in some ranges of the model's parameters is consistent
with observational data if $N\geq53.1$. Also, the DBI model with
$\phi^{4}$ potential and $\phi^{-4}$ GB coupling, in some ranges
of the model's parameters is consistent with observation if
$N\geq59.6$. Here also, we have studied the cases with $N=50$,
$N=60$ and $N=70$ for both $r-n_{s}$ and $r-n_{T}$ trajectories.
The constraints obtained from these studied cases are as
$0.613\leq \beta\leq 0.948$ for $n=2$ and $0.683\leq \beta\leq
0.915$ for $n=4$. Analyzing the reheating phase shows that, in
this case, we can not get the viable reheating process. In fact,
with both potentials the effective equation of state parameter
parameter doesn't reach $\frac{1}{3}$. Also, with $\phi^{4}$
potential, this parameter starts with positive values which is not
the case.

\item Considering the GB effect with $e^{-\lambda\phi}$ coupling
for the DBI model also, leads to the observational viability of
the model with both $\phi^{2}$ and $\phi^{4}$ potentials. As the
previous cases, the GB effects gives the smaller tensor-to-scalar
ratio which is consistent with the different data sets. By
exploring both $r-n_{s}$ and $r-n_{T}$ trajectories for $f=4$,
$f=15$, $f=35$ and $f=60$, we have found $0.332\leq \beta \leq
0.986$ for $n=2$ and $0.683\leq \beta \leq 0.948$ for $n=4$. The
DBI model with $\phi^{2}$ potential and dilaton-like GB coupling
looses its viability when we study the reheating phase. This is
because the effective equation of state parameter in this case,
doesn't reach $\frac{1}{3}$. However, the DBI model with
$\phi^{4}$ potential and dilaton-like GB coupling has capability
to explain the reheating phase.

\end{itemize}

{\bf Acknowledgement}\\
We thank the referee for the very insightful comments that have
improved the quality of the paper considerably. This work has been
supported financially by Research Institute for Astronomy $\&$
Astrophysics of Maragha (RIAAM) under research
project number 1/6025-8.\\

\end{document}